# Multiple tests of association with biological annotation metadata

**Sandrine Dudoit[1], Sündüz Keleş[2] and Mark J. van der Laan[1]**

*University of California, Berkeley, University of Wisconsin, Madison and University of California, Berkeley*

**Abstract:** We propose a general and formal statistical framework for multiple tests of association between known fixed features of a genome and unknown parameters of the distribution of variable features of this genome in a population of interest. The known gene-annotation profiles, corresponding to the fixed features of the genome, may concern Gene Ontology (GO) annotation, pathway membership, regulation by particular transcription factors, nucleotide sequences, or protein sequences. The unknown gene-parameter profiles, corresponding to the variable features of the genome, may be, for example, regression coefficients relating possibly censored biological and clinical outcomes to genome-wide transcript levels, DNA copy numbers, and other covariates. A generic question of great interest in current genomic research regards the detection of associations between biological annotation metadata and genome-wide expression measures. This biological question may be translated as the test of multiple hypotheses concerning association measures between gene-annotation profiles and gene-parameter profiles. A general and rigorous formulation of the statistical inference question allows us to apply the multiple hypothesis testing methodology developed in [*Multiple Testing Procedures with Applications to Genomics* (2008) Springer, New York] and related articles, to control a broad class of Type I error rates, defined as generalized tail probabilities and expected values for arbitrary functions of the numbers of Type I errors and rejected hypotheses. The resampling-based single-step and stepwise multiple testing procedures of [*Multiple Testing Procedures with Applications to Genomics* (2008) Springer, New York] take into account the joint distribution of the test statistics and provide Type I error control in testing problems involving general data generating distributions (with arbitrary dependence structures among variables), null hypotheses, and test statistics.

The proposed statistical and computational methods are illustrated using the acute lymphoblastic leukemia (ALL) microarray dataset of [*Blood* **103** (2004) 2771–2778], with the aim of relating GO annotation to differential gene expression between B-cell ALL with the BCR/ABL fusion and cytogenetically normal NEG B-cell ALL. The sensitivity of the identified lists of GO terms to the choice of association parameter between GO annotation and differential gene expression demonstrates the importance of translating the biological question in terms of suitable gene-annotation profiles, gene-parameter profiles, and association measures. In particular, the results reveal the limitations of binary gene-parameter profiles of differential expression indicators, which are still the norm for combined GO annotation and microarray data analyses. Procedures based on such binary gene-parameter profiles tend to be conservative and lack robustness with respect to the estimator for the set of differentially expressed genes. Our proposed statistical framework, with general definitions for the gene-annotation and gene-parameter profiles, allows consideration of a much broader class of inference problems, that extend beyond GO annotation and microarray data analysis.

---

[1]Division of Biostatistics and Department of Statistics, University of California, Berkeley, 101 Haviland Hall #7358, Berkeley, CA 94720-7358, USA, e-mail: sandrine@stat.berkeley.edu; laan@stat.berkeley.edu.

[2]Departments of Statistics and of Biostatistics & Medical Informatics, University of Wisconsin, Madison, 1300 University Avenue, 1245B Medical Sciences Center, Madison, WI 53705, USA, e-mail: keles@stat.wisc.edu.

**Contents**

## 1. Introduction

### *1.1. Motivation*

Experimental data, such as microarray gene expression measures, gain much in relevance from their association with *biological annotation metadata*, i.e., data on data, such as, GenBank sequences, Gene Ontology terms, KEGG pathways, and PubMed abstracts. A challenging and fascinating area of research for statisticians concerns the development of methods for relating experimental data to the wealth of metadata available publicly on the WWW. Tasks include accessing and pre-processing the data, making inference from these data, and summarizing and interpreting the results.

In this context, an important class of statistical problems involves testing for associations between known fixed features of a genome and unknown parameters of the distribution of variable features of this genome in a population of interest. Here, features of a genome are said to be *fixed*, if they remain constant among population units. In contrast, *variable* features are allowed to differ among population units. Fixed features typically consist of gene annotation metadata, that reflect current knowledge on gene properties, such as, nucleotide and protein sequences, regulation, and function. Variable features often consist of gene expression measures, that reflect cellular type/state/activity under particular conditions. The fixed and variable features define, respectively, *gene-annotation profiles* and *gene-parameter profiles*; the parameter of interest then corresponds to *measures of association between known gene-annotation profiles and unknown gene-parameter profiles*.

For instance, for the yeast *Saccharomyces cerevisiae* (in short, *S. cerevisiae*), one may be interested in detecting associations between the vector of mean transcript (i.e., mRNA) levels for all (approximately 6,500) genes under heat-shock conditions and *Gene Ontology* (GO) annotation for these genes. The reader is referred to the Gene Ontology Consortium website (www.geneontology.org) and to Section 4, below, for more information on gene ontologies, and to the Saccharomyces Genome Database (SGD) website (www.yeastgenome.org), for details on *S. cerevisiae*. In this example, the population of interest may consist of all heat-shocked cells from well-defined cultures of a particular strain of *S. cerevisiae* (e.g., strain S288C). For each of the three gene ontologies (BP, CC, and MF, as described in Section 4.1), each gene is annotated with a fixed set of GO terms (i.e., this set is constant across population units for a given version of the GO Database). Thus, for a given GO term, one may define a gene-annotation profile as a known, fixed binary vector indicating for each gene whether it is annotated or not with the particular GO term. The transcript levels, however, vary among population units and the gene-parameter profile, i.e., the vector of genome-wide mean transcript levels in the population of heat-shocked yeast cells, is unknown and may be estimated, for example, from a microarray experiment involving a sample of yeast cells from the population. The association parameter of interest, between GO annotation and transcript levels, is then a vector of association measures (e.g., two-sample *t*-statistics) between the known binary gene-annotation profiles and the unknown continuous gene-parameter profile.

Similar inference questions arise in many other contexts and involve a variety of definitions for the gene-annotation profiles, the gene-parameter profiles, and the association parameters of interest. For example, in cancer microarray studies, one may seek associations between GO gene-annotation profiles and a gene-parameter profile of regression coefficients relating (censored) patient survival data to genome-wide transcript levels or DNA copy numbers. Furthermore, gene-annotation profiles need not be binary or even polychotomous, and may correspond to pathway membership, regulation by particular transcription factors, nucleotide sequences, and protein sequences.

Note that, for the sake of illustration, we focus on gene-level features. However, our proposed methodology is generic and may be applied to other types of features, such as those concerning gene isoforms and proteins. For instance, as in alternative splicing microarray analysis, one may collect data at the finer level of gene isoforms, where one gene may have multiple isoforms [10]. In this context, *isoform-parameter profiles* may refer to the distribution of microarray isoform expression measures in a well-defined population, while *isoform-annotation profiles* may consist of exon/intron counts/lengths/nucleotide distributions. One may also consider protein-level features, where, for example, *protein-parameter profiles* correspond to antibody microarray expression measures and *protein-annotation profiles* refer to protein function, domain structure, and post-translational modification (e.g., Swiss-Prot, www.expasy.org/sprot).

## 1.2. Contrast with other approaches

Existing approaches for tests of association with biological annotation metadata focus primarily on relating microarray gene expression measures and GO annotation. Relevant articles and software packages include: `FatiGO` from the `BABELOMICS` suite ([1, 2]; www.babelomics.org); `GOstat` ([4]; gostat.wehi.edu.au); `Ontologizer` ([22]; www.charite.de/ch/medgen/ontologizer); `CSEPCT` ([28]; genome3.ucsf.edu:8080/cgi-bin/compareExp.cgi); `GSEA-P` ([29], [36]; www.broad.mit.edu/gsea/doc/doc_index.html); [37].

Methods proposed thus far suffer from a number of limitations, related, to a large extent, to the absence of a clear and precise statement of the statistical inference question. As a result, the analyses often lack statistical rigor and tend to be ad hoc and dataset-specific.

One of our main contributions is the systematic and precise translation of a general class of biological questions into a corresponding class of multiple hypothesis testing problems. A key step in this process is the proper definition of the gene-annotation profiles, gene-parameter profiles, and association parameters of interest. This general formulation then allows us to apply the multiple hypothesis testing methodology developed in [14] and related articles [8, 15, 16, 31, 32, 33, 34, 39, 40, 41, 42], to control a broad class of Type I error rates, defined as generalized tail probabilities (gTP), $gTP(q,g) = \Pr(g(V_n, R_n) > q)$, and generalized expected values (gEV), $gEV(g) = \mathrm{E}[g(V_n, R_n)]$, for arbitrary functions $g(V_n, R_n)$ of the numbers of false positives $V_n$ and rejected hypotheses $R_n$.

We wish to emphasize the crucial and often ignored distinction between: (i) *defining a parameter* of interest, measuring the association between gene-annotation and gene-parameter profiles, i.e., the statistical formulation of the biological question; (ii) *making inferences*, i.e., *estimating* and *testing hypotheses* concerning this parameter, based on a sample drawn from the population under consideration. Most methods proposed to date focus on (ii), without providing a clear statement of the question being answered in (i), that is, various estimation and testing procedures are proposed for an undefined parameter of interest.

Due to its general and rigorous statistical framework, our approach to multiple tests of association with biological annotation metadata differs in a number of important ways from current approaches, such as those developed for inference with Gene Ontology metadata and implemented in the software packages listed on the Gene Ontology Tools webpage (www.geneontology.org/GO.tools.shtml).

**General gene-annotation profiles.** Existing approaches typically consider binary gene-annotation profiles, e.g., vectors of indicators of GO term annotation. Our general definition of gene-annotation profiles allows consideration of arbitrary qualitative and quantitative fixed features of a genome, e.g., membership of genes to any number of pathways or clusters, exon/intron counts/lengths/nucleotide distributions, mean transcript levels.

**General gene-parameter profiles.** Existing approaches typically consider binary gene-parameter profiles, e.g., vectors of indicators of differential expression. Our general definition of gene-parameter profiles allows consideration of a much broader class of testing problems, concerning arbitrary qualitative and quantitative parameters, such as, differences in mean expression levels or regression coefficients relating expression levels to clinical outcomes.

**Estimated gene-parameter profiles.** Existing approaches typically assume known gene-parameter profiles. For example, the list of differentially expressed genes from a microarray experiment is usually treated as known and fixed in subsequent analyses with GO, while in fact it corresponds to an unknown and estimated parameter. Distinguishing between the definition of a parameter and inference concerning this parameter, as in Section 3, provides a more rigorous and general formulation of the statistical question.

**General tests of association.** Common approaches to tests of association with GO annotation are typically limited to tests of independence in $2 \times 2$ contingency tables (e.g., based on the hypergeometric distribution, Fisher's exact test). As in Table 2, rows correspond to gene annotation with a given GO term (fixed binary gene-annotation profile) and columns to a gene property of interest, such as differential expression (treated as a fixed binary gene-parameter profile). The approach proposed in Section 3 allows consideration of a broader class of biological testing problems, while properly accounting for the fact that gene-parameter profiles are usually unknown and replaced by a random (i.e., data-driven) estimator.

### *1.3. Outline*

This article proposes a general and formal statistical framework for multiple tests of association with biological annotation metadata, using the multiple hypothesis testing methodology developed in [14] and related articles.

Section 2 provides an introduction to multiple hypothesis testing. Section 3 presents the proposed statistical framework for multiple tests of association with biological annotation metadata and discusses in detail the main components of the inference problem, namely, the gene-annotation profiles, the gene-parameter profiles, and the association parameters. Multiple testing procedures (MTP) for tests of association between gene-annotation profiles and gene-parameter profiles are outlined. Section 4 gives an overview of the Gene Ontology (GO) and R software for accessing and analyzing GO annotation metadata (e.g., for assembling GO gene-annotation profiles). The proposed statistical and computational methods are illustrated in Section 5, using the acute lymphoblastic leukemia (ALL) microarray dataset of [13], with the aim of relating GO annotation to differential gene expression between B-cell ALL with the BCR/ABL fusion and cytogenetically normal NEG B-cell ALL. Finally, Section 6 summarizes our findings and outlines ongoing work.

## **2. Multiple hypothesis testing**

This section, based on Chapter 1 of [14], introduces a general statistical framework for multiple hypothesis testing and summarizes in turn the main ingredients of a multiple testing problem, including: the data generating distribution; the parameters of interest; the null and alternative hypotheses; the test statistics; the rejection regions (i.e., cut-offs) for the test statistics; errors in multiple hypothesis testing; Type I error rate and power; adjusted *p*-values.

The section also provides an overview of multiple testing procedures developed in [14, Chapters 2–7], for controlling generalized tail probability and expected value error rates, including the key choices of a joint null distribution and rejection regions for the test statistics.

The reader is referred to our book and articles for further detail on the multiple testing methodology, its software implementation, and its application to a variety of testing problems in biomedical and genomic research [5, 6, 7, 8, 9, 14, 15, 16, 26, 31, 32, 33, 34, 39, 40, 41, 42].

## *2.1. Null and alternative hypotheses*

*Hypothesis testing* is concerned with using observed data to make decisions regarding properties of (i.e., hypotheses for) the unknown data generating distribution.

Let $\mathcal{X}_n \equiv \{X_i : i = 1, \ldots, n\}$ denote a *random sample* of $n$ independent and identically distributed (IID) random variables from a *data generating distribution* $P$, i.e., $X_i \overset{IID}{\sim} P$, $i = 1, \ldots, n$. Suppose that the data generating distribution $P$ is an element of a particular statistical *model* $\mathcal{M}$, i.e., a set of possibly non-parametric distributions. Let $P_n$ denote the *empirical distribution* corresponding to the sample $\mathcal{X}_n$, i.e., the distribution which places probability $1/n$ on each realization of $X$.

In order to cover a broad class of testing problems, define $M$ pairs of null and alternative hypotheses in terms of a collection of $M$ *submodels*, $\mathcal{M}(m) \subseteq \mathcal{M}$, $m = 1, \ldots, M$, for the data generating distribution $P$. Specifically, the $M$ *null hypotheses* and corresponding *alternative hypotheses* are defined as

$$(1) \qquad H_0(m) \equiv \mathrm{I}\left(P \in \mathcal{M}(m)\right) \quad \text{and} \quad H_1(m) \equiv \mathrm{I}\left(P \notin \mathcal{M}(m)\right),$$

respectively.

The general submodel representation accommodates tests of means, quantiles, covariances, correlation coefficients, and regression coefficients in linear and non-linear models (e.g., logistic, survival, time-series models).

In many testing problems, the submodels concern *parameters*, i.e., functions $\Psi(P) = \psi = (\psi(m) : m = 1, \ldots, M) \in \mathbb{R}^M$ of the data generating distribution $P$, and each null hypothesis $H_0(m)$ refers to a single parameter, $\psi(m) = \Psi(P)(m) \in \mathbb{R}$. One distinguishes between two types of testing problems for such parametric hypotheses: *one-sided tests*,

$$(2) \qquad H_0(m) = \mathrm{I}\left(\psi(m) \leq \psi_0(m)\right)$$
$$\text{vs. } H_1(m) = \mathrm{I}\left(\psi(m) > \psi_0(m)\right), \quad m = 1, \ldots, M,$$

and *two-sided tests*,

$$(3) \qquad H_0(m) = \mathrm{I}\left(\psi(m) = \psi_0(m)\right)$$
$$\text{vs. } H_1(m) = \mathrm{I}\left(\psi(m) \neq \psi_0(m)\right), \quad m = 1, \ldots, M.$$

The hypothesized *null values*, $\psi_0(m)$, are frequently zero.

Let

$$(4) \qquad \mathcal{H}_0 = \mathcal{H}_0(P) \equiv \{m : H_0(m) = 1\} = \{m : P \in \mathcal{M}(m)\}$$

denote the set of $h_0 \equiv |\mathcal{H}_0|$ *true null hypotheses*, where the longer notation $\mathcal{H}_0(P)$ emphasizes the dependence of this set on the data generating distribution $P$. Likewise, let

$$(5) \qquad \mathcal{H}_1 = \mathcal{H}_1(P) \equiv \{m : H_1(m) = 1\} = \{m : P \notin \mathcal{M}(m)\} = \mathcal{H}_0^c(P)$$

be the set of $h_1 \equiv |\mathcal{H}_1| = M - h_0$ *false null hypotheses*.

The goal of a multiple testing procedure is to accurately estimate, i.e., *reject*, the set $\mathcal{H}_1$, while probabilistically controlling false positives.

## 2.2. Test statistics

A *testing procedure* is a *random* or *data-driven rule* for estimating the set of false null hypotheses $\mathcal{H}_1 = \{m : H_0(m) = 0\} = \{m : P \notin \mathcal{M}(m)\}$, i.e., for deciding which null hypotheses should be *rejected*.

The decisions to reject or not the null hypotheses are based on an $M$-vector of *test statistics*, $T_n = (T_n(m) : m = 1, \ldots, M)$, that are functions $T_n(m) = T(m; \mathcal{X}_n) = T(m; P_n)$ of the data $\mathcal{X}_n$, i.e., of the empirical distribution $P_n$. Denote the typically unknown (finite sample) *joint distribution of the test statistics* $T_n$ by $Q_n = Q_n(P)$.

As in [14, Chapter 1], for the test of single-parameter null hypotheses of the form $H_0(m) = \mathrm{I}\,(\psi(m) \leq \psi_0(m))$ or $H_0(m) = \mathrm{I}\,(\psi(m) = \psi_0(m))$, $m = 1, \ldots, M$, consider two main types of test statistics, *difference statistics*,

$$(6) \qquad T_n(m) \equiv \text{Estimator} - \text{Null value} = \sqrt{n}(\psi_n(m) - \psi_0(m)),$$

and *t-statistics* (i.e., standardized differences),

$$(7) \qquad T_n(m) \equiv \frac{\text{Estimator} - \text{Null value}}{\text{Standard error}} = \sqrt{n}\frac{\psi_n(m) - \psi_0(m)}{\sigma_n(m)}.$$

Here, $\hat{\Psi}(P_n) = \psi_n = (\psi_n(m) : m = 1, \ldots, M)$ denotes an *estimator* for the parameter $\Psi(P) = \psi = (\psi(m) : m = 1, \ldots, M)$ and $(\sigma_n(m)/\sqrt{n} : m = 1, \ldots, M)$ denote the estimated *standard errors* for elements $\psi_n(m)$ of $\psi_n$.

Further suppose an *asymptotically linear* estimator $\psi_n$ of the parameter $\psi$, with $M$-dimensional vector *influence curve* (IC) $IC(X|P) = (IC(X|P)(m) : m = 1, \ldots, M)$, such that

$$(8) \qquad \psi_n(m) - \psi(m) = \frac{1}{n}\sum_{i=1}^{n} IC(X_i|P)(m) + o_P(1/\sqrt{n})$$

and $\mathrm{E}[IC(X|P)(m)] = 0$, for each $m = 1, \ldots, M$. Let $\Sigma(P) = \sigma = (\sigma(m, m') : m, m' = 1, \ldots, M)$ denote the $M \times M$ covariance matrix of the vector influence curve $IC(X|P)$, where $\sigma(m, m') \equiv \mathrm{E}[IC(X|P)(m)IC(X|P)(m')]$ and we may adopt the shorter notation $\sigma^2(m) = \sigma(m, m) = \mathrm{E}[IC^2(X|P)(m)]$ for variances. Assume that $\sigma_n^2(m)$ are *consistent* estimators of the IC variances $\sigma^2(m)$.

The influence curve of a given estimator can be derived as its mean-zero-centered functional derivative (as a function of the empirical distribution $P_n$ for the entire sample of size $n$), applied to the empirical distribution for a sample of size one [19, 20].

This general representation for the test statistics covers standard one-sample and two-sample $t$-statistics for testing hypotheses concerning mean parameters, but also test statistics for correlation coefficients and regression coefficients in linear and nonlinear models. Test statistics for other types of null hypotheses include $F$-statistics, $\chi^2$-statistics, and likelihood ratio statistics.

## 2.3. Rejection regions

A *multiple testing procedure* (MTP) provides *rejection regions* $\mathcal{C}_n(m)$, i.e., sets of values for each test statistic $T_n(m)$ that lead to the decision to reject the corresponding null hypothesis $H_0(m)$ and declare that $P \notin \mathcal{M}(m)$, $m = 1, \ldots, M$. In other words, a MTP produces a random (i.e., data-dependent) set $\mathcal{R}_n$ of rejected hypotheses that estimates the set $\mathcal{H}_1$ of false null hypotheses,

$$(9) \quad \mathcal{R}_n = \mathcal{R}(T_n, Q_{0n}, \alpha) \equiv \{m : T_n(m) \in \mathcal{C}_n(m)\} = \{m : H_0(m) \text{ is rejected}\},$$

where $\mathcal{C}_n(m) = \mathcal{C}(m; T_n, Q_{0n}, \alpha)$, $m = 1, \ldots, M$, denote possibly random rejection regions.

The long notation $\mathcal{R}(T_n, Q_{0n}, \alpha)$ and $\mathcal{C}(m; T_n, Q_{0n}, \alpha)$ emphasizes that the MTP depends on the following three ingredients:

- the *data*, $\mathcal{X}_n = \{X_i : i = 1, \ldots, n\}$, through the $M$-vector of *test statistics*, $T_n = (T_n(m) : m = 1, \ldots, M)$ (Section 2.2);
- an (estimated) $M$-variate *test statistics null distribution*, $Q_{0n}$, which replaces the unknown true test statistics distribution $Q_n = Q_n(P)$ for the purpose of deriving rejection regions, confidence regions, and adjusted $p$-values (Section 2.7);
- the *nominal Type I error level* $\alpha$, i.e., a user-supplied upper bound for a suitably defined Type I error rate (Section 2.5).

Given a proper test statistics null distribution $Q_0$ (or estimator thereof, $Q_{0n}$), the main task is to specify rejection regions for each null hypothesis, so that the resulting procedure probabilistically controls Type I errors. We consider MTPs based on *nested* rejection regions, that is,

$$(10) \quad \mathcal{C}(m; T_n, Q_{0n}, \alpha_1) \subseteq \mathcal{C}(m; T_n, Q_{0n}, \alpha_2), \qquad \text{whenever } \alpha_1 \leq \alpha_2.$$

Rejection regions are typically defined in terms of intervals, such as, $\mathcal{C}_n(m) = (u_n(m), +\infty)$, $\mathcal{C}_n(m) = (-\infty, l_n(m))$, or $\mathcal{C}_n(m) = (-\infty, l_n(m)) \cup (u_n(m), +\infty)$, where $l_n(m) = l(m; T_n, Q_{0n}, \alpha)$ and $u_n(m) = u(m; T_n, Q_{0n}, \alpha)$ are to-be-determined lower and upper *critical values*, or *cut-offs*, computed under the null distribution $Q_{0n}$ for the test statistics $T_n$. Two-sided rejection regions of the form $\mathcal{C}_n(m) = (-\infty, l_n(m)) \cup (u_n(m), +\infty)$ allow the use of asymmetric cut-offs for two-sided tests.

Unless specified otherwise, we assume that large values of the test statistic $T_n(m)$ provide evidence against the corresponding null hypothesis $H_0(m)$, that is, we consider one-sided rejection regions of the form $\mathcal{C}_n(m) = (c_n(m), +\infty)$, where $c_n(m) = c(m; T_n, Q_{0n}, \alpha)$. For two-sided tests of single-parameter null hypotheses using difference or $t$-statistics, as in Equations (6) and (7), one could take absolute values of the test statistics.

Among the different approaches for defining rejection regions, we distinguish the following.

**Marginal vs. joint multiple testing procedures.** In *marginal* multiple testing procedures, rejection regions are based solely on the marginal distributions of the test statistics (e.g., FWER-controlling single-step Bonferroni procedure [12]). In contrast, *joint* procedures take into account the dependence structure of the test statistics (e.g., FWER-controlling single-step maxT Procedure 1). Joint MTPs tend to be more powerful than marginal MTPs.

Note that while a procedure may be marginal, proof of Type I error control by this MTP may require certain assumptions on the dependence structure of the test statistics (e.g., FWER-controlling step-up Hochberg procedure [23]).

**Single-step vs. stepwise multiple testing procedures.** In *single-step* multiple testing procedures, each null hypothesis $H_0(m)$ is tested using a rejection region that is independent of the results of the tests of other hypotheses and is not a function of the data $\mathcal{X}_n$ (unless these data are used to estimate the null distribution). In contrast, in *stepwise* procedures, the decision to reject a particular null hypothesis depends on the outcome of the tests of other hypotheses. That is, the (single-step) testing procedure is applied to a *sequence of successively smaller nested random (i.e., data-dependent) subsets of null hypotheses*, defined by the *ordering* of the test statistics (common-cut-off MTPs) or unadjusted $p$-values (common-quantile MTPs). The rejection regions are therefore allowed to depend on the data $\mathcal{X}_n$ via the test statistics $T_n$.

Stepwise procedures are of two main types, depending on the order in which the null hypotheses are tested. In *step-down* procedures, the *most significant* null hypotheses (i.e., the null hypotheses with the largest test statistics for common-cut-off MTPs or smallest unadjusted $p$-values for common-quantile MTPs) are considered successively, with further tests depending on the outcome of earlier ones. As soon as one fails to reject a null hypothesis, no further hypotheses are rejected. In contrast, for *step-up* procedures, the *least significant* null hypotheses are considered successively, again with further tests depending on the outcome of earlier ones. As soon as one null hypothesis is rejected, all remaining more significant hypotheses are rejected.

Stepwise MTPs tend to be more powerful than single-step MTPs.

**Common-cut-off vs. common-quantile multiple testing procedures.** In *common-cut-off* multiple testing procedures, the same cut-off $c_0$ is used for each test statistic (e.g., FWER-controlling single-step and step-down maxT procedures, based on maxima of test statistics). In contrast, in *common-quantile* procedures, the cut-offs are the $\delta_0$-quantiles of the marginal null distributions of the test statistics (e.g., FWER-controlling single-step and step-down minP procedures, based on minima of unadjusted $p$-values).

The latter $p$-value-based procedures place the null hypotheses on an "equal footing", i.e., are more balanced than their common-cut-off counterparts, and may therefore be preferable. However, this comes at the expense of increased computational complexity.

### 2.4. Errors in multiple hypothesis testing

In any testing problem, two types of errors can be committed. A *Type I error*, or *false positive*, is committed by rejecting a true null hypothesis ($\mathcal{R}_n \cap \mathcal{H}_0$). A *Type II error*, or *false negative*, is committed by failing to reject a false null hypothesis ($\mathcal{R}_n^c \cap \mathcal{H}_1$).

The situation can be summarized as in Table 1, where the number of rejected null hypotheses is

$$(11) \qquad R_n \equiv |\mathcal{R}_n| = \sum_{m=1}^{M} \mathrm{I}\left(T_n(m) \in \mathcal{C}_n(m)\right),$$

the number of Type I errors or false positives is

$$(12) \qquad V_n \equiv |\mathcal{R}_n \cap \mathcal{H}_0| = \sum_{m \in \mathcal{H}_0} \mathrm{I}\left(T_n(m) \in \mathcal{C}_n(m)\right),$$

the number of Type II errors or false negatives is

$$(13) \qquad U_n \equiv |\mathcal{R}_n^c \cap \mathcal{H}_1| = \sum_{m \in \mathcal{H}_1} \mathrm{I}\left(T_n(m) \notin \mathcal{C}_n(m)\right),$$

TABLE 1. *Type I and Type II errors in multiple hypothesis testing.* This table summarizes the different types of decisions and errors in multiple hypothesis testing. The number of rejected null hypotheses is $R_n = |\mathcal{R}_n|$, the number of Type I errors or false positives is $V_n = |\mathcal{R}_n \cap \mathcal{H}_0|$, the number of Type II errors or false negatives is $U_n = |\mathcal{R}_n^c \cap \mathcal{H}_1|$, the number of true negatives is $W_n = |\mathcal{R}_n^c \cap \mathcal{H}_0|$, and the number of true positives is $S_n = |\mathcal{R}_n \cap \mathcal{H}_1|$. Cells corresponding to errors are enclosed in boxes.

|  |  | Null hypotheses | | |
|---|---|---|---|---|
|  |  | Non-rejected, $\mathcal{R}_n^c$ | Rejected, $\mathcal{R}_n$ |  |
| **Null hypotheses** | **True, $\mathcal{H}_0$** | $W_n = |\mathcal{R}_n^c \cap \mathcal{H}_0|$ | $V_n = |\mathcal{R}_n \cap \mathcal{H}_0|$ | $h_0$ |
|  | **False, $\mathcal{H}_1$** | $U_n = |\mathcal{R}_n^c \cap \mathcal{H}_1|$ | $S_n = |\mathcal{R}_n \cap \mathcal{H}_1|$ | $h_1$ |
|  |  | $M - R_n$ | $R_n$ | $M$ |

the number of *true negatives* is

$$(14) \quad W_n \equiv |\mathcal{R}_n^c \cap \mathcal{H}_0| = \sum_{m \in \mathcal{H}_0} \mathrm{I}\left(T_n(m) \notin \mathcal{C}_n(m)\right) = M - R_n - U_n = h_0 - V_n,$$

and the number of *true positives* is

$$(15) \quad S_n \equiv |\mathcal{R}_n \cap \mathcal{H}_1| = \sum_{m \in \mathcal{H}_1} \mathrm{I}\left(T_n(m) \in \mathcal{C}_n(m)\right) = R_n - V_n = h_1 - U_n.$$

Note that $S_n$, $U_n$, $V_n$, and $W_n$ each depend on the unknown data generating distribution $P$ through the unknown set of true null hypotheses $\mathcal{H}_0 = \mathcal{H}_0(P)$. Therefore, the numbers $h_0 = |\mathcal{H}_0|$ and $h_1 = |\mathcal{H}_1| = M - h_0$ of true and false null hypotheses are *unknown parameters* (row margins of Table 1), the number of rejected hypotheses $R_n$ is an *observable random variable* (column margins of Table 1), and $S_n$, $U_n$, $V_n$, and $W_n$ are *unobservable random variables* (cells of Table 1).

Ideally, one would like to simultaneously minimize both the number of Type I errors and the number of Type II errors. Unfortunately, this is not feasible and one seeks a *trade-off* between the two types of errors. A standard approach is to specify an acceptable level $\alpha$ for a suitably defined Type I error rate and derive testing procedures (i.e., rejection regions) that aim to minimize a Type II error rate (i.e., maximize power) within the class of tests with Type I error level at most $\alpha$.

### 2.5. Type I error rate and power

When testing multiple hypotheses, there are many possible definitions for the Type I error rate and power of a testing procedure. Accordingly, we define a *Type I error rate* as a parameter $\theta_n = \Theta(F_{V_n,R_n})$ of the joint distribution $F_{V_n,R_n}$ of the numbers of Type I errors $V_n = |\mathcal{R}_n \cap \mathcal{H}_0|$ and rejected hypotheses $R_n = |\mathcal{R}_n|$. Likewise, we define *power* as a parameter $\vartheta_n = \Theta(F_{U_n,R_n})$ of the joint distribution $F_{U_n,R_n}$ of the numbers of Type II errors $U_n = |\mathcal{R}_n^c \cap \mathcal{H}_1|$ and rejected hypotheses $R_n = |\mathcal{R}_n|$. We focus primarily on mappings such that $\theta_n \in [0,1]$ and $\vartheta_n \in [0,1]$.

This parametric representation covers a broad class of Type I error rates, including *generalized tail probability* (gTP) error rates,

$$(16) \quad gTP(q,g) \equiv \Pr(g(V_n, R_n) > q),$$

and *generalized expected value* (gEV) error rates,

$$(17) \qquad gEV(g) \equiv \mathrm{E}[g(V_n, R_n)],$$

for arbitrary functions $g(V_n, R_n)$ of the numbers of Type I errors $V_n$ and rejected hypotheses $R_n$ and user-supplied bounds $q$.

Generalized tail probability and expected value error rates include as special cases the following commonly-used Type I error rates.

The *generalized family-wise error rate* (gFWER), corresponding to $g(v, r) = v$ and $q \in \{0, \ldots, (h_0 - 1)\}$, is the probability of at least $(q + 1)$ Type I errors,

$$(18) \qquad gFWER(q) \equiv \Pr(V_n > q) = 1 - F_{V_n}(q).$$

When $q = 0$, the gFWER reduces to the usual *family-wise error rate* (FWER), controlled by the classical Bonferroni procedure.

The *tail probability for the proportion of false positives* (TPPFP) *among the rejected hypotheses*, corresponding to $g(v, r) = v/r$ and $q \in (0, 1)$, is defined as

$$(19) \qquad TPPFP(q) \equiv \Pr\left(\frac{V_n}{R_n} > q\right) = 1 - F_{V_n/R_n}(q),$$

with the convention that $V_n/R_n \equiv 0$ if $R_n = 0$.

The *false discovery rate* (FDR), corresponding to $g(v, r) = v/r$, is the expected proportion of false positives among the rejected hypotheses,

$$(20) \qquad FDR \equiv \mathrm{E}\left[\frac{V_n}{R_n}\right] = \int q dF_{V_n/R_n}(q),$$

again with the convention that $V_n/R_n \equiv 0$ if $R_n = 0$.

Error rates $\Theta(F_{V_n/R_n})$, based on the *proportion* of false positives (e.g., TPPFP and FDR), are especially appealing for the large-scale testing problems encountered in genomics, compared to error rates $\Theta(F_{V_n})$, based on the *number* of false positives (e.g., gFWER), as they do not increase exponentially with the number $M$ of tested hypotheses. However, error rates $\Theta(F_{V_n/R_n})$ tend to be more difficult to control than error rates $\Theta(F_{V_n})$, as they are based on the joint distribution of $V_n$ and $R_n$, rather than only the marginal distribution of $V_n$.

### 2.6. Adjusted p-values

As in the case of single hypothesis testing, one can report the results of a multiple testing procedure in terms of the following quantities: rejection regions for the test statistics, confidence regions for the parameters of interest, and adjusted $p$-values.

Adjusted $p$-values, for the test of multiple hypotheses, are defined as straightforward extensions of unadjusted $p$-values, for the test of individual hypotheses. Consider any multiple testing procedure $\mathcal{R}_n(\alpha) = \mathcal{R}(T_n, Q_0, \alpha)$, with rejection regions $\mathcal{C}_n(m; \alpha) = \mathcal{C}(m; T_n, Q_0, \alpha)$. Then, one can define an $M$-vector of *adjusted p-values*, $\widetilde{P}_{0n} = (\widetilde{P}_{0n}(m) : m = 1, \ldots, M) = \widetilde{P}(T_n, Q_0) = \widetilde{P}(\mathcal{R}(T_n, Q_0, \alpha) : \alpha \in [0, 1])$, as

$$\begin{aligned}
(21) \quad \widetilde{P}_{0n}(m) &\equiv \inf\{\alpha \in [0, 1] : \text{Reject } H_0(m) \text{ at nominal MTP level } \alpha\} \\
&= \inf\{\alpha \in [0, 1] : m \in \mathcal{R}_n(\alpha)\} \\
&= \inf\{\alpha \in [0, 1] : T_n(m) \in \mathcal{C}_n(m; \alpha)\}, \qquad m = 1, \ldots, M.
\end{aligned}$$

That is, the adjusted $p$-value $\widetilde{P}_{0n}(m)$, for null hypothesis $H_0(m)$, is the smallest nominal Type I error level (e.g., gFWER, TPPFP, or FDR) of the multiple hypothesis testing procedure at which one would reject $H_0(m)$, given $T_n$.

As in single hypothesis tests, the smaller the adjusted $p$-value $\widetilde{P}_{0n}(m)$, the stronger the evidence against the corresponding null hypothesis $H_0(m)$. Thus, one rejects $H_0(m)$ for small adjusted $p$-values $\widetilde{P}_{0n}(m)$.

For instance, the adjusted $p$-values for the classical FWER-controlling marginal single-step common-quantile Bonferroni procedure are $\widetilde{P}_{0n}(m) = \min\{MP_{0n}(m), 1\}$. Adjusted $p$-values for FWER-controlling joint single-step common-cut-off maxT Procedure 1 are given in Equation (26).

Under the nestedness assumption of Equation (10), one has two equivalent representations for a MTP, in terms of rejection regions for the test statistics and in terms of adjusted $p$-values. Specifically, the set of rejected null hypotheses at multiple test nominal Type I error level $\alpha$ is

$$(22) \quad \mathcal{R}_n(\alpha) = \{m : T_n(m) \in \mathcal{C}_n(m; \alpha)\} = \left\{m : \widetilde{P}_{0n}(m) \leq \alpha\right\}.$$

Reporting the results of a MTP in terms of adjusted $p$-values, as opposed to only rejection or not of the null hypotheses, offers several advantages, including the following.

- Adjusted $p$-values can be defined for *any Type I error rate* (e.g., gFWER, TPPFP, or FDR).
- They reflect the strength of the evidence against each null hypothesis in terms of the *Type I error rate for the entire MTP*.
- They are *flexible summaries* of a MTP, in the sense that results are supplied for *all Type I error levels* $\alpha$, i.e., the level $\alpha$ need not be chosen ahead of time.
- They provide convenient *benchmarks to compare different MTPs*, whereby smaller adjusted $p$-values indicate a less conservative procedure.
- *Plots of sorted adjusted p-values* allow investigators to examine sets of rejected hypotheses associated with various Type I error rates (e.g., gFWER, TPPFP, or FDR) and nominal levels $\alpha$. Such plots provide tools to decide on an appropriate combination of the number of rejected hypotheses and tolerable false positive rate for a particular experiment and available resources.

### 2.7. Test statistics null distribution

As detailed in Chapter 2 of [14], a key feature of our proposed multiple testing procedures is the *test statistics null distribution* (rather than data generating null distribution) used to obtain rejection regions for the test statistics, confidence regions for the parameters of interest, and adjusted $p$-values. Indeed, whether testing single or multiple hypotheses, one needs the (joint) distribution of the test statistics in order to derive a procedure that probabilistically controls Type I errors. In practice, however, the true distribution $Q_n = Q_n(P)$ of the test statistics $T_n$ is unknown and replaced by a null distribution $Q_0$. The choice of a proper null distribution is crucial, in order to ensure that (finite sample or asymptotic) control of the Type I error rate under the *assumed null distribution* does indeed provide the desired control under the *true distribution*. This issue is particularly relevant for large-scale testing problems, such as those involving biological annotation metadata, which concern high-dimensional multivariate distributions, with complex and unknown dependence structures among variables.

Chapter 2 of [14] provides a general characterization of a proper test statistics null distribution in terms of *null domination* conditions for the joint distribution of the test statistics $(T_n(m) : m \in \mathcal{H}_0)$ for the true null hypotheses $\mathcal{H}_0$ (Section 2.2). This general characterization leads to the explicit proposal of the following two main types of test statistics null distributions: a *null shift and scale-transformed test statistics null distribution*, based on user-supplied upper bounds for the means and variances of the test statistics for the true null hypotheses (Section 2.3), and a *null quantile-transformed test statistics null distribution*, based on user-supplied marginal test statistics null distributions (Section 2.4).

In practice, the test statistics null distribution $Q_0 = Q_0(P)$ is unknown, as it depends on the unknown data generating distribution $P$. Resampling procedures based on the *bootstrap* are provided to conveniently obtain consistent estimators of the null distribution and of the corresponding test statistic cut-offs, parameter confidence regions, and adjusted $p$-values [14, Sections 2.3.2, 2.4.2, 4.4, 5].

As argued in [14, Chapter 2], the following two main points distinguish our approach from existing approaches to Type I error control and the choice of a test statistics null distribution (e.g., [24] and [45]). Firstly, we are only concerned with *control of the Type I error rate under the true data generating distribution $P$*, i.e., under the joint distribution $Q_n = Q_n(P)$, implied by $P$, for the test statistics $T_n$. The concepts of weak and strong control of a Type I error rate and the related restrictive assumption of subset pivotality are therefore irrelevant in our context [45, p. 9–10, 42–43]. Secondly, we propose a *null distribution for the test statistics* $(T_n \sim Q_0)$ rather than a data generating null distribution $(X \sim P_0 \in \cap_{m=1}^{M} \mathcal{M}(m))$. The latter practice does not necessarily provide proper Type I error control under the true distribution $P$. Indeed, the test statistics assumed null distribution $Q_n(P_0)$ and their true distribution $Q_n(P)$ may have different dependence structures for the true null hypotheses $\mathcal{H}_0$ and, as a result, may violate the required null domination conditions for Type I error control.

We stress the generality of our proposed test statistics null distributions: Type I error control does not rely on restrictive assumptions such as subset pivotality and holds for general data generating distributions (with arbitrary dependence structures among variables), null hypotheses (defined in terms of submodels for the data generating distribution), and test statistics (e.g., $t$-statistics, $\chi^2$-statistics, $F$-statistics).

*2.7.1. Null shift and scale-transformed test statistics null distribution*

The first original null distribution of [16, 33, 41], is defined as the asymptotic distribution $Q_0 = Q_0(P)$ of the $M$-vector $Z_n$ of *null shift and scale-transformed test statistics*,

$$(23) \qquad Z_n(m) \equiv \sqrt{\min\left\{1, \frac{\tau_0(m)}{\text{Var}[T_n(m)]}\right\}} (T_n(m) - \text{E}[T_n(m)]) + \lambda_0(m),$$

where $\lambda_0(m)$ and $\tau_0(m)$ are, respectively, user-supplied upper bounds for the means and variances of the $\mathcal{H}_0$-specific test statistics.

In this construction, the null shift values $\lambda_0(m)$ are chosen so that the $\mathcal{H}_0$-specific statistics $(Z_n(m) : m \in \mathcal{H}_0)$ are asymptotically stochastically greater than the original test statistics $(T_n(m) : m \in \mathcal{H}_0)$. The resulting null distribution therefore satisfies the required null domination conditions for Type I error control.

In contrast, the null scale values $\tau_0(m)$ are not needed for Type I error control. The purpose of $\tau_0(m)$ is to avoid a degenerate null distribution and infinite cut-offs for the false null hypotheses ($m \in \mathcal{H}_1$), an important property for power considerations. This scaling is needed, in particular, for $F$-statistics that have asymptotically infinite means and variances for non-local alternative hypotheses.

For a broad class of testing problems, such as the test of single-parameter null hypotheses using *t-statistics* (Equation (7)), the null distribution $Q_0$ is an $M$-variate Gaussian distribution, with mean vector zero and covariance matrix $\sigma^* = \Sigma^*(P)$, that is, $Q_0 = N(0, \sigma^*)$. For tests where the parameter of interest is the $M$-dimensional mean vector $\Psi(P) = \psi = E[X]$, the estimator $\psi_n$ is simply the $M$-vector of empirical means and $\sigma^* = \Sigma^*(P) = \text{Cor}[X]$ is the correlation matrix of $X \sim P$, that is, $Q_0 = N(0, \text{Cor}[X])$. More generally, for an asymptotically linear estimator $\psi_n$, $\Sigma^*(P)$ is the correlation matrix of the vector influence curve. This situation covers standard one-sample and two-sample $t$-statistics for tests of means, but also test statistics for correlation coefficients and regression coefficients in linear and non-linear models.

For testing the equality of $K$ population mean vectors using *F-statistics*, an $F$-statistic-specific null distribution may be defined as the joint distribution of an $M$-vector of quadratic forms of Gaussian random variables.

*2.7.2. Null quantile-transformed test statistics null distribution*

The second and most recent proposal of [42] is defined as the asymptotic distribution $Q_0 = Q_0(P)$ of the $M$-vector $\breve{Z}_n$ of *null quantile-transformed test statistics*,

$$(24) \qquad \breve{Z}_n(m) \equiv q_{0,m}^{-1} Q_{n,m}^\Delta(T_n(m)),$$

where $q_{0,m}$ are user-supplied marginal test statistics null distributions that satisfy marginal null domination conditions. According to the *generalized quantile-quantile function transformation* of [46], define $Q_{n,m}^\Delta(z) \equiv \Delta Q_{n,m}(z) + (1 - \Delta)Q_{n,m}(z^-)$, where $Q_{n,m}$ are the marginal distributions of the test statistics $T_n(m)$ and the random variable $\Delta$ is uniformly distributed on the interval $[0, 1]$, independently of the data $\mathcal{X}_n$.

This latest proposal has the additional advantage that the marginal test statistics null distributions may be set to the optimal (i.e., most powerful) null distributions one would use in single hypothesis testing (e.g., permutation marginal null distributions, Gaussian or other parametric marginal null distributions).

## *2.8. Overview of multiple testing procedures*

Having identified a suitable test statistics null distribution $Q_0$ (or estimator thereof, $Q_{0n}$), there remains the main task of specifying rejection regions (i.e., cut-offs) for the test statistics, confidence regions for the parameters of interest, and adjusted $p$-values.

As detailed in [14, Chapters 3–7], we have developed resampling-based single-step and stepwise multiple testing procedures for controlling a broad class of Type I error rates, defined as generalized tail probabilities, $gTP(q, g) = \Pr(g(V_n, R_n) > q)$, and generalized expected values, $gEV(g) = E[g(V_n, R_n)]$, for arbitrary functions $g(V_n, R_n)$ of the numbers of false positives $V_n$ and rejected hypotheses $R_n$. Our proposed procedures take into account the joint distribution of the test statistics and provide Type I error control in testing problems involving general data generating distributions (with arbitrary dependence structures among variables), null hypotheses (defined in terms of submodels for the data generating distribution), and test statistics (e.g., $t$-statistics, $\chi^2$-statistics, $F$-statistics).

An overview of available MTPs is provided in Chapter 3 of [14]. Core methodological Chapters 4–7 discuss the following main approaches for deriving rejection regions.

*Joint single-step common-cut-off* and *common-quantile procedures* for controlling *general Type I error rates* $\Theta(F_{V_n})$, defined as arbitrary parameters of the distribution of the number of Type I errors $V_n$ (Chapter 4 in [14], [16, 33]). Error rates of the form $\Theta(F_{V_n})$ include the generalized family-wise error rate, $gFWER(q) = 1 - F_{V_n}(q) = \Pr(V_n > q)$, i.e., the chance of at least $(q+1)$ Type I errors.

*Joint step-down common-cut-off (maxT)* and *common-quantile (minP) procedures* for controlling the *family-wise error rate*, $FWER = gFWER(0) = 1 - F_{V_n}(0) = \Pr(V_n > 0)$ (Chapter 5 in [14], [41]).

(Marginal/joint single-step/stepwise common-cut-off/common-quantile) *augmentation multiple testing procedures* (AMTP) for controlling *generalized tail probability* error rates, based on an initial gFWER-controlling procedure (Chapter 6 in [14], [15, 40]).

*Joint resampling-based empirical Bayes procedures* for controlling *generalized tail probability* error rates (Chapter 7 in [14], [39]).

The above multiple testing procedures are implemented in the Bioconductor R package multtest ([14, Section 13.1]; [32]; www.bioconductor.org).

### *2.9. FWER-controlling single-step common-cut-off maxT procedure*

This section focusses on control of the family-wise error rate, using the single-step maxT procedure, a common-cut-off procedure exploiting the joint distribution of the test statistics. The method is summarized below; details are given in [14, Chapter 4] and [16].

---

**Procedure 1 [FWER-controlling single-step common-cut-off maxT procedure].**

*Given an $M$-variate test statistics null distribution $Q_0$, the single-step common-cut-off maxT procedure is based on the distribution of the maximum test statistic, $\max_m Z(m)$, for the $M$-vector $Z = (Z(m) : m = 1, \ldots, M) \sim Q_0$. For controlling the FWER at nominal level $\alpha \in (0, 1)$, the common cut-off $c(Q_0, \alpha)$ is defined as the $(1-\alpha)$-quantile of the distribution of $\max_m Z(m)$, that is,*

$$(25) \qquad c(Q_0, \alpha) \equiv \inf\left\{z \in \mathbb{R} : \Pr_{Q_0}\left(\max_{m=1,\ldots,M} Z(m) \leq z\right) \geq (1-\alpha)\right\}.$$

*The adjusted p-value $\widetilde{p}_{0n}(m)$ for null hypothesis $H_0(m)$ is the probability, under $Q_0$, that $\max_m Z(m)$ exceeds the corresponding observed test statistic $t_n(m)$, that is,*

$$(26) \qquad \widetilde{p}_{0n}(m) = \Pr_{Q_0}\left(\max_{m=1,\ldots,M} Z(m) \geq t_n(m)\right), \qquad m = 1, \ldots, M.$$

---

Procedure 1 provides proper FWER control when based on either of the two null-transformed test statistics null distributions $Q_0$ introduced in Section 2.7. Consistent estimators $Q_{0n}$ of the null distribution $Q_0$ and corresponding single-step maxT cut-offs and adjusted *p*-values may be obtained using the bootstrap, as in Procedure 2.9, below, for the null shift and scale-transformed test statistics null distribution [14, Section 4.4].

**Procedure 2  [FWER-controlling bootstrap-based single-step common-cut-off maxT procedure].**

1. *Let $P_n^\star$ denote a bootstrap estimator of the data generating distribution $P$. For the non-parametric bootstrap, $P_n^\star$ is simply the empirical distribution $P_n$, that is, samples of size $n$ are drawn at random, with replacement from the observed data $\mathcal{X}_n = \{X_i : i = 1, \ldots, n\}$. For the model-based bootstrap, $P_n^\star$ belongs to a model $\mathcal{M}$ for the data generating distribution $P$, such as a family of multivariate Gaussian distributions.*
2. *Generate $B$ bootstrap samples, $\mathcal{X}_n^b \equiv \{X_i^b : i = 1, \ldots, n\}$, $b = 1, \ldots, B$. For the bth sample, the $X_i^b$, $i = 1, \ldots, n$, are IID according to $P_n^\star$.*
3. *For each bootstrap sample $\mathcal{X}_n^b$, compute an $M$-vector of test statistics, $T_n^B(\cdot, b) = (T_n^B(m, b) : m = 1, \ldots, M)$, that can be arranged in an $M \times B$ matrix, $\mathbf{T}_n^B = \left(T_n^B(m, b) : m = 1, \ldots, M; b = 1, \ldots, B\right)$, with rows corresponding to the $M$ null hypotheses and columns to the $B$ bootstrap samples.*
4. *Compute row means and variances of the matrix $\mathbf{T}_n^B$, to yield estimators of the means, $\mathrm{E}[T_n(m)]$, and variances, $\mathrm{Var}[T_n(m)]$, of the test statistics under the data generating distribution $P$.*
   *That is, compute*

$$(27) \qquad \mathrm{E}[T_n^B(m, \cdot)] \equiv \frac{1}{B} \sum_{b=1}^{B} T_n^B(m, b),$$

$$\mathrm{Var}[T_n^B(m, \cdot)] \equiv \frac{1}{B} \sum_{b=1}^{B} (T_n^B(m, b) - \mathrm{E}[T_n^B(m, \cdot)])^2.$$

5. *Obtain an $M \times B$ matrix, $\mathbf{Z}_n^B = \left(Z_n^B(m, b) : m = 1, \ldots, M; b = 1, \ldots, B\right)$, of null shift and scale-transformed bootstrap test statistics $Z_n^B(m, b)$, by row-shifting and scaling the matrix $\mathbf{T}_n^B$ using the bootstrap estimators of $\mathrm{E}[T_n(m)]$ and $\mathrm{Var}[T_n(m)]$ and the user-supplied null values $\lambda_0(m)$ and $\tau_0(m)$. That is, define*
(28)

$$Z_n^B(m, b) \equiv \sqrt{\min\left\{1, \frac{\tau_0(m)}{\mathrm{Var}[T_n^B(m, \cdot)]}\right\}} \left(T_n^B(m, b) - \mathrm{E}[T_n^B(m, \cdot)]\right) + \lambda_0(m).$$

   *For t-statistics as in Equation (7), the null values are $\lambda_0(m) = 0$ and $\tau_0(m) = 1$.*
6. *The bootstrap estimator $Q_{0n}$ of the null shift and scale-transformed null distribution $Q_0$ is the empirical distribution of the $B$ columns $\{Z_n^B(\cdot, b) : b = 1, \ldots, B\}$ of matrix $\mathbf{Z}_n^B$.*
7. *For each column $b$ of the matrix $\mathbf{Z}_n^B$ (i.e., bootstrap sample $b$), compute the maximum statistic, $\max_m Z_n^B(m, b)$, $b = 1, \ldots, B$.*

> 8. *For controlling the FWER at nominal level $\alpha \in (0,1)$, the bootstrap single-step maxT common cut-off $c(Q_{0n}, \alpha)$ is the $(1-\alpha)$-quantile of the empirical distribution of the $B$ maxima $\{\max_m Z_n^B(m,b) : b = 1, \ldots, B\}$, that is,*
>
> $$(29) \quad c(Q_{0n}, \alpha) \equiv \inf\left\{z \in \mathbb{R} : \frac{1}{B}\sum_{b=1}^{B} I\left(\max_{m=1,\ldots,M} Z_n^B(m,b) \leq z\right) \geq (1-\alpha)\right\}.$$
>
> 9. *The bootstrap single-step maxT adjusted p-value $\widetilde{p}_{0n}(m)$ for null hypothesis $H_0(m)$ is the proportion of maxima $\{\max_m Z_n^B(m,b) : b = 1, \ldots, B\}$ that exceed the corresponding observed test statistic $t_n(m)$, that is,*
>
> $$(30) \quad \widetilde{p}_{0n}(m) = \frac{1}{B}\sum_{b=1}^{B} I\left(\max_{m=1,\ldots,M} Z_n^B(m,b) \geq t_n(m)\right), \quad m = 1, \ldots, M.$$

## 3. Statistical framework for multiple tests of association with biological annotation metadata

Sections 3.1–3.3 introduce the main components of our approach to multiple tests of association with biological annotation metadata, namely, the gene-annotation profiles $A$, the gene-parameter profiles $\lambda$, and the association measures $\psi = \rho(A, \lambda)$ between gene-annotation and gene-parameter profiles. We stress that the choice of a suitable association parameter $\psi$ is perhaps the most important and hardest aspect of the inference problem, as this parameter represents the statistical translation of the biological question of interest. Once the association parameter $\psi$ is appropriately and precisely defined, one can rely on a variety of statistical methods to estimate and test hypotheses concerning this parameter. Section 3.4 describes how the multiple testing methodology of [14] and related articles may be used to detect associations between gene-annotation and gene-parameter profiles.

Note that, for the sake of illustration, we focus on gene-level features. However, as mentioned in Section 1.1, the methodology is generic and may be applied to other types of features, such as those concerning gene isoforms and proteins.

### 3.1. Gene-annotation profiles

Gene-annotation profiles refer to features of a genome that are assumed to be known and constant among units in a population of interest. Such features typically consist of gene annotation metadata, that reflect current knowledge on gene properties, such as, nucleotide and protein sequences, regulation, and function.

Specifically, let $A = (A(g, m) : g = 1, \ldots, G; m = 1, \ldots, M)$ denote a $G \times M$ *gene-annotation matrix*, providing data on $M$ features for $G$ genes in an organism of interest. Thus, row $A(g, \cdot) = (A(g, m) : m = 1, \ldots, M)$ denotes an $M$-dimensional gene-specific feature vector for the $g$th gene, $g = 1, \ldots, G$, and column $A(\cdot, m) = (A(g, m) : g = 1, \ldots, G)$ denotes a $G$-dimensional *gene-annotation profile* for the $m$th feature, $m = 1, \ldots, M$.

In many applications, the element $A(g, m)$ is a binary indicator, coding the YES/NO answer to the $m$th question, among a collection of $M$ questions one may ask about gene $g$. For example, $A(g, m)$ could indicate whether gene $g$ is annotated with a particular GO term $m$, among $M$ terms in one of the three ontologies (BP, CC, or MF), i.e., whether gene $g$ is an element of the node corresponding to the $m$th term in the GO directed acyclic graph (DAG). Other gene-annotation profiles of interest may refer to exon/intron counts/lengths/nucleotide distributions, gene pathway membership (e.g., from the Kyoto Encyclopedia of Genes and Genomes, KEGG, www.genome.ad.jp/kegg), or gene regulation by particular transcription factors. Regarding transcription regulation, one could use data from the Transcription Factor DataBase (TRANSFAC, www.gene-regulation.com) to generate gene-annotation profiles as follows. For a given transcription factor binding motif, a binary gene-annotation profile could consist of indicators for the presence or absence of the motif in the upstream control region of each gene. A continuous gene-annotation profile could be based on the position weight matrix of the binding motif.

Note that the aforementioned features are only *fixed in time* for a given version/release of the corresponding database(s), i.e., such biological data are constantly evolving as our knowledge of the roles of genes and proteins is accumulating and changing. The dynamic nature of biological annotation metadata is an important issue in terms of software design (Section 4.2; [18]).

Note also that gene-annotation profiles are not restricted to be binary or even polychotomous and, in particular, could be continuous gene-parameter profiles, suitably estimated from previous studies.

The main point, regarding the formulation of the statistical inference question, is that gene-annotation profiles are *known* and *constant among population units*.

## 3.2. *Gene-parameter profiles*

Gene-parameter profiles are generally unknown and concern the distribution of variable features of a genome in a well-defined population. Gene-specific variables of interest reflect cellular type/state/activity under particular conditions and include microarray measures of transcript levels and comparative genomic hybridization (CGH) measures of DNA copy numbers.

Specifically, let $X = (X(j) : j = 1, \ldots, J)$ be a $J$-dimensional random vector, containing $G$ *gene-specific random variables* $(X(g) : g = 1, \ldots, G)$. In addition to the $G$ gene-specific variables, $X$ may include various biological and clinical covariates (e.g., age, sex, treatment, timepoint) and outcomes (e.g., survival time, response to treatment, tumor class). Let $P$ denote the data generating distribution for the random $J$-vector $X$ and suppose that $P$ belongs to a (possibly non-parametric) model $\mathcal{M}$.

Let the parameter mapping $\Lambda : \mathcal{M} \to \mathbb{R}^G$ define a $G$-dimensional *gene-parameter profile*, $\Lambda(P) = \lambda = (\lambda(g) : g = 1, \ldots, G) \in \mathbb{R}^G$, where each $\lambda(g) = \Lambda(P)(g) \in \mathbb{R}$ is a gene-specific real-valued parameter. For example, $\lambda(g)$ could be the mean expression measure $\mathrm{E}[X(g)]$ of gene $g$ or a regression coefficient relating an outcome component of $X$ to the expression measure $X(g)$ of gene $g$, $g = 1, \ldots, G$.

While gene-annotation profiles are known and fixed, gene-parameter profiles are typically *unknown* and need to be *estimated*, e.g., from a microarray experiment involving a sample of population units. The sample $\mathcal{X}_n = \{X_i : i = 1, \ldots, n\}$ is assumed to consist of $n$ independent and identically distributed (IID) copies of $X \sim P$, corresponding to $n$ randomly sampled population units.

### 3.3. Association measures for gene-annotation and gene-parameter profiles

Let the parameter mapping $\Psi : \mathcal{M} \to \mathbb{R}^M$ specify an $M$-dimensional *association parameter vector*,

(31) $$\Psi(P) = \psi = (\psi(m) : m = 1, \ldots, M) \equiv \rho(A, \Lambda(P)),$$

defined in terms of an *association measure* $\rho : \mathbb{R}^{G \times M} \times \mathbb{R}^G \to \mathbb{R}^M$, *known fixed gene-annotation profiles* $A$, and an *unknown gene-parameter profile* $\lambda = \Lambda(P)$.

The choice of a suitable association parameter is subject matter-dependent and requires careful consideration. For instance, for Gene Ontology annotation, it is desirable that the association parameter reflect the structure of the GO directed acyclic graph (Section 4.1). In principle, the dimension of the association parameter vector $\psi$ could differ from the number $M$ of features under consideration. In addition, one could accommodate several gene-parameter profiles $\lambda$.

The various quantities in the inference problem are summarized in Figure 1; examples of association parameters are given next and in Section 5.

#### 3.3.1. Univariate association measures

In the simplest case, one could define the $M$ association parameters univariately, i.e., define $\psi(m)$ based only on the $m$th gene-annotation profile $A(\cdot, m)$, $m = 1, \ldots, M$.

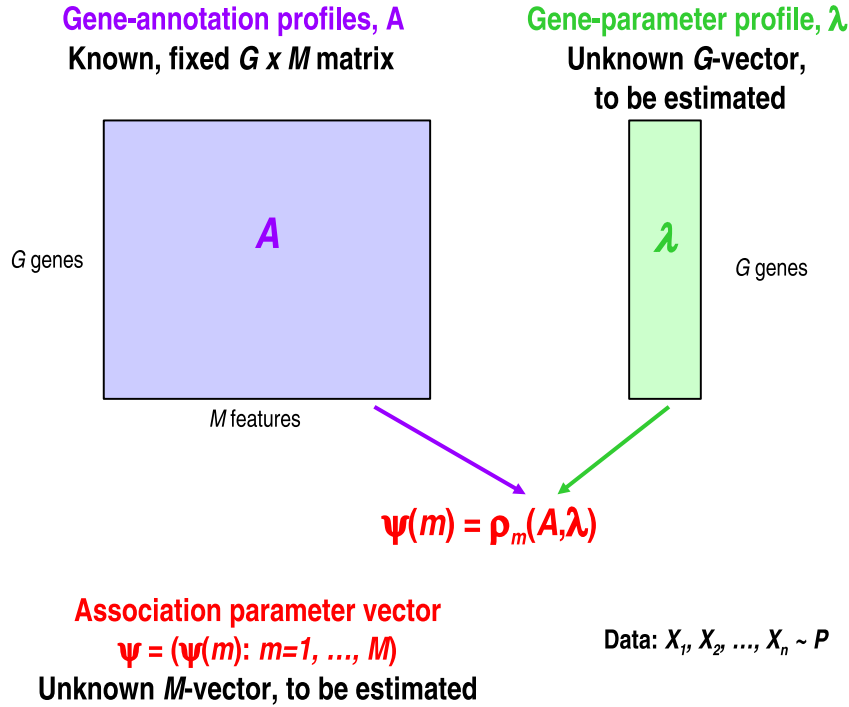

FIG 1. *Parameters for tests of association with biological annotation metadata. This figure represents the main ingredients involved in multiple tests of association with biological annotation metadata: the gene-annotation profiles, the gene-parameter profile, and the association parameters. (Color version on website companion.)*

Specifically, for the $m$th feature, let

$$\Psi(P)(m) = \psi(m) \equiv \rho_m(A(\cdot, m), \Lambda(P)), \tag{32}$$

where $\rho_m : \mathbb{R}^G \times \mathbb{R}^G \to \mathbb{R}$ provides a measure of association (e.g., correlation coefficient) between the $G$-dimensional gene-annotation profile $A(\cdot, m)$ and gene-parameter profile $\lambda = \Lambda(P)$. In many situations, the same association measure $\rho_m$ may be used for each of the $M$ features.

**Continuous gene-annotation profiles and continuous gene-parameter profiles** For continuous gene-annotation and gene-parameter profiles, one may use as association measure the Pearson *correlation coefficient* between two $G$-vectors. That is,

$$\psi(m) = \frac{\sum_{g=1}^G (A(g,m) - \bar{A}(m))(\lambda(g) - \bar{\lambda})}{\sqrt{\sum_{g=1}^G (A(g,m) - \bar{A}(m))^2} \sqrt{\sum_{g=1}^G (\lambda(g) - \bar{\lambda})^2}}, \tag{33}$$

where $\bar{A}(m) \equiv \sum_g A(g,m)/G$ and $\bar{\lambda} \equiv \sum_g \lambda(g)/G$ denote, respectively, the averages of the $G$ elements of the gene-annotation profile $A(\cdot, m)$ and gene-parameter profile $\lambda$.

**Binary gene-annotation profiles and binary gene-parameter profiles** For binary gene-annotation and gene-parameter profiles, one may build $2 \times 2$ contingency Table 2 and use as association measure the $\chi^2$-*statistic* (or corresponding $p$-value) for the test of independence of rows and columns. That is,

$$\psi(m) = \frac{G(g_{00}(m)g_{11}(m) - g_{01}(m)g_{10}(m))^2}{g_{0\cdot}(m)g_{\cdot 0}(m)g_{\cdot 1}(m)g_{1\cdot}(m)}, \tag{34}$$

where $g_{kk'}(m) \equiv \sum_g I(A(g,m) = k) I(\lambda(g) = k')$, $g_{k\cdot}(m) \equiv g_{k0}(m) + g_{k1}(m) = \sum_g I(A(g,m) = k)$, and $g_{\cdot k'}(m) \equiv g_{0k'}(m) + g_{1k'}(m) = \sum_g I(\lambda(g) = k')$, $k, k' \in \{0, 1\}$. Note that in this context *the $\chi^2$-statistic $\psi(m)$ is a parameter*, i.e., it is a function of the data generating distribution $P$, via the gene-parameter profile $\lambda = \Lambda(P)$, and is therefore *unknown* and *constant* among population units.

TABLE 2. *Binary gene-annotation and gene-parameter profiles.* Given a binary gene-annotation profile $A(\cdot, m)$ and a binary gene-parameter profile $\lambda$, one may build a $2 \times 2$ contingency table, with rows corresponding to the gene-annotation profile and columns to the gene-parameter profile. Cell counts are defined as $g_{kk'}(m) = \sum_g I(A(g,m) = k) I(\lambda(g) = k')$, $k, k' \in \{0, 1\}$. For example, for tests of association between GO annotation and differential gene expression, $g_{11}(m)$ could correspond to the number of genes that are annotated with GO term $m$ and differentially expressed.

| | | Gene-parameter profile, $\lambda$ | | |
|---|---|---|---|---|
| | | **1** | **0** | |
| Gene-annotation profile, $A(\cdot, m)$ | **1** | $g_{11}(m) = \sum_{g=1}^G A(g,m)\lambda(g)$ | $g_{10}(m) = \sum_{g=1}^G A(g,m)(1-\lambda(g))$ | $A_1(m) = \sum_{g=1}^G A(g,m)$ |
| | **0** | $g_{01}(m) = \sum_{g=1}^G (1-A(g,m))\lambda(g)$ | $g_{00}(m) = \sum_{g=1}^G (1-A(g,m))(1-\lambda(g))$ | $A_0(m) = \sum_{g=1}^G (1-A(g,m))$ |
| | | $G\bar{\lambda} = \sum_{g=1}^G \lambda(g)$ | $G(1-\bar{\lambda}) = \sum_{g=1}^G (1-\lambda(g))$ | $G$ |

**Binary gene-annotation profiles** For binary gene-annotation profiles, one may consider association parameter vectors of the form

$$\psi = A^\top \lambda. \tag{35}$$

That is, the association parameter for the $m$th feature is the *sum*,

$$\psi(m) = \sum_{g=1}^{G} A(g,m)\lambda(g) = \sum_{g=1}^{G} \mathrm{I}\left(A(g,m) = 1\right)\lambda(g),$$

of the parameters $\lambda(g)$ for genes $g$ that have the property of interest, i.e., such that $A(g,m) = 1$. Such an association parameter is considered by [37], to relate continuous microarray differential expression gene-parameter profiles to binary pathway gene-annotation profiles.

The following standardized association parameters, corresponding to association measures based on *two-sample t-statistics*, may also be considered,

$$\psi(m) = \frac{\bar{\lambda}_1(m) - \bar{\lambda}_0(m)}{\sqrt{\frac{v[\lambda]_1(m)}{A_1(m)} + \frac{v[\lambda]_0(m)}{A_0(m)}}}, \tag{36}$$

where, for the $m$th feature, $A_k(m) \equiv \sum_g \mathrm{I}\left(A(g,m) = k\right)$,

$$\bar{\lambda}_k(m) \equiv \sum_g \mathrm{I}\left(A(g,m) = k\right)\lambda(g)/A_k(m),$$

and $v[\lambda]_k(m) \equiv \sum_g \mathrm{I}\left(A(g,m) = k\right)(\lambda(g) - \bar{\lambda}_k(m))^2/(A_k(m) - 1)$ denote, respectively, the numbers, averages, and variances of annotated ($k = 1$) and unannotated ($k = 0$) gene-parameters $\lambda(g)$.

In commonly-encountered combined GO annotation and microarray data analyses, a binary gene-parameter profile could indicate whether genes are differentially expressed in two populations of cells, a continuous gene-parameter profile could consist of coefficients for the regression of a (censored) clinical outcome on gene expression measures, and binary gene-annotation profiles could denote whether genes are annotated with particular GO terms (Section 5; [1, 2, 4, 22]).

*3.3.2. Multivariate association measures*

More generally, the $m$th association parameter could be based on the entire gene-annotation matrix $A$ or a subset of columns thereof, that is, $\Psi(P)(m) = \psi(m) \equiv \rho_m(A, \Lambda(P))$, for an association measure $\rho_m : \mathbb{R}^{G \times M} \times \mathbb{R}^G \to \mathbb{R}$.

Association parameters of interest include: linear combinations of association parameters for several features, partial correlation coefficients, $\chi^2$-statistics for higher-dimensional contingency tables (e.g., with one dimension corresponding to a gene-parameter profile $\lambda$ and other dimensions to several gene-annotation profiles $A(\cdot, m)$), and (contrasts of) regression coefficients of a gene-parameter profile $\lambda$ on several gene-annotation profiles $A(\cdot, m)$.

In the case of Gene Ontology annotation, the association parameter $\psi$ should preferably reflect the structure of the GO directed acyclic graph, by taking into account, for instance, annotation information for ancestor (i.e., less specific) or offspring (i.e., more specific) terms (Section 4.1). Specifically, let $\mathcal{P}(m)$ denote the set of (immediate) parents of a term $m$. As the genes annotated by the child term $m$ are subsets of the genes annotated by the parent terms $\mathcal{P}(m)$, then $A(g, m) = 1$ implies $A(g, p) = 1$ for $p \in \mathcal{P}(m)$.

Following the causal inference literature [38, 43], an association parameter of interest for GO term $m$ is the *marginal causal effect parameter*, defined as

$$(37) \quad \psi(m) = \mathrm{E}[\mathrm{E}[\lambda | A(\cdot, m) = 1, A(\cdot, \mathcal{P}(m))]] - \mathrm{E}[\mathrm{E}[\lambda | A(\cdot, m) = 0, A(\cdot, \mathcal{P}(m))]],$$

where $A(\cdot, \mathcal{P}(m))$ denotes the submatrix of gene-annotation profiles for parent terms $\mathcal{P}(m)$ and the expected values are defined with respect to the empirical distribution of $\{(A(g, m), A(g, \mathcal{P}(m)), \lambda(g)) : g = 1, \ldots, G\}$.

In the special case of binary (differential expression) gene-parameter profiles, the so-called parent-child method of [22] takes into account the structure of the GO DAG by testing for associations between gene-annotation and gene-parameter profiles using hypergeometric $p$-values computed conditionally on the annotation status of parent terms.

One could also consider Boolean combinations of annotation indicators for multiple features, that is, a transformed gene-annotation matrix whose columns are Boolean combinations of the columns of the original gene-annotation matrix. Such an approach would be particularly relevant in the context of transcription regulation, where individual features correspond to single transcription factor binding motifs and Boolean combinations to binding modules for multiple transcription factors.

### *3.4. Multiple hypothesis testing*

#### *3.4.1. Null and alternative hypotheses*

Certain biological annotation metadata analyses may involve the *two-sided test* of the $M$ null hypotheses of no association between gene-annotation profiles $A(\cdot, m)$, $m = 1, \ldots, M$, and a gene-parameter profile $\lambda$, i.e., the test of

$$(38) \qquad H_0(m) = \mathrm{I}\left(\psi(m) = \psi_0(m)\right) \qquad \text{vs.} \qquad H_1(m) = \mathrm{I}\left(\psi(m) \neq \psi_0(m)\right).$$

Other analyses may call for the *one-sided test* of

$$(39) \qquad H_0(m) = \mathrm{I}\left(\psi(m) \leq \psi_0(m)\right) \qquad \text{vs.} \qquad H_1(m) = \mathrm{I}\left(\psi(m) > \psi_0(m)\right).$$

One-sided tests are appropriate if, for example, one is interested in determining whether differentially expressed genes are enriched regarding GO annotation.

The $M$-vector $\psi_0 = (\psi_0(m) : m = 1, \ldots, M)$, of *null values* for the association parameter $\psi$, is determined by the biological question. For example, if $\psi(m) = \rho_m(A(\cdot, m), \lambda)$ is the Pearson correlation coefficient between the gene-annotation profile $A(\cdot, m)$ and the gene-parameter profile $\lambda$, then one may set $\psi_0(m) = 0$.

Note that in many situations, the same association measure $\rho_m$ is used for each of the $M$ features and one only has a single, common null value $\psi_0(m)$.

#### *3.4.2. Test statistics*

As in Section 2.2, above, and Chapter 1 of [14], consider the general situation where, given a random sample $\mathcal{X}_n$ from the data generating distribution $P$, one has an *asymptotically linear* estimator $\psi_n = \hat{\Psi}(P_n)$ of the association parameter vector $\psi = \Psi(P)$, with $M$-dimensional vector *influence curve* $IC(X|P)$. Let $\hat{\Sigma}(P_n) = \sigma_n = (\sigma_n(m, m') : m, m' = 1, \ldots, M)$ denote a consistent estimator of the covariance matrix $\Sigma(P) = \sigma = (\sigma(m, m') : m, m' = 1, \ldots, M)$ of the vector influence curve $IC(X|P)$. For example, $\sigma_n$ could be a bootstrap-based estimator of the covariance matrix $\sigma$ or could be computed from an estimator $IC_n(X)$ of the influence curve $IC(X|P)$.

A broad range of association parameters $\psi$ and corresponding estimators $\psi_n$ satisfy the above conditions. In particular, suppose $\lambda_n = \hat{\Lambda}(P_n)$ is an asymptotically linear estimator of the gene-parameter profile $\lambda = \Lambda(P)$, based on a random sample $\mathcal{X}_n$ from $P$. Let $\psi_n \equiv \rho(A, \lambda_n)$ denote the corresponding *resubstitution*, or *plug-in*, estimator of the association parameter vector $\psi = \rho(A, \lambda)$. Then, if the function $\rho(A, \lambda)$ is differentiable with respect to $\lambda$, the resubstitution estimator $\psi_n$ is also asymptotically linear.

One can therefore handle tests where the gene-parameter profiles $\lambda$ are (functions of) means, variances, correlation coefficients, and regression coefficients, and where the association measures $\rho$ are correlation coefficients, two-sample $t$-statistics, and $\chi^2$-statistics. Examples are provided in Section 5, in the context of tests of association between differential gene expression in ALL and GO annotation.

Each null hypothesis $H_0(m)$ may then be tested using a (unstandardized) *difference statistic*,

$$(40) \qquad T_n(m) = \sqrt{n}\left(\psi_n(m) - \psi_0(m)\right),$$

or a (standardized) *t-statistic*,

$$(41) \qquad T_n(m) = \sqrt{n}\frac{\psi_n(m) - \psi_0(m)}{\sigma_n(m)},$$

where we adopt the shorter notation $\sigma_n^2(m) = \sigma_n(m, m)$ for variances.

Certain testing problems may call for other test statistics $T_n$, such as, $F$-statistics, $\chi^2$-statistics, and likelihood ratio statistics.

Let $Q_n = Q_n(P)$ denote the typically unknown (finite sample) joint distribution of the $M$-vector of test statistics $T_n = (T_n(m) : m = 1, \ldots, M)$, under the data generating distribution $P$.

### 3.4.3. Multiple testing procedures

As mentioned in Section 2.7, above, and detailed in [14, Chapter 2], a key feature of our proposed multiple testing procedures is the *test statistics null distribution* $Q_0$ used in place of the unknown true test statistics distribution $Q_n = Q_n(P)$, for the purpose of obtaining rejection regions for the test statistics, confidence regions for the parameters of interest, and adjusted $p$-values.

Given a suitable test statistics null distribution $Q_0$ (or estimator thereof, $Q_{0n}$), the multiple testing procedures developed in [14] and related articles may be applied to control a broad class of Type I error rates, defined as generalized tail probabilities, $gTP(q, g) = \Pr(g(V_n, R_n) > q)$, and generalized expected values, $gEV(g) = \mathrm{E}[g(V_n, R_n)]$, for arbitrary functions $g(V_n, R_n)$ of the numbers of false positives $V_n$ and rejected hypotheses $R_n$ (Section 2.8).

For the purpose of illustration, we focus, as in Section 2.9, on control of the family-wise error rate, using the single-step common-cut-off maxT procedure, based on a non-parametric bootstrap estimator of the null shift and scale-transformed test statistics null distribution (Procedures 1 and 2.9).

## 4. The Gene Ontology

### *4.1. Overview of the Gene Ontology*

The *Gene Ontology* (GO) Consortium (`www.geneontology.org`) provides *ontologies*, i.e., structured and controlled vocabularies, to describe gene products in terms of their associated biological processes, cellular components, and molecular functions. The ontologies specify terminologies and relationships among terms. They are organism-independent and can be applied even as our knowledge of the roles of genes and proteins is accumulating and changing.

The GO Consortium and other organizations supply *mappings* between GO terms and genes in various organisms.

Detailed documentation is available on the Gene Ontology Documentation webpage (`www.geneontology.org/GO.contents.doc.html`).

#### *4.1.1. The three gene ontologies: BP, CC, and MF*

The GO Consortium provides three ontologies, each consisting of a structured network of terms describing gene products.

**Biological Process** (BP or P). The Biological Process ontology refers to series of biological events that are accomplished by one or more ordered assemblies of molecular functions. Examples of broad BP terms are *cellular physiological process* (`GO:0050875`) and *signal transduction* (`GO:0007165`); examples of more specific BP terms are *pyrimidine base metabolism* (`GO:0006206`) and *alpha-glucoside transport* (`GO:0000017`).

**Cellular Component** (CC or C). The Cellular Component ontology refers to subcellular structures, with the proviso that the components be part of some larger object, which may be an anatomical structure (e.g., *rough endoplasmic reticulum* (`GO:0005791`), *nucleus* (`GO:0005634`)) or a gene product group (e.g., *ribosome* (`GO:0005840`)).

**Molecular Function** (MF or F). The Molecular Function ontology refers to tasks or activities performed by individual (or assembled complexes of) gene products. Examples of broad MF terms are *catalytic activity* (`GO:0003824`), *transporter activity* (`GO:0005215`), and *binding* (`GO:0005488`); examples of more specific MF terms are *adenylate cyclase activity* (`GO:0004016`) and *Toll binding* (`GO:0005121`).

A gene product may be used in one or more biological processes, may be associated with one or more cellular components, and may have one or more molecular functions.

**Example 1. Gene product** *ABL1_HUMAN*. The *Homo sapiens* gene product *Splice Isoform IA of Proto-oncogene tyrosine-protein kinase ABL1* (*ABL1_HUMAN*) can be described by the following terms in each of the three gene ontologies (AmiGO browser, Last updated: 2006-05-25, `www.godatabase.org/cgi-bin/amigo/go.cgi?view=details&search_constraint=gp&session_id=6973b1139030258&gp=P00519`).

**Biological Process:**
    *regulation of progression through cell cycle* (`GO:0000074`);
    *S-phase-specific transcription in mitotic cell cycle* (`GO:0000115`);
    *mismatch repair* (`GO:0006298`);

*regulation of transcription, DNA-dependent* (`GO:0006355`);
*DNA damage response, signal transduction resulting in induction of apoptosis* (`GO:0008630`).

**Cellular Component:**
*nucleus* (`GO:0005634`).

**Molecular Function:**
*DNA binding* (`GO:0003677`);
*protein-tyrosine kinase activity* (`GO:0004713`);
*protein binding* (`GO:0005515`).

### 4.1.2. GO directed acyclic graphs

For each of the three gene ontologies, GO terms are organized in a *directed acyclic graph* (DAG), where a *directed* graph has one-way edges and an *acyclic* graph has no path starting and ending at the same vertex. Each GO term is associated with a single vertex, or node, in the DAG. The words *term*, *node*, and *vertex*, may therefore be used interchangeably.

For a given GO term, an *ancestor* refers to a less specialized term; an *offspring* refers to a more specialized term. A *parent* is an immediate/direct ancestor of a term; a *child* is an immediate/direct offspring of a term. A *root* node has no parents (i.e., no incoming edges); a *leaf* node has no children (i.e., no outgoing edges). In a DAG, a child may have several parents.

GO terms must obey the so-called *true path rule*: if a (child) term describes a gene product, then all its immediate parent and more distant ancestor terms must also apply to the gene product.

The DAG structure of GO terms and corresponding true path rule are germane to the definition of a suitable association measure between gene-annotation profiles and gene-parameter profiles (Section 3.3). Furthermore, as discussed in Sections 4.2–4.5, in the context of Bioconductor annotation software, the true path rule is also relevant when assembling gene-annotation matrices.

### 4.1.3. GO software tools

Many software tools have been developed to deal with GO annotation metadata. The Gene Ontology Tools webpage (`www.geneontology.org/GO.tools.shtml`) provides a list of consortium and non-consortium software for searching and browsing the three gene ontologies, for annotating genes and gene products using GO, and for combined GO and gene expression microarray data analysis.

For instance, the AmiGO browser (`www.godatabase.org`) allows: searching for a GO term and viewing all gene products annotated with this term; searching for a gene product and viewing all its associated GO terms; and browsing the ontologies to view relationships among terms and gene products annotated with a given term.

The QuickGO browser (`www.ebi.ac.uk/ego`), developed by the European Bioinformatics Institute (EBI), permits searches and graphical displays of the Gene Ontology by GO term, GO term identifier (ID), gene product, and other identifiers.

Software packages developed as part of the Bioconductor Project are discussed in Sections 4.2–4.5.

**Example 2. GO term *protein-tyrosine kinase activity*.** To get a sense of the information provided by the GO Consortium, consider the Molecular Function ontology and the GO term *protein-tyrosine kinase activity*, with GO term ID `GO:0004713`.

Go to the AmiGO browser (www.godatabase.org), enter the GO term ID `GO:0004713` in the `Search GO` box, select `Exact Match`, select `Terms`, and click on the `Submit Query` button. There are two main options for displaying information on a GO term: a "tree view" and a "graphical view". Click on the small tree-like icon (top-left corner of the table) to display the tree view with all ancestors (i.e., less specific terms) of the GO term *protein-tyrosine kinase activity*. Click on the `Graphical View` button to display the portion of the MF DAG corresponding to the GO term. Additional information may be obtained by clicking on the hyperlinked text `protein-tyrosine kinase activity`.

The GO term *protein-tyrosine kinase activity* has one (immediate) parent, *protein kinase activity* (`GO:0004672`), which itself has two parents, *kinase activity* (`GO:0016301`) and *phosphotransferase activity, alcohol group as acceptor* (`GO:0016773`). Altogether, the term *protein-tyrosine kinase activity* has 7 ancestors. According to the true path rule, any gene annotated with the GO term *protein-tyrosine kinase activity* should also be annotated with all of its less specific ancestor terms.

The portion of the MF DAG for the GO term *protein-tyrosine kinase activity* is displayed in Figure 2 using the QuickGO browser.

### 4.1.4. GO gene-annotation matrices

For each of the three gene ontologies, one may define a $G \times M$ *binary gene-annotation matrix* $A$, indicating for each gene $g$ whether it is annotated with each GO term $m$,

$$(42) \qquad A(g,m) \equiv \begin{cases} 1, & \text{if gene } g \text{ is annotated with GO term } m, \\ 0, & \text{otherwise} \end{cases}$$

$$g = 1, \ldots, G, \ m = 1, \ldots, M.$$

Section 4.5 provides sample R code for assembling GO gene-annotation matrices using Bioconductor annotation metadata packages.

As detailed in Section 3, detecting associations between GO annotation and other interesting features of a genome may be viewed as the multiple test of the null hypotheses of no association between gene-annotation profiles $A(\cdot, m)$ and a gene-parameter profile $\lambda = \Lambda(P)$. The multiple testing methodology proposed in [14] and related articles is well-suited to handle the complex and unknown dependence structure among test statistics implied by the DAG structure of GO terms. The methodology is summarized in Section 2 and illustrated in Section 5, for tests of association between differential gene expression in ALL and GO annotation.

### 4.2. Overview of R and Bioconductor software for GO annotation metadata analysis

As discussed in [18], the *Bioconductor Project* provides R packages for accessing and performing statistical inference with GO annotation metadata (www.bioconductor.org; www.r-project.org). The packages include:

- a general annotation software package: annotate;
- packages for graph theoretical analyses: e.g., graph, Rgraphviz;
- a GO-specific metadata package for navigating the three GO DAGs: GO;

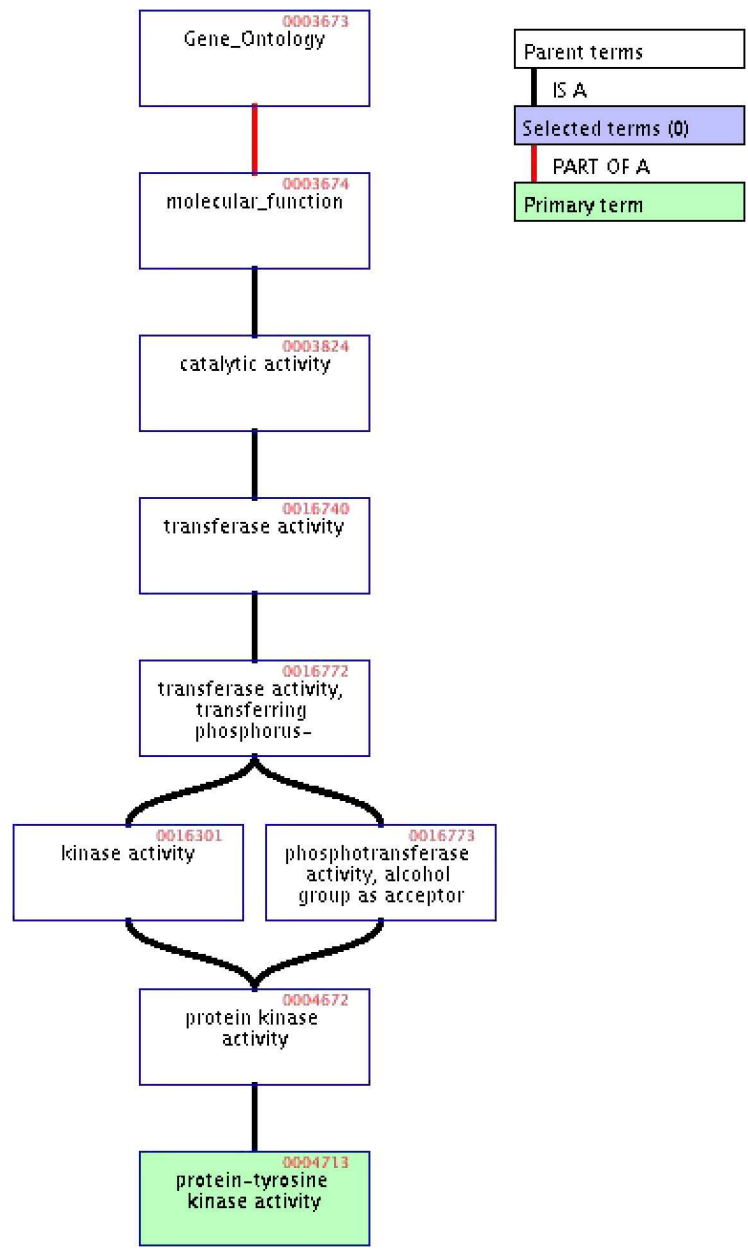

Fig 2. *DAG for MF GO term* `GO:0004713`, *QuickGO*. Portion of the directed acyclic graph for the GO term *protein-tyrosine kinase activity* (`GO:0004713`), in the Molecular Function ontology. This display, obtained using the EBI QuickGO browser (Last updated 2001-03-30 04:29:44.0, <www.ebi.ac.uk/ego>), shows the nodes corresponding to all (less specific) ancestors of the term *protein-tyrosine kinase activity*. (Higher-resolution color version on website companion.)

- an Entrez Gene[1]-specific metadata package, providing bi-directional mappings between Entrez Gene IDs and GO term IDs: humanLLMappings (<www.ncbi.nlm.nih.gov/entrez/query.fcgi?db=gene>);

---

[1]N.B. The NCBI LocusLink database has been superseded by the Entrez Gene database.

*Multiple tests of association with biological annotation metadata* 181- various Affymetrix chip-specific metadata packages, providing bi-directional mappings between Affymetrix probe[2] IDs and GO term IDs: e.g., hgu95av2, hu6800 (www.affymetrix.com);
- a package for annotating and generating HTML reports for Affymetrix chip data: annaffy.

Bioconductor metadata packages are updated regularly to reflect the evolving nature of biological annotation metadata; it is therefore crucial to keep track of *version* numbers. For information on Bioconductor software, please consult [17] and the Documentation (www.bioconductor.org/docs) and Workshops (www.bioconductor.org/workshops) sections of the Bioconductor Project website, in addition to the standard R help facilities (e.g., help function, manuals, etc.).

The remainder of this section provides sample R code demonstrating Bioconductor software (results reported for R Release 2.2.1 and Bioconductor Release 1.7). In order to run through the examples, one needs to install and load the Bioconductor packages annotate, GO, and hgu95av2. The annotation metadata used in the examples correspond to the following package versions.

```
> library(annotate)
> library(GO)
> library(hgu95av2)
>
> packageDescription("annotate")$Version
[1] "1.8.0"
> packageDescription("GO")$Version
[1] "1.10.0"
> packageDescription("hgu95av2")$Version
[1] "1.10.0"
```

Accessing and analyzing annotation metadata from databases such as GenBank (www.ncbi.nlm.nih.gov/Genbank), GO (www.geneontology.org), and PubMed (www.pubmed.gov), presupposes the ability to perform the following essential bookkeeping task: *mapping between different identifiers* (ID) for a given gene/probe. Bioconductor annotation metadata packages consist of *environment* objects that provide *key-value mappings* between different sets of gene/probe identifiers.

For instance, in the annotation metadata package hgu95av2, for the Affymetrix chip series HG-U95Av2, the hgu95av2PMID environment provides mappings from Affymetrix probe IDs (keys) to PubMed IDs (values); similarly, the hgu95av2GO environment provides mappings from Affymetrix probe IDs (keys) to GO term IDs (values).

**Example 3. Affymetrix probe ID 1635_at.** As of Version 1.10.0 of the hgu95av2 package, the Affymetrix probe with ID 1635_at corresponds to the gene with symbol ABL1 and long name v-abl Abelson murine leukemia viral oncogene homolog 1, located on the long arm of chromosome 9. This probe maps to one GenBank accession number, one Entrez Gene ID, 14 distinct GO term IDs, and 160 distinct PubMed IDs.

```
> probe <- "1635_at"
```

---

[2] N.B. In the context of Affymetrix oligonucleotide chips, we use the shorter term *probe* to refer to a *probe-pair-set*, i.e., a collection of *perfect match* (PM) and *mismatch* (MM) *probe-pairs* that map to a particular gene.

```
> get(probe, env=hgu95av2SYMBOL)
[1] "ABL1"
> get(probe, env=hgu95av2GENENAME)
[1] "v-abl Abelson murine leukemia viral oncogene homolog 1"
> get(probe, env=hgu95av2MAP)
[1] "9q34.1"
> get(probe, env=hgu95av2ACCNUM)
[1] "U07563"
> get(probe, env=hgu95av2LOCUSID )
[1] 25
> unique(names(get(probe, env=hgu95av2GO)))
 [1] "GO:0000074" "GO:0000115" "GO:0000166" "GO:0003677"
 [5] "GO:0004713" "GO:0005515" "GO:0005524" "GO:0005634"
 [9] "GO:0006298" "GO:0006355" "GO:0006468" "GO:0007242"
[13] "GO:0008630" "GO:0016740"
> length(get(probe, env=hgu95av2PMID))
[1] 160
```

The remainder of this section gives a brief overview of two main types of Bioconductor annotation metadata packages: the GO package (Section 4.3) and the hgu95av2 package for the Affymetrix chip series HG-U95Av2 (Section 4.4). Section 4.5 illustrates how these two packages may be used to assemble a GO gene-annotation matrix.

### 4.3. The annotation metadata package GO

The GO package provides environment objects containing key-value pairs for mappings between GO term IDs, GO terms, GO term ancestors, GO term parents, GO term children, GO term offspring, and Entrez Gene IDs. The GO() command lists all environments available in the GO package.

```
> GO()

Quality control information for GO
Date built: Created: Fri Sep 30 03:02:24 2005

Mappings found for non-probe based rda files:
         GOALLLOCUSID found 9556
         GOBPANCESTOR found 9888
         GOBPCHILDREN found 4989
         GOBPOFFSPRING found 4989
         GOBPPARENTS found 9888
         GOCCANCESTOR found 1612
         GOCCCHILDREN found 578
         GOCCOFFSPRING found 578
         GOCCPARENTS found 1612
         GOLOCUSID2GO found 70818
         GOLOCUSID found 8017
         GOMFANCESTOR found 7334
         GOMFCHILDREN found 1403
         GOMFOFFSPRING found 1403
         GOMFPARENTS found 7334
```

```
                GOOBSOLETE found 1032
                GOTERM found 18834
```

For information on any of the GO environments, use the `help` function, e.g., `help(GOTERM)` or `?GOBPPARENTS`. For instance, the environment `GOTERM` provides mappings from *GO term IDs* (keys) to *GO terms* (values); the environments `GOBPPARENTS`, `GOCCPARENTS`, and `GOMFPARENTS`, provide ontology-specific mappings from *GO term IDs* (keys) to *GO term parent IDs* (values). The environments `GOALLLOCUSID`, `GOLOCUSID2GO`, and `GOLOCUSID`, provide mappings between *GO term IDs* and *Entrez Gene IDs* and are used in Section 4.5, below, to assemble an Entrez Gene ID-by-GO term ID gene-annotation matrix for the MF gene ontology.

**Example 4. GO term ID** `GO:0004713`. Let us use the GO package to obtain information on (all) ancestors, (immediate) parents, (immediate) children, and (all) offspring of the term corresponding to the GO term ID `GO:0004713`.

```
> ## List all GO IDs
> GOID <- ls(env = GOTERM)
> length(GOID)
[1] 18834
> GOID[1:10]
 [1] "GO:0000001" "GO:0000002" "GO:0000003" "GO:0000004"
 [5] "GO:0000006" "GO:0000007" "GO:0000009" "GO:0000010"
 [9] "GO:0000011" "GO:0000012"
>
> ## Get information on GO term corresponding to GO ID
> ## GO:0004713
> GOID <- "GO:0004713"
> term <- get(GOID,env=GOTERM)
> class(term)
[1] "GOTerms"
attr(,"package")
[1] "annotate"
> slotNames(term)
[1] "GOID"       "Term"       "Synonym"       "Secondary"
[5] "Definition" "Ontology"
> term
GOID = GO:0004713
Term = protein-tyrosine kinase activity
Synonym = protein tyrosine kinase activity
Definition = Catalysis of the reaction: ATP + a protein
    tyrosine = ADP + protein tyrosine phosphate.
Ontology = MF
>
> ## Get GO IDs of parents
> parents <- get(GOID,env=GOMFPARENTS)
> parents
         isa
"GO:0004672"
> mget(parents,env=GOTERM)
$"GO:0004672"
GOID = GO:0004672
```

```
Term = protein kinase activity
Definition = Catalysis of the transfer of a phosphate
     group, usually from ATP, to a protein substrate.
Ontology = MF

>
> ## Get GO IDs of ancestors
> ancestors <- get(GOID,env=GOMFANCESTOR)
> ancestors
[1] "all"       "GO:0003674" "GO:0003824" "GO:0016740"
[5] "GO:0016772" "GO:0016773" "GO:0016301" "GO:0004672"
>
> ## Get GO IDs of children
> children <- get(GOID,env=GOMFCHILDREN)
> children
[1] "GO:0004714" "GO:0004715" "GO:0004716"
>
> ## Get GO IDs of offspring
> offspring <- get(GOID,env=GOMFOFFSPRING)
> offspring
 [1] "GO:0004714" "GO:0004715" "GO:0004716" "GO:0005020"
 [5] "GO:0005021" "GO:0005023" "GO:0005010" "GO:0005011"
 [9] "GO:0005017" "GO:0005003" "GO:0005006" "GO:0005007"
[13] "GO:0005008" "GO:0005009" "GO:0008288" "GO:0005018"
[17] "GO:0005019" "GO:0005004" "GO:0005005" "GO:0008313"
[21] "GO:0004718"
```

As already noted in Example 2 and Figure 2, the term corresponding to the GO term ID GO:0004713 is *protein-tyrosine kinase activity*, in the Molecular Function ontology. It has one (immediate) parent term, *protein kinase activity*.

### 4.4. *Affymetrix chip-specific annotation metadata packages: The* hgu95av2 *package*

The Bioconductor Project provides *Affymetrix chip-specific annotation metadata packages* for the main chip series for the human, mouse, rat, and other genomes (e.g., HG-U133, HG-U95, HU-6800, MG-U74, and RG-U34 series). These packages, built using the infrastructure package AnnBuilder, contain environment objects for mappings between Affymetrix probe IDs and other types of gene/probe identifiers.

Note that analogous packages are not supplied for two-color spotted microarrays, as there is no standard microarray design for this type of platform and specialized annotation metadata packages may have to be created for each microarray facility (e.g., using AnnBuilder). Once annotation metadata packages are available to provide mappings between different sets of gene/probe identifiers, the tools in annotate and related packages may be used in a similar manner for any type of microarray platform.

Consider the hgu95av2 package, for the Affymetrix chip series HG-U95Av2. This package provides the following environments.

```
> ? hgu95av2
```

```
> hgu95av2()

Quality control information for hgu95av2
Date built: Created: Tue Oct 4 21:31:35 2005

Number of probes: 12625
Probe number missmatch: None
Probe missmatch: None
Mappings found for probe based rda files:
        hgu95av2ACCNUM found 12625 of 12625
        hgu95av2CHRLOC found 11673 of 12625
        hgu95av2CHR found 12145 of 12625
        hgu95av2ENZYME found 1886 of 12625
        hgu95av2GENENAME found 11418 of 12625
        hgu95av2GO found 9942 of 12625
        hgu95av2LOCUSID found 12203 of 12625
        hgu95av2MAP found 12109 of 12625
        hgu95av2OMIM found 9881 of 12625
        hgu95av2PATH found 3928 of 12625
        hgu95av2PMID found 12086 of 12625
        hgu95av2REFSEQ found 12008 of 12625
        hgu95av2SUMFUNC found 0 of 12625
        hgu95av2SYMBOL found 12159 of 12625
        hgu95av2UNIGENE found 12118 of 12625
Mappings found for non-probe based rda files:
        hgu95av2CHRLENGTHS found 25
        hgu95av2ENZYME2PROBE found 643
        hgu95av2GO2ALLPROBES found 5480
        hgu95av2GO2PROBE found 3890
        hgu95av2ORGANISM found 1
        hgu95av2PATH2PROBE found 155
        hgu95av2PFAM found 10439
        hgu95av2PMID2PROBE found 98214
        hgu95av2PROSITE found 8249
```

For information on any of these environments, use the `help` function, e.g., `help(hgu95av2GO)` or `?hgu95av2GO`. We focus on the three environments related to GO: `hgu95av2GO`, `hgu95av2GO2ALLPROBES`, and `hgu95av2GO2PROBE`.

The HG-U95Av2 chip contains 12,625 probes (corresponding to the keys in the `hgu95av2GO` environment), with the first 10 Affymetrix probe IDs listed below.

```
> ## List all Affymetrix IDs
> AffyID <- ls(env = hgu95av2GO)
> length(AffyID)
[1] 12625
> AffyID[1:10]
  [1] "1000_at"    "1001_at"    "1002_f_at"  "1003_s_at"  "1004_at"
  [6] "1005_at"    "1006_at"    "1007_s_at"  "1008_f_at"  "1009_at"
```

*4.4.1. Probes-to-most specific GO terms mappings: The* `hgu95av2GO` *environment*

The `hgu95av2GO` environment provides key-value pairs for the mappings from *Affymetrix probe IDs* (keys) to *GO term IDs* (values). Each Affymetrix probe ID is mapped to a list of one or more elements, where each element corresponds to a particular GO term and is itself a list with the following three elements.

- `"GOID"`: A GO term ID corresponding to the Affymetrix probe ID (key).
- `"Evidence"`: A code for the evidence supporting the association of the GO term to the Affymetrix probe.
- `"Ontology"`: An abbreviation for the name of the ontology to which the GO term belongs: BP (Biological Process), CC (Cellular Component), or MF (Molecular Function).

Note that only the *directly associated terms* or *most specific terms* (i.e., not their less specific ancestor terms) a probe is annotated with are returned as values in `hgu95av2GO`. The GO package may be used to obtain more information on the GO term IDs, e.g., GO term, (all) ancestors, (immediate) parents, (immediate) children, and (all) offspring (Section 4.3).

**Example 5. GO terms directly associated with Affymetrix probe ID** `1635_at`. Let us obtain GO annotation information for the probe with Affymetrix ID `1635_at`, corresponding to the `ABL1` gene. The code below shows that probe `1635_at` is directly annotated with 14 distinct GO terms (the same GO term ID may be returned multiple times with a different evidence code). As already noted in Example 1, one of these terms, with GO term ID `GO:0004713`, is *protein-tyrosine kinase activity*, in the Molecular Function ontology.

```
> probe <- "1635_at"
> probe2GO <- get(probe, env = hgu95av2GO)
> length(probe2GO)
[1] 14
> unique(names(probe2GO))
 [1] "GO:0000074" "GO:0000115" "GO:0000166" "GO:0003677"
 [5] "GO:0004713" "GO:0005515" "GO:0005524" "GO:0005634"
 [9] "GO:0006298" "GO:0006355" "GO:0006468" "GO:0007242"
[13] "GO:0008630" "GO:0016740"
> probe2GO[[5]]
$GOID
[1] "GO:0004713"

$Evidence
[1] "TAS"

$Ontology
[1] "MF"

> get(probe2GO[[5]]$GOID, env=GOTERM)
GOID = GO:0004713
Term = protein-tyrosine kinase activity
Synonym = protein tyrosine kinase activity
Definition = Catalysis of the reaction: ATP + a protein
    tyrosine = ADP + protein tyrosine phosphate.
Ontology = MF
```

The `hgu95av2GO` environment (and analogous environments for other chip series) may be used to assemble an Affymetrix probe ID-by-GO term ID gene-annotation matrix, row by row. This may entail, however, a number of data processing steps. Firstly, only the most specific terms a probe is annotated with are returned as values in `hgu95av2GO`. One therefore needs to add all (less specific) ancestor terms in order to comply with the true path rule. Secondly, several probes may correspond to the same gene, i.e., several Affymetrix probe IDs may map to the same Entrez Gene ID according to the environment `hgu95av2LOCUSID`. Thirdly, the `hgu95av2GO` environment returns GO terms for all three gene ontologies at once. One may need to separate terms according to membership in the BP, CC, and MF ontologies (e.g., using the `GOTERM` environment from the `GO` package).

Alternately, one may assemble an Affymetrix probe ID-by-GO term ID gene-annotation matrix, column by column, using the `hgu95av2GO2ALLPROBES` environment described below.

*4.4.2. GO terms-to-directly annotated probes mappings: The `hgu95av2GO2PROBE` environment*

The `hgu95av2GO2PROBE` environment provides key-value pairs for the mappings from *GO term IDs* (keys) to *Affymetrix probe IDs* (values). Values are vectors of length one or greater depending on whether a given GO term ID is mapped to one or more Affymetrix probe IDs. The value names are evidence codes for the GO term IDs.

Note that the probes a particular GO term is mapped to are only those associated *directly* with the GO term (vs. indirectly via its immediate children or more distant offspring). For a list of all probes associated directly or indirectly with a particular GO term, one may use the `hgu95av2GO2ALLPROBES` environment.

**Example 6. Affymetrix probes directly associated with GO term ID** `GO:0004713`. In the following example, 205 distinct Affymetrix probe IDs are associated directly with the GO term *protein-tyrosine kinase activity* (`GO:0004713`). The Affymetrix probe IDs include `1635_at`, corresponding to the `ABL1` gene.

```
> GOID <- "GO:0004713"
> GO2Probes <- get(GOID, env = hgu95av2GO2PROBE)
> length(unique(GO2Probes))
[1] 205
> GO2Probes[1:10]
       <NA>      <NA>       <NA>       <NA>       <NA>
    "1635_at" "1636_g_at" "1656_s_at" "2040_s_at" "2041_i_at"
        TAS       IEA        IEA        IEA         TAS
    "39730_at"  "1084_at" "35162_s_at" "1564_at"    "854_at"
> is.element("1635_at", GO2Probes)
[1] TRUE
```

*4.4.3. GO terms-to-all annotated probes mappings: The `hgu95av2GO2ALLPROBES` environment*

The `hgu95av2GO2ALLPROBES` environment provides key-value pairs for the mappings from *GO term IDs* (keys) to *Affymetrix probe IDs* (values). Values are vectors of length one or greater depending on whether a given GO term ID is mapped to one or more Affymetrix probe IDs. The value names are evidence codes for the GO term IDs.

Note that, in accordance with the true path rule, the probes a particular GO term is mapped to are associated either *directly* with the GO term or *indirectly* via any of its immediate children or more distant offspring. The main difference between the `hgu95av2GO2PROBE` and `hgu95av2GO2ALLPROBES` environments is that the former considers only the GO term itself, whereas the latter considers the GO term and any of its offspring. Thus, the Affymetrix probe IDs returned by `hgu95av2GO2PROBE` are a subset of the probe IDs returned by `hgu95av2GO2ALLPROBES`.

**Example 7. Affymetrix probes directly or indirectly associated with GO term ID `GO:0004713`.** In the following example, 319 distinct Affymetrix probe IDs (some with multiple evidence codes) are associated either directly or indirectly with the GO term *protein-tyrosine kinase activity* (`GO:0004713`). This list of 319 Affymetrix probe IDs indeed includes the list of 205 probe IDs associated directly with the GO term ID `GO:0004713`.

```
> GOID <- "GO:0004713"
> GO2AllProbes <- get(GOID, env = hgu95av2GO2ALLPROBES)
> length(GO2AllProbes)
[1] 370
> length(unique(GO2AllProbes))
[1] 319
> sum(is.element(GO2Probes,GO2AllProbes))
[1] 205
```

The `hgu95av2GO2ALLPROBES` environment immediately yields an Affymetrix probe ID-by-GO term ID gene-annotation matrix, column by column. However, as with the `hgu95av2GO` environment, a number of data processing steps may be required, concerning, for example, uniqueness of Entrez Gene IDs and membership in the BP, CC, and MF ontologies.

### 4.5. Assembling a GO gene-annotation matrix

This section provides R code for assembling an Entrez Gene ID-by-GO term ID gene-annotation matrix $A$, column by column. Specifically, rows correspond to (unique) Entrez Gene IDs mapping to probes on the HG-U95Av2 chip and columns to terms in the Molecular Function ontology mapping directly or indirectly to at least 10 Entrez Gene IDs for the HG-U95Av2 chip.

In practice, it may not be desirable to build the full $G \times M$ gene-annotation matrix, as this matrix could potentially be very large and sparse (padded with zeros). Rather, we assemble a (smaller) *gene-annotation list*, that provides, for each GO term ID, a list of Entrez Gene IDs annotated with the GO term.

**Example 8. Entrez Gene ID-by-GO term ID gene-annotation matrix for MF ontology.**

```
> ## List all Affymetrix IDs for HG-U95Av2 chip
> AffyID <- ls(env=hgu95av2GO)
> length(AffyID)
[1] 12625
>
> ## Get all unique Entrez Gene IDs for HG-U95Av2 chip
> LLID <- as.character(unique(unlist(mget(AffyID,
```

```
+     env=hgu95av2LOCUSID))))
> length(LLID)
[1] 9085
>
> ## Get MF GO IDs
> GOID <- ls(env=GOTERM)
> O <- unlist(lapply(mget(GOID, env=GOTERM),
+     function(z) z@Ontology))
> table(O)
O
  BP    CC    MF
9888  1612  7334
> MFID <- GOID[O=="MF"]
>
> ## For each MF GO ID, get all Entrez Gene IDs for genes
> ## annotated directly or indirectly with the GO term
> allMFLLID <- mget(MFID, env=GOALLLOCUSID)
>
> ## For each MF GO ID, get HG-U95Av2-specific Entrez Gene IDs
> ## for genes annotated directly or indirectly with the GO term
> MFLLID <- lapply(allMFLLID, function(z) intersect(z, LLID))
> numMFLLID <- unlist(lapply(MFLLID, length))
> summary(numMFLLID)
    Min.  1st Qu.  Median    Mean  3rd Qu.    Max.
   0.000    1.000   1.000    9.539   1.000  6762.000
>
> ## Retain only MF GO IDs that map to at least 10
> ## Entrez Gene IDs for the HG-U95Av2 chip
> MFAnnotList <- MFLLID[numMFLLID > 9]
> length(MFAnnotList)
[1] 466
> summary(unlist(lapply(MFAnnotList, length)))
    Min.  1st Qu. Median  Mean 3rd Qu.   Max.
    10.0     16.0   27.5  132.2   70.0 6762.0
> MFAnnotList[1]
$"GO:0000146"
 [1] "4620"  "4621"  "4624"  "4625"  "4640"  "4643"  "4644"
 [8] "4646"  "4647"  "4650"  "58498"

>
> ## Get Entrez Gene IDs for probes annotated with GO ID
> ## GO:0004713
> is.element("GO:0004713",names(MFAnnotList))
[1] TRUE
> length(MFAnnotList["GO:0004713"][[1]])
[1] 180
```

## 5. Tests of association between GO annotation and differential gene expression in ALL

### 5.1. Acute lymphoblastic leukemia study of Chiaretti et al. [13]

Our proposed approach to tests of association with biological annotation metadata is illustrated using the *acute lymphoblastic leukemia* (ALL) microarray dataset of [13] and Gene Ontology (GO) annotation metadata.

#### 5.1.1. Bioconductor experimental data R package ALL

The ALL dataset is available in the Bioconductor experimental data R package ALL (Version 1.0.2, Bioconductor Release 1.7). The main object in this package is ALL, an instance of the class *exprSet*. The exprs slot of ALL provides a matrix of 12,625 *microarray expression measures* (Affymetrix chip series HG-U95Av2) for each of 128 ALL cell samples. The phenoData slot contains 21 *phenotypes* (i.e., covariates and outcomes) for each of the 128 cell samples. Phenotypes of interest include: ALL$BT, the type and stage of the cancer (i.e., B-cell ALL or T-cell ALL, of stage 1, 2, 3, or 4), and ALL$mol.biol, the molecular class of the cancer (i.e., BCR/ABL, NEG, ALL1/AF4, E2A/PBX1, p15/p16, or NUP-98).

The expression measures have been obtained using the three-step robust multichip average (RMA) pre-processing method, implemented in the Bioconductor R package affy [11], and have been subjected to a base 2 logarithmic transformation.

For greater detail on the ALL dataset, please consult the ALL package documentation.

#### 5.1.2. The BCR/ABL fusion

A number of recent articles have investigated the prognostic relevance of the *BCR/ABL fusion* in adult ALL of the B-cell lineage [21].

The BCR/ABL fusion is the molecular analogue of the *Philadelphia chromosome*, one of the most frequent cytogenetic abnormalities in human leukemias. This *t(9;22) translocation* leads to a head-to-tail fusion of the v-abl Abelson murine leukemia viral oncogene homolog 1 (ABL1) from chromosome 9 with the 5' half of the breakpoint cluster region (BCR) on chromosome 22 (Figure 3). The ABL1 proto-oncogene encodes a cytoplasmic and nuclear protein tyrosine kinase that has been implicated in processes of cell differentiation, cell division, cell adhesion, and stress response. Although the BCR/ABL fusion protein, encoded by sequences from both the ABL1 and BCR genes, has been extensively studied, the function of the normal product of the BCR gene is not clear. The BCR/ABL proto-oncogene has been found to be highly expressed in chronic myeloid leukemia (CML) and acute myeloid leukemia (AML) cells [30].

An interesting question is therefore the identification of genes that are differentially expressed between B-cell ALL with the BCR/ABL fusion and cytogenetically normal NEG B-cell ALL.

In order to address this question, we consider gene expression measures for the $n = 79$ B-cell ALL cell samples (ALL$BT equal to B, B1, B2, B3, or B4), of the BCR/ABL or NEG molecular types (ALL$mol.biol equal to BCR/ABL or NEG).

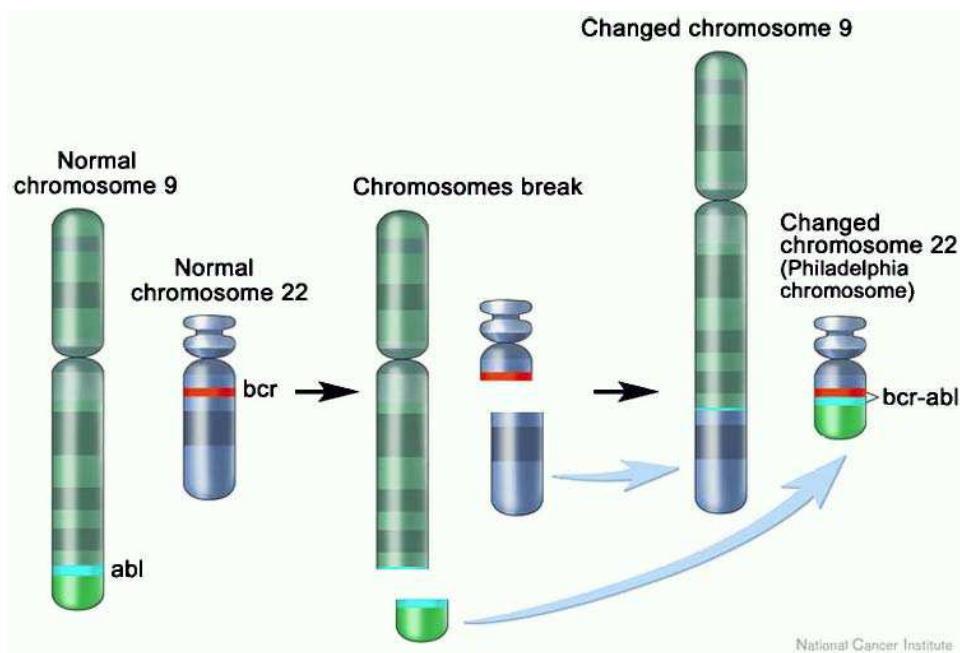

FIG 3. *The Philadelphia chromosome and the BCR/ABL fusion.* The BCR/ABL fusion is the molecular analogue of the Philadelphia chromosome. This t(9;22) translocation leads to a head-to-tail fusion of the `v-abl Abelson murine leukemia viral oncogene homolog 1` (`ABL1`) from chromosome 9 with the 5' half of the `breakpoint cluster region` (`BCR`) on chromosome 22. Figure obtained from the National Cancer Institute website (`www.cancer.gov/Templates/db_alpha.aspx?CdrID=44179`). (Color version on website companion.)

### 5.1.3. Gene filtering

Many of the genes represented by the 12,625 probes are not expressed in B-cell lymphocytes. Accordingly, as in [44], we only retain the 2,391 probes that meet the following two criteria: (i) fluorescence intensities greater than 100 (absolute scale) for at least 25% of the 79 cell samples; (ii) interquartile range (IQR) of the fluorescence intensities for the 79 cell samples greater than 0.5 (log base 2 scale).

Furthermore, different probes may correspond to the same gene, i.e., map to the same Entrez Gene ID, according to the environment `hgu95av2LOCUSID` from the `hgu95av2` package. In order to obtain one expression measure per gene, we choose to average the expression measures of multiple probes mapping to the same gene.

These various pre-processing steps lead to $G = 2,071$ genes with unique Entrez Gene IDs.

### 5.1.4. Reduced ALL dataset: Genotypes and phenotypes of interest

The combined genotypic and phenotypic data for the $n = 79$ B-cell ALL cell samples of the BCR/ABL and NEG molecular types may be summarized by the set $\mathcal{XY}_n \equiv \{(X_i, Y_i) : i = 1, \ldots, n\}$, of $n$ pairs of $G$-dimensional gene expression profiles $X_i = (X_i(g) : g = 1, \ldots, G)$, $G = 2,071$, and cancer class labels $Y_i \in \{NEG, BCR/ABL\}$. Among the $n = 79$ B-cell ALL cell samples, there are $n_{BCR/ABL} \equiv \sum_i I(Y_i = BCR/ABL) = 37$ BCR/ABL samples and $n_{NEG} \equiv \sum_i I(Y_i = NEG) = 42$ NEG samples.

### 5.2. Multiple hypothesis testing framework

Our primary question of interest is the identification of genes that are *differentially expressed* (DE) between BCR/ABL and NEG B-cell ALL. A subsequent question involves *relating differential gene expression to GO annotation*.

As detailed below, GO annotation metadata for the filtered list of $G = 2,071$ unique genes from the HG-U95Av2 chip may be summarized by binary gene-annotation profiles.

The gene-parameter profiles of interest concern differential gene expression between BCR/ABL and NEG B-cell ALL, i.e., the association between microarray gene expression measures and cancer class. Continuous gene-parameter profiles of unstandardized and standardized measures of differential expression are estimated, respectively, by (unstandardized) differences of empirical means and (standardized) two-sample $t$-statistics. Binary gene-parameter profiles, indicating whether genes are differentially expressed, are estimated by imposing cut-off rules on two-sample $t$-statistics or adjusted $p$-values.

The following association measures between GO gene-annotation profiles and DE gene-parameter profiles are considered: two-sample $t$-statistics for tests of association between binary GO gene-annotation profiles and continuous DE gene-parameter profiles; $\chi^2$-statistics for tests of association between binary GO gene-annotation profiles and binary DE gene-parameter profiles.

Significant associations between differential gene expression and GO annotation are identified by applying FWER-controlling single-step maxT Procedure 1, based on the non-parametric bootstrap null shift and scale-transformed test statistics null distribution of Procedure 2.

#### 5.2.1. Gene-annotation profiles

Gene Ontology annotation metadata for the HG-U95Av2 chip series are obtained as described in Sections 4.2–4.5, from the following Bioconductor R packages: the GO-specific metadata package GO (Version 1.10.0, Bioconductor Release 1.7) and the Affymetrix chip-specific metadata package hgu95av2 (Version 1.10.0, Bioconductor Release 1.7).

For each of the three gene ontologies, *binary gene-annotation matrices* $A_{BP}$, $A_{CC}$, and $A_{MF}$, are assembled for the GO terms annotating at least 10 of the $G = 2,071$ filtered genes (sample R code provided in Section 4.5). Specifically, for gene ontology $o \in \{BP, CC, MF\}$, $A_o = (A_o(g,m) : g = 1, \ldots, G; m = 1, \ldots, M_o)$ is a $G \times M_o$ matrix, with element $A_o(g,m)$ indicating whether gene $g$ is annotated by GO term $m$ and such that $\sum_g A_o(g,m) \geq 10$ for each term $m$. The numbers of terms considered in each gene ontology are $M_{BP} = 367$, $M_{CC} = 81$, and $M_{MF} = 185$.

#### 5.2.2. Gene-parameter profiles

**Definition of gene-parameter profiles**   Consider a data structure $(X, Y) \sim P$, where $X = (X(g) : g = 1, \ldots, G)$ is a $G = 2,071$-dimensional vector of microarray gene expression measures and $Y \in \{NEG, BCR/ABL\}$ is a binary cancer class label. Let $\eta_k \equiv \Pr(Y = k)$ denote the proportion of cancers of class $k \in \{NEG, BCR/ABL\}$. Define conditional $G$-dimensional mean vectors and $G \times G$ covariance matrices for the expression measures of class $k \in \{NEG, BCR/ABL\}$ cancers by

$$\mu_k \equiv \mathrm{E}[X|Y=k] \quad \text{and} \quad \sigma_k \equiv \mathrm{Cov}[X|Y=k], \tag{43}$$

respectively.

*Gene-parameter profiles*, concerning differential gene expression between BCR/ABL and NEG B-cell ALL, may be specified in various ways. *Continuous* DE gene-parameter profiles may be defined in terms of the following *unstandardized* and *standardized* measures of differential gene expression between BCR/ABL and NEG B-cell ALL,

$$\lambda^d(g) \equiv \mu_{BCR/ABL}(g) - \mu_{NEG}(g) \tag{44}$$

and

$$\lambda^t(g) \equiv \frac{\mu_{BCR/ABL}(g) - \mu_{NEG}(g)}{\sqrt{\frac{\sigma_{BCR/ABL}(g,g)}{\eta_{BCR/ABL}} + \frac{\sigma_{NEG}(g,g)}{\eta_{NEG}}}}.$$

Absolute values of $\lambda^d(g)$ and $\lambda^t(g)$ may be used for measuring two-sided differential expression, i.e., either over- or under-expression in BCR/ABL compared to NEG B-cell ALL.

*Binary* DE gene-parameter profiles may be defined in terms of indicators for two-sided and one-sided differential expression.

$$\begin{aligned}
\lambda^{\neq}(g) &\equiv \mathrm{I}\left(\mu_{BCR/ABL}(g) \neq \mu_{NEG}(g)\right) \\
&= \mathrm{I}\left(\lambda^d(g) \neq 0\right) = \mathrm{I}\left(\lambda^t(g) \neq 0\right), \\
\lambda^+(g) &\equiv \mathrm{I}\left(\mu_{BCR/ABL}(g) > \mu_{NEG}(g)\right) \\
&= \mathrm{I}\left(\lambda^d(g) > 0\right) = \mathrm{I}\left(\lambda^t(g) > 0\right), \\
\lambda^-(g) &\equiv \mathrm{I}\left(\mu_{BCR/ABL}(g) < \mu_{NEG}(g)\right) \\
&= \mathrm{I}\left(\lambda^d(g) < 0\right) = \mathrm{I}\left(\lambda^t(g) < 0\right).
\end{aligned} \tag{45}$$

**Estimation of gene-parameter profiles** The above DE gene-parameter profiles may be estimated as follows, based on the sample $\mathcal{XY}_n$ of gene expression measures for the $n = 79$ B-cell ALL cell samples of the BCR/ABL and NEG molecular types.

The *resubstitution estimators* of the continuous gene-parameter profiles of Equation (44) are based, respectively, on differences of empirical means and two-sample Welch $t$-statistics (up to the multiplier $1/\sqrt{n}$). That is,

$$\lambda_n^d(g) \equiv \mu_{BCR/ABL,n}(g) - \mu_{NEG,n}(g) \tag{46}$$

and

$$\lambda_n^t(g) \equiv \frac{1}{\sqrt{n}} \frac{\mu_{BCR/ABL,n}(g) - \mu_{NEG,n}(g)}{\sqrt{\frac{\sigma_{BCR/ABL,n}(g,g)}{n_{BCR/ABL}} + \frac{\sigma_{NEG,n}(g,g)}{n_{NEG}}}},$$

where $\mu_{k,n}(g) \equiv \sum_i \mathrm{I}(Y_i = k) X_i(g)/n_k$ and $\sigma_{k,n}(g,g) \equiv \sum_i \mathrm{I}(Y_i = k)(X_i(g) - \mu_{k,n}(g))^2/(n_k - 1)$ denote, respectively, the empirical means and variances of the gene expression measures for cancers of class $k \in \{NEG, BCR/ABL\}$.

Estimating the two-sided binary gene-parameter profile $\lambda^{\neq}$ of Equation (45) involves the *two-sided test* of the $G$ null hypotheses $H_0(g) = \mathrm{I}(\mu_{BCR/ABL}(g) = \mu_{NEG}(g))$, of no differences in mean gene expression measures between BCR/ABL and NEG B-cell ALL. Likewise, estimating the one-sided binary gene-parameter profiles $\lambda^+$ and $\lambda^-$ involves, respectively, the *one-sided test* of the $G$ null hypotheses of no over-expression ($H_0(g) = \mathrm{I}(\mu_{BCR/ABL}(g) \leq \mu_{NEG}(g))$) and no under-expression ($H_0(g) = \mathrm{I}(\mu_{BCR/ABL}(g) \geq \mu_{NEG}(g))$) in BCR/ABL compared to NEG B-cell ALL. For single-step common-cut-off maxT Procedure 1, adjusted $p$-values produce the same gene ranking as the test statistics defined in Equation (46). Simple and naive estimators of the three sets of differentially expressed genes (i.e., false null hypotheses), represented by the gene-parameter profiles $\lambda^{\neq}$, $\lambda^+$, and $\lambda^-$, are therefore given, respectively, by the sets of genes with the largest $\gamma G$ values of $|\lambda_n^t(g)|$, $\lambda_n^t(g)$, and $-\lambda_n^t(g)$. That is,

$$
\begin{aligned}
(47) \quad \lambda_{n,\gamma G}^{\neq}(g) &\equiv \mathrm{I}\left(\sum_{g'=1}^{G} \mathrm{I}\left(|\lambda_n^t(g)| \geq |\lambda_n^t(g')|\right) > (1-\gamma)G\right), \\
\lambda_{n,\gamma G}^{+}(g) &\equiv \mathrm{I}\left(\sum_{g'=1}^{G} \mathrm{I}\left(\lambda_n^t(g) \geq \lambda_n^t(g')\right) > (1-\gamma)G\right), \\
\lambda_{n,\gamma G}^{-}(g) &\equiv \mathrm{I}\left(\sum_{g'=1}^{G} \mathrm{I}\left(-\lambda_n^t(g) \geq -\lambda_n^t(g')\right) > (1-\gamma)G\right).
\end{aligned}
$$

Analogous estimators may also be based on other test statistics, such as unstandardized difference statistics $\lambda_n^d$.

More sophisticated estimators, that translate the proportion $\gamma$ of rejected null hypotheses into a Type I error rate such as the gFWER, TPPFP, or FDR, could be based on adjusted $p$-values for the multiple test of the $G$ null hypotheses $H_0(g)$. For example, one could estimate $\lambda^{\neq}$ by

$$
(48) \qquad \lambda_{n,\alpha}^{\neq}(g) \equiv \mathrm{I}\left(\widetilde{P}_{0n}^{\neq}(g) \leq \alpha\right),
$$

where $\widetilde{P}_{0n}^{\neq}(g)$ are adjusted $p$-values for a suitably chosen multiple testing procedure, such as, FWER-controlling single-step maxT Procedure 1 or a TPPFP-controlling augmentation multiple testing procedure (Chapter 6 in [14], [40]). One-sided binary gene-parameter profiles $\lambda^+$ and $\lambda^-$ could be estimated likewise.

*5.2.3. Association measures for gene-annotation and gene-parameter profiles*

The association between continuous DE gene-parameter profiles, as in Equation (44), and binary GO gene-annotation profiles may be measured by *two-sample Welch t-statistics* (or corresponding $p$-values). Specifically, given a continuous $G$-vector $x$ and a binary $G$-vector $y$, define the following association measure,

$$
(49) \qquad \rho^t(x,y) \equiv \frac{\bar{x}_1 - \bar{x}_0}{\sqrt{\frac{v[x]_1}{y_1} + \frac{v[x]_0}{y_0}}},
$$

where $y_k \equiv \sum_g \mathrm{I}(y(g) = k)$, $\bar{x}_k \equiv \sum_g \mathrm{I}(y(g) = k)\, x(g)/y_k$, and $v[x]_k \equiv \sum_g \mathrm{I}(y(g) = k)\,(x(g) - \bar{x}_k)^2/(y_k - 1)$, $k \in \{0,1\}$.

The association between binary DE gene-parameter profiles, as in Equation (45), and binary GO gene-annotation profiles may be measured by $\chi^2$-*statistics* (or corresponding *p*-values) for the test of independence of rows and columns in a $2 \times 2$ contingency table, such as Table 2. Specifically, given binary $G$-vectors $x$ and $y$, define the following association measure,

$$(50) \qquad \rho^\chi(x,y) \equiv \frac{G(g_{00}g_{11} - g_{01}g_{10})^2}{(g_{00}+g_{01})(g_{00}+g_{10})(g_{11}+g_{01})(g_{11}+g_{10})},$$

where $g_{kk'} \equiv \sum_g \mathrm{I}(x(g)=k)\mathrm{I}(y(g)=k')$, $k,k' \in \{0,1\}$.

Given an association measure[3] $\rho : \mathbb{R}^{G \times M} \times \mathbb{R}^G \to \mathbb{R}^M$, a $G \times M$ GO gene-annotation matrix $A$, and a $G$-dimensional DE gene-parameter profile $\lambda = \Lambda(P)$, the $M$-dimensional *association parameter* vector $\psi = \Psi(P)$ of primary interest is defined as

$$(51) \qquad \psi \equiv \rho(A, \lambda).$$

The corresponding *resubstitution estimator* $\psi_n = \hat{\Psi}(P_n)$ is simply obtained by replacing the gene-parameter profile $\lambda$ by an estimator thereof $\lambda_n = \hat{\Lambda}(P_n)$, that is,

$$(52) \qquad \psi_n \equiv \rho(A, \lambda_n).$$

### 5.2.4. Null and alternative hypotheses

For the *t*-statistic-based association measure $\rho^t$ of Equation (49), the identification of GO terms $m$ that are significantly (positively or negatively) associated with BCR/ABL vs. NEG differential gene expression involves the *two-sided test* of the $M$ null hypotheses $H_0(m) = \mathrm{I}(\psi(m) = \psi_0(m))$ against the alternative hypotheses $H_1(m) = \mathrm{I}(\psi(m) \neq \psi_0(m))$, with null values $\psi_0(m) = 0$. In some contexts, one may be interested in identifying positive (negative) associations, i.e., in the *one-sided test* of the $M$ null hypotheses $H_0(m) = \mathrm{I}(\psi(m) \leq \psi_0(m))$ ($H_0(m) = \mathrm{I}(\psi(m) \geq \psi_0(m))$) against the alternative hypotheses $H_1(m) = \mathrm{I}(\psi(m) > \psi_0(m))$ ($H_1(m) = \mathrm{I}(\psi(m) < \psi_0(m))$).

For the $\chi^2$-statistic-based association measure $\rho^\chi$ of Equation (50), the identification of GO terms $m$ that are significantly (positively or negatively) associated with BCR/ABL vs. NEG differential gene expression involves the *one-sided test* of the $M$ null hypotheses $H_0(m) = \mathrm{I}(\psi(m) \leq \psi_0(m))$ against the alternative hypotheses $H_1(m) = \mathrm{I}(\psi(m) > \psi_0(m))$. A natural choice for the null values is the mean of the $\chi^2(1)$-distribution, $\psi_0(m) = 1$.

### 5.2.5. Test statistics

One-sided and two-sided tests of null hypotheses concerning any of the association parameters defined above may be based on (unstandardized) *difference statistics* $T_n(m)$, defined as in Equation (40).

For one-sided tests, large values of the test statistics $T_n(m)$ provide evidence against the corresponding null hypotheses $H_0(m)$, that is, rejection regions are of the form $\mathcal{C}_n(m) = (c_n(m), +\infty)$. For two-sided tests, large values of the absolute test statistics $|T_n(m)|$ provide evidence against the corresponding null hypotheses $H_0(m)$.

---

[3]N.B. For ease of notation, $\rho^t$ and $\rho^\chi$, defined in Equations (49) and (50) as real-valued association measures, may also refer loosely to $\mathbb{R}^M$-valued association measures, defined as $\rho^t(X,y) \equiv (\rho^t(X(\cdot,m),y) : m = 1,\ldots,M)$ and $\rho^\chi(X,y) \equiv (\rho^\chi(X(\cdot,m),y) : m = 1,\ldots,M)$ for $X \in \mathbb{R}^{G \times M}$ and $y \in \mathbb{R}^G$.

### 5.2.6. Multiple testing procedures

For the purpose of illustration, we focus on control of the *family-wise error rate*, using *single-step maxT Procedure 2.9*, based on the *non-parametric bootstrap null shift-transformed test statistics null distribution* of Procedure 2.9 (null shift values $\lambda_0(m) = 0$ and no scaling).

Let $O_n(m)$ denote indices for the ordered adjusted $p$-values, so that $\widetilde{P}_{0n}(O_n(1)) \leq \cdots \leq \widetilde{P}_{0n}(O_n(M))$. GO terms with adjusted $p$-values less than or equal to $\alpha$ are declared significantly associated with differential gene expression at nominal FWER level $\alpha$. That is, the list of GO terms found to be associated with differential gene expression is

$$(53) \qquad \mathcal{R}_n(\alpha) \equiv \left\{m : \widetilde{P}_{0n}(m) \leq \alpha\right\} = \{O_n(1), \ldots, O_n(R_n(\alpha))\},$$

where $R_n(\alpha) \equiv |\mathcal{R}_n(\alpha)|$ denotes the number of identified GO terms.

### 5.2.7. Summary of testing scenarios

This section summarizes our approach for identifying GO terms associated with BCR/ABL vs. NEG differential gene expression. For each of the three gene ontologies (i.e., BP, CC, and MF), we consider the following three types of testing scenarios, each corresponding to a different association parameter $\psi = \rho(A, \lambda)$ for GO annotation and BCR/ABL vs. NEG differential gene expression. Scenarios MT$[t, t]$ and MT$[d, t]$ are very similar and correspond, respectively, to *continuous* gene-parameter profiles of *standardized* and *unstandardized* measures of differential gene expression. In contrast, Scenario MT$[\neq, \chi]$ corresponds to a *binary* gene-parameter profile of differential gene expression indicators.

**Scenario MT$[t, t]$: Association parameter $\psi^{t,t} = \rho^t(A, |\lambda^t|)$, for standardized continuous DE gene-parameter profile $\lambda^t$.** Consider the two-sided test of

$$H_0^{t,t}(m) \equiv \mathrm{I}\left(\psi^{t,t}(m) = \psi_0^{t,t}(m)\right)$$

vs.

$$H_1^{t,t}(m) \equiv \mathrm{I}\left(\psi^{t,t}(m) \neq \psi_0^{t,t}(m)\right),$$

where the association parameter vector of interest is defined as

$$\psi^{t,t} \equiv \rho^t(A, |\lambda^t|),$$

based on Equations (44) and (49), and the null values are $\psi_0^{t,t}(m) \equiv 0$. The continuous DE gene-parameter profile $\lambda^t$ is estimated by $\lambda_n^t$, as in Equation (46), and the association parameter $\psi^{t,t}$ is estimated by the resubstitution estimator $\psi_n^{t,t} \equiv \rho^t(A, |\lambda_n^t|)$, as in Equation (52). The test statistics are defined as (unstandardized) difference statistics,

$$T_n^{t,t}(m) \equiv \sqrt{n}(\psi_n^{t,t}(m) - \psi_0^{t,t}(m)),$$

and the null hypotheses $H_0^{t,t}(m)$ are rejected for large absolute values of $T_n^{t,t}(m)$.

**Scenario MT[$d,t$]: Association parameter $\psi^{d,t} = \rho^t(A, |\lambda^d|)$, for unstandardized continuous DE gene-parameter profile $\lambda^d$.** Consider the two-sided test of

$$H_0^{d,t}(m) \equiv \mathrm{I}\left(\psi^{d,t}(m) = \psi_0^{d,t}(m)\right)$$

vs.

$$H_1^{d,t}(m) \equiv \mathrm{I}\left(\psi^{d,t}(m) \neq \psi_0^{d,t}(m)\right),$$

where the association parameter vector of interest is defined as

$$\psi^{d,t} \equiv \rho^t(A, |\lambda^d|),$$

based on Equations (44) and (49), and the null values are $\psi_0^{d,t}(m) \equiv 0$. The continuous DE gene-parameter profile $\lambda^d$ is estimated by $\lambda_n^d$, as in Equation (46), and the association parameter $\psi^{d,t}$ is estimated by the resubstitution estimator $\psi_n^{d,t} \equiv \rho^t(A, |\lambda_n^d|)$, as in Equation (52). The test statistics are defined as (unstandardized) difference statistics,

$$T_n^{d,t}(m) \equiv \sqrt{n}(\psi_n^{d,t}(m) - \psi_0^{d,t}(m)),$$

and the null hypotheses $H_0^{d,t}(m)$ are rejected for large absolute values of $T_n^{d,t}(m)$.

**Scenario MT[$\neq, \chi$]: Association parameter $\psi^{\neq,\chi} = \rho^\chi(A, \lambda^{\neq})$, for binary DE gene-parameter profile $\lambda^{\neq}$.** Consider the one-sided test of

$$H_0^{\neq,\chi}(m) \equiv \mathrm{I}\left(\psi^{\neq,\chi}(m) \leq \psi_0^{\neq,\chi}(m)\right)$$

vs.

$$H_1^{\neq,\chi}(m) \equiv \mathrm{I}\left(\psi^{\neq,\chi}(m) > \psi_0^{\neq,\chi}(m)\right),$$

where the association parameter vector of interest is defined as

$$\psi^{\neq,\chi} \equiv \rho^\chi(A, \lambda^{\neq}),$$

based on Equations (45) and (50), and the null values are $\psi_0^{\neq,\chi}(m) \equiv 1$ (the mean of the $\chi^2(1)$-distribution). The following two types of estimators $\lambda_n^{\neq}$ are considered for the binary DE gene-parameter profile $\lambda^{\neq}$: $\lambda_{n,\gamma G}^{\neq}$, with numbers of DE genes $\gamma G = 20, 50, 100$ (Equation (47)); $\lambda_{n,\alpha}^{\neq}$, defined in terms of adjusted $p$-values for FWER-controlling permutation-based single-step maxT Procedure 2.9 ($B = 1,000$ permutations of the cancer class labels) and nominal FWER level $\alpha = 0.05$ (Equation (48)). Given an estimator $\lambda_n^{\neq}$ of $\lambda^{\neq}$, the association parameter $\psi^{\neq,\chi}$ is estimated by the resubstitution estimator $\psi_n^{\neq,\chi} \equiv \rho^\chi(A, \lambda_n^{\neq})$, as in Equation (52). The test statistics are defined as (unstandardized) difference statistics,

$$T_n^{\neq,\chi}(m) \equiv \sqrt{n}(\psi_n^{\neq,\chi}(m) - \psi_0^{\neq,\chi}(m)),$$

and the null hypotheses $H_0^{\neq,\chi}(m)$ are rejected for large values of $T_n^{\neq,\chi}(m)$.

For each of the three testing scenarios, the null shift-transformed test statistics null distribution $Q_0$ is estimated as in Procedure 2, with $B = 5,000$ non-parametric bootstrap samples of the data $\mathcal{X}\mathcal{Y}_n$ and $Z_n^B(m,b) = T_n^B(m,b) - \mathrm{E}[T_n^B(m,\cdot)]$ (i.e., null shift values $\lambda_0(m) = 0$ and no scaling). For one-sided testing Scenario MT[$\neq, \chi$], bootstrap-based single-step maxT adjusted $p$-values $\widetilde{P}_{0n}(m)$ are computed as in Procedure 2. For two-sided testing Scenarios MT$[t,t]$ and MT$[d,t]$, adjusted $p$-values are computed based on absolute values of $Z_n^B(m,b)$ and $T_n(m)$.

In what follows, the $G$-dimensional gene-parameter profiles $\lambda$ correspond to the $G = 2{,}071$ genes with unique Entrez Gene IDs, obtained as described in Section 5.1. For each of the three gene ontologies, binary gene-annotation matrices are assembled for the GO terms annotating at least 10 of the $G = 2{,}071$ genes of interest: $G = 2{,}071 \times M_{BP} = 367$ gene-annotation matrix $A_{BP}$ for the BP ontology, $G = 2{,}071 \times M_{CC} = 81$ gene-annotation matrix $A_{CC}$ for the CC ontology, and $G = 2{,}071 \times M_{MF} = 185$ gene-annotation matrix $A_{MF}$ for the MF ontology.

### 5.3. Results

#### 5.3.1. Differentially expressed genes between BCR/ABL and NEG B-cell ALL

In order to identify differentially expressed genes between BCR/ABL and NEG B-cell ALL, two-sided tests of the $G$ null hypotheses

$$H_0(g) = \mathrm{I}\left(\mu_{BCR/ABL}(g) = \mu_{NEG}(g)\right)$$

are performed using the two-sample $t$-statistics $\lambda_n^t(g)$ of Equation (46) and FWER-controlling bootstrap-based single-step maxT Procedure 2. Adjusted $p$-values $\widetilde{P}_{0n}^{\neq}(g)$ are obtained using the MTP function from the multtest package (Version 1.8.0, Bioconductor Release 1.7), with $B = 5{,}000$ non-parametric bootstrap samples and other arguments set to their default values.

Figure 4 displays a normal quantile-quantile plot of the test statistics $\lambda_n^t(g)$ (Panel (a)) and a plot of the sorted bootstrap-based single-step maxT adjusted $p$-values $\widetilde{P}_{0n}^{\neq}(g)$ (Panel (b)). A handful of genes stand out in terms of their large absolute test statistics and small adjusted $p$-values.

For control of the FWER at nominal level $\alpha = 0.05$, Procedure 2.9 identifies 16 differentially expressed genes, i.e., 16 genes with $\widetilde{P}_{0n}^{\neq}(g) \leq \alpha$. Table 3 provides

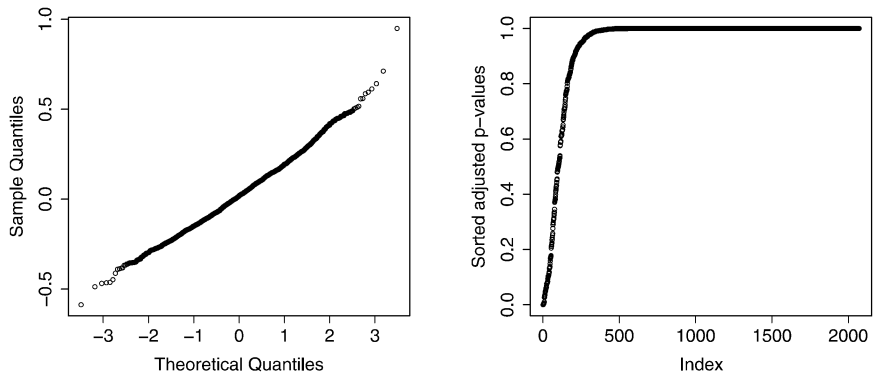

Panel (a): Test statistics  Panel (b): Adjusted $p$-values

FIG 4. *Differentially expressed genes between BCR/ABL and NEG B-cell ALL.* Panel (a): Normal quantile-quantile plot of two-sample $t$-statistics $\lambda_n^t(g)$. Panel (b): Plot of sorted bootstrap-based single-step maxT adjusted $p$-values $\widetilde{P}_{0n}^{\neq}(g)$.

TABLE 3. *Differentially expressed genes between BCR/ABL and NEG B-cell ALL.* This table provides the Affymetrix probe IDs, Entrez Gene IDs (`hgu95av2LOCUSID` environment in `hgu95av2` package), gene symbols (`hgu95av2SYMBOL` environment), gene names (`hgu95av2GENENAME` environment), test statistics $\lambda_n^t(g)$ (Equation (46)), and adjusted $p$-values $\widetilde{P}_{0n}^{\neq}(g)$, for the 16 genes found to be significantly differentially expressed between BCR/ABL and NEG B-cell ALL, at nominal FWER level $\alpha = 0.05$, according to the bootstrap-based single-step maxT procedure, with two-sample $t$-statistics $\lambda_n^t(g)$ and $B = 5,000$ bootstrap samples. A more detailed hyperlinked table, including information on gene function, chromosomal location, links to GenBank, Entrez Gene, NCBI Map Viewer, UniGene, PubMed, AmiGO, and KEGG, is provided on the website companion (Supplementary Table 10.1).

| Probe ID | Entrez Gene ID | Symbol | $\lambda_n^t(g)$ | $\widetilde{P}_{0n}^{\neq}(g)$ |
|---|---|---|---|---|
| 1635_at | 25 | ABL1 | 8.44 | 0.0e+00 |
| v-abl Abelson murine leukemia viral oncogene homolog 1 | | | | |
| 40202_at | 687 | KLF9 | 6.33 | 0.0e+00 |
| Kruppel-like factor 9 | | | | |
| 37027_at | 79026 | AHNAK | 5.71 | 1.4e-03 |
| AHNAK nucleoprotein (desmoyokin) | | | | |
| 39837_s_at | 168544 | ZNF467 | 5.45 | 3.4e-03 |
| zinc finger protein 467 | | | | |
| 33774_at | 841 | CASP8 | 5.29 | 4.2e-03 |
| caspase 8, apoptosis-related cysteine peptidase | | | | |
| 37014_at | 4599 | MX1 | −5.23 | 5.0e-03 |
| myxovirus (influenza virus) resistance 1, interferon-inducible protein p78 (mouse) | | | | |
| 2039_s_at | 2534 | FYN | 5.21 | 5.0e-03 |
| FYN oncogene related to SRC, FGR, YES | | | | |
| 39329_at | 87 | ACTN1 | 4.97 | 9.6e-03 |
| actinin, alpha 1 | | | | |
| 32542_at | 2273 | FHL1 | 4.96 | 1.0e-02 |
| four and a half LIM domains 1 | | | | |
| 40051_at | 9697 | TRAM2 | 4.59 | 2.7e-02 |
| translocation associated membrane protein 2 | | | | |
| 38032_at | 9900 | SV2A | 4.54 | 3.1e-02 |
| synaptic vesicle glycoprotein 2A | | | | |
| 39319_at | 3937 | LCP2 | 4.50 | 3.5e-02 |
| lymphocyte cytosolic protein 2 (SH2 domain containing leukocyte protein of 76 kDa) | | | | |
| 33232_at | 1396 | CRIP1 | 4.46 | 3.7e-02 |
| cysteine-rich protein 1 (intestinal) | | | | |
| 36591_at | 7277 | TUBA1 | 4.37 | 4.4e-02 |
| tubulin, alpha 1 (testis specific) | | | | |
| 38994_at | 8835 | SOCS2 | 4.35 | 4.7e-02 |
| suppressor of cytokine signaling 2 | | | | |
| 40076_at | 7165 | TPD52L2 | −4.33 | 4.8e-02 |
| tumor protein D52-like 2 | | | | |

the test statistics, adjusted $p$-values, and various identifiers for these 16 genes. A more detailed hyperlinked table is posted on the website companion (Supplementary Table 10.1; www.stat.berkeley.edu/~sandrine/MTBook/BAM/BAM.html).

Only 2 of the 16 identified genes have a negative test statistic (`MX1` and `TPD52L2`), suggesting that most DE genes tend to be *over-expressed* in cell samples with the BCR/ABL fusion.

Unsurprisingly, the gene showing the most over-expression in BCR/ABL cell samples, as measured by the $t$-statistics $\lambda_n^t$, is the `ABL1` gene (`v-abl Abelson murine leukemia viral oncogene homolog 1`), located on the long arm of chro-mosome 9 (9q34.1). As mentioned in Section 5.1, the BCR/ABL phenotype is indeed defined in terms of the ABL1 gene.

Furthermore, many of the DE genes seem to be related to apoptosis or oncogenesis. For example, the Kruppel-like factor 9 (KLF9) gene encodes a transcription factor that binds GC-box elements in gene promoter regions. The Krüppel-like factor (KLF) family is comprised of highly related zinc-finger proteins, that are important components of the eukaryotic cellular transcriptional machinery and that take part in a wide range of cellular functions (e.g., cell proliferation, apoptosis, differentiation, and neoplastic transformation). In particular, KLFs have been linked to various cancers [25]. The intron-less gene AHNAK nucleoprotein (desmoyokin) (AHNAK), located on the long arm of chromosome 11 (11q12.2), encodes an unusually large protein ($\approx$ 700 kiloDalton (kDa)) that is typically repressed in cell lines derived from human neuroblastomas and several other types of tumors [35]. Yet another example, the caspase 8, apoptosis-related cysteine peptidase (CASP8) gene, encodes a key enzyme at the top of the apoptotic cascade and has been linked to neuroblastoma [3]. Likewise, other genes listed in Table 3, including MX1, FYN, ACTN1, FHL1, and TRAM2, appear to be related to the molecular biology of cancer.

Our results are in general agreement with those of [44], slight differences being due, most likely, to our preliminary gene filtering, which involves averaging the expression measures of multiple probes mapping to the same Entrez Gene ID.

For greater detail, the interested reader is invited to consult Supplementary Table 10.1 on the website companion and follow links to PubMed and other databases. Further exploration of the DE genes may be performed in R using the Bioconductor packages annotate and annaffy.

### 5.3.2. *GO terms associated with differential gene expression between BCR/ABL and NEG B-cell ALL*

Figure 5 displays, for each of the three gene ontologies and each of the three testing scenarios, plots of the sorted adjusted $p$-values, $\widetilde{P}_{0n}(O_n(1)) \leq \cdots \leq \widetilde{P}_{0n}(O_n(M))$, for FWER-controlling bootstrap-based single-step maxT Procedure 2 ($B = 5,000$ bootstrap samples). The smaller the adjusted $p$-values, the less conservative the procedure, and the longer the list $\mathcal{R}_n(\alpha) = \{m : \widetilde{P}_{0n}(m) \leq \alpha\}$ of identified GO terms at any given nominal Type I error level $\alpha$.

Table 4 summarizes the results in terms of the numbers $R_n(\alpha) = |\mathcal{R}_n(\alpha)|$ of GO terms found to be significantly associated with BCR/ABL vs. NEG differential gene expression for different nominal FWER levels $\alpha$.

In general, adjusted $p$-values tend to be quite large, with only a handful of GO terms identified as being significantly associated with BCR/ABL vs. NEG differential gene expression for nominal FWER level $\alpha \in \{0.05, 0.10, 0.20\}$. The adjusted $p$-values for Scenarios $\mathsf{MT}[t,t]$ and $\mathsf{MT}[d,t]$ (red and blue plotting symbols), corresponding, respectively, to standardized and unstandardized continuous DE gene-parameter profiles, are similar. For the BP and MF gene ontologies, Scenario $\mathsf{MT}[t,t]$ seems to be slightly more conservative than Scenario $\mathsf{MT}[d,t]$; however, this does not hold for the CC ontology. Scenario $\mathsf{MT}[\neq,\chi]$, with four different estimators of the binary DE gene-parameter profile $\lambda^{\neq}$, tends to be more conservative than either Scenario $\mathsf{MT}[t,t]$ or $\mathsf{MT}[d,t]$. Furthermore, the choice of parameter $\gamma G$, for the number of genes called differentially expressed, can have a substantial impact on the adjusted *p*-values for Scenario MT$[\neq, \chi : \gamma G]$. There are some indications, especially for the CC ontology, that greater values of the parameter $\gamma G$ lead to greater numbers of identified GO terms. Note that for Scenario MT$[\neq, \chi]$, the *p*-value-based estimator $\lambda^{\neq}_{n,\alpha}$, with $\alpha = 0.05$, and the naive estimator $\lambda^{\neq}_{n,\gamma G}$, with $\gamma G = 20$, yield very similar results (green and purple plotting symbols). Indeed, when applied to the entire dataset for the $n = 79$ cell samples, permutation-based single-step maxT Procedure 1 identifies 20 genes as being differentially expressed between BCR/ABL and NEG B-cell ALL at nominal FWER level $\alpha = 0.05$. In other words, $\lambda^{\neq}_{n,0.05}$ and $\lambda^{\neq}_{n,20}$ yield the same estimate of the binary gene-parameter profile $\lambda^{\neq}$ for the set of DE genes. Minor discrepancies between the results of Scenarios MT$[\neq, \chi : \alpha = 0.05]$ and $\mathsf{MT}[\neq,\chi:\gamma G = 20]$ are due to the fact that while the estimators $\lambda_{n,0.05}^{\neq}$ and $\lambda_{n,20}^{\neq}$ coincide on the original sample, they may differ on bootstrap samples of these data.

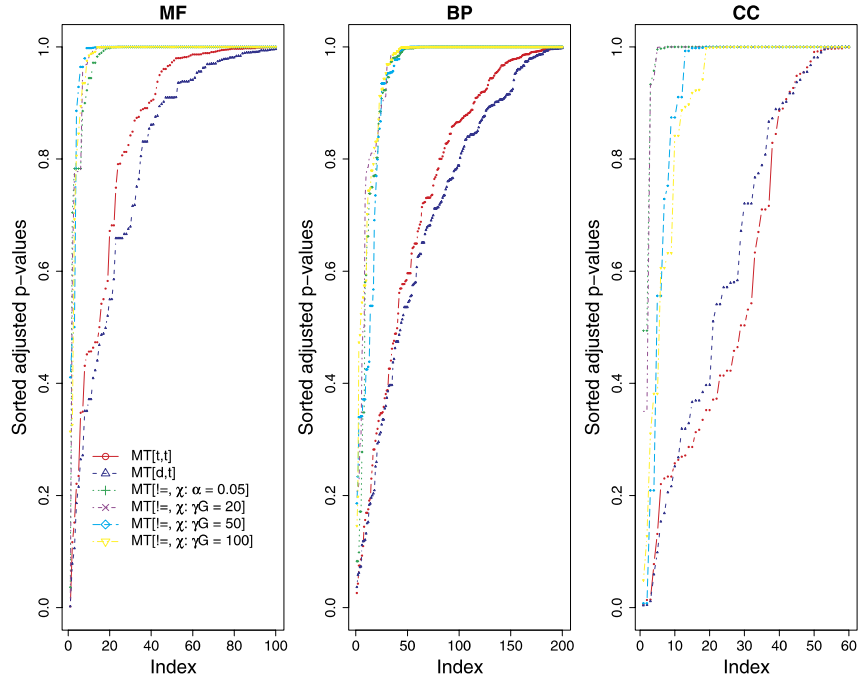

Fig 5. *GO terms associated with differential gene expression between BCR/ABL and NEG B-cell ALL, adjusted p-values.* Plots of sorted bootstrap-based single-step maxT adjusted *p*-values $\widetilde{P}_{0n}(m)$, for each of the three gene ontologies and each of the three testing scenarios.

Table 4. *GO terms associated with differential gene expression between BCR/ABL and NEG B-cell ALL.* This table reports, for each of the three gene ontologies and each of the three testing scenarios, the numbers $R_n(\alpha) = |\mathcal{R}_n(\alpha)|$ of GO terms found to be significantly associated with BCR/ABL vs. NEG differential gene expression for different nominal FWER levels $\alpha$.

| | **Nominal FWER level, $\alpha$** | | | | | | | | |
|---:|:---:|:---:|:---:|:---:|:---:|:---:|:---:|:---:|:---:|
| | 0.05 | 0.10 | 0.20 | 0.05 | 0.10 | 0.20 | 0.05 | 0.10 | 0.20 |
| **MT**$[t,t]$ | 2 | 6 | 14 | 3 | 4 | 5 | 1 | 1 | 3 |
| **MT**$[d,t]$ | 1 | 5 | 16 | 3 | 5 | 7 | 1 | 2 | 4 |
| **MT**$[\neq,\chi : \alpha = 0.05]$ | 0 | 3 | 5 | 0 | 0 | 0 | 1 | 1 | 1 |
| **MT**$[\neq,\chi : \gamma G = 20]$ | 0 | 0 | 0 | 0 | 0 | 0 | 1 | 1 | 1 |
| **MT**$[\neq,\chi : \gamma G = 50]$ | 0 | 0 | 1 | 2 | 2 | 2 | 0 | 0 | 0 |
| **MT**$[\neq,\chi : \gamma G = 100]$ | 0 | 0 | 2 | 1 | 1 | 2 | 0 | 0 | 0 |
| | **BP** | | | **CC** | | | **MF** | | |

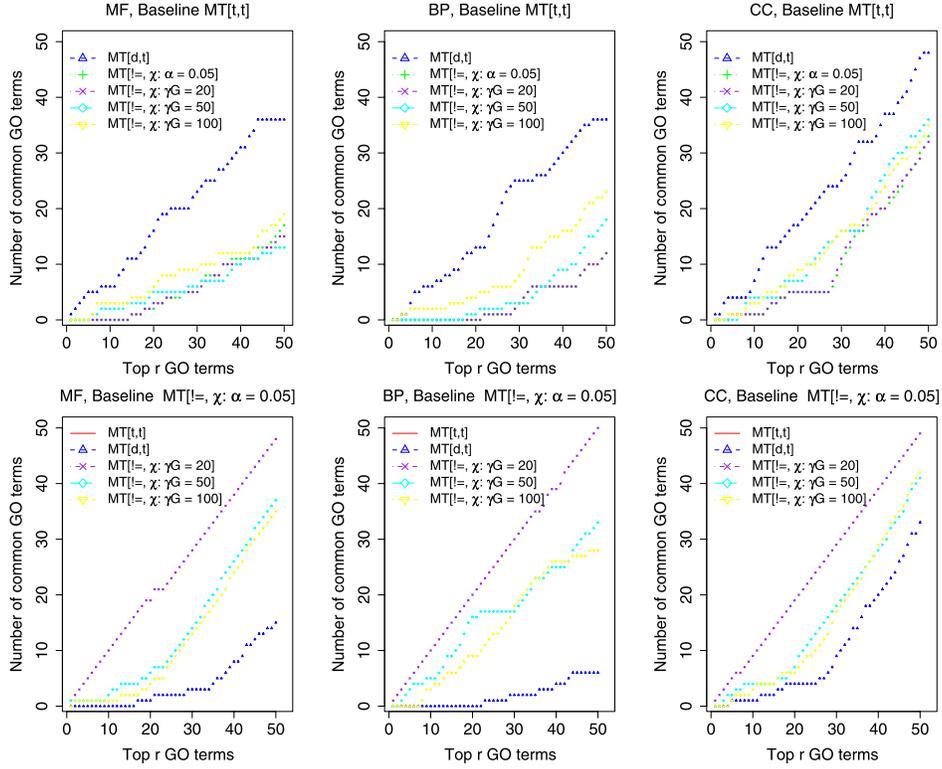

FIG 6. *GO terms associated with differential gene expression between BCR/ABL and NEG B-cell ALL, common terms between testing scenarios.* Plots of numbers of common GO terms among sets of ordered GO terms $\mathcal{O}_n(r)$ of various cardinality $r$ for pairs of testing scenarios. Scenario $\mathsf{MT}[t,t]$ is used as the baseline in the top panels and Scenario $\mathsf{MT}[\neq,\chi:\alpha = 0.05]$, with adjusted $p$-value-based estimator $\lambda_{n,\alpha}^{\neq}$, $\alpha = 0.05$, for the binary DE gene-parameter profile $\lambda^{\neq}$, is used as the baseline in the bottom panels. For example, the blue curve in the top-left panel is a plot of $\left|\mathcal{O}_n^{d,t}(r) \cap \mathcal{O}_n^{t,t}(r)\right|$ vs. $r$ for the MF gene ontology, i.e., of the overlap between the $r$ most significant MF GO terms according to Scenarios $\mathsf{MT}[d,t]$ and $\mathsf{MT}[t,t]$.

Next, the three testing scenarios are compared in terms of the contents of the lists $\mathcal{R}_n(\alpha)$ of identified GO terms. Specifically, let $\mathcal{O}_n(r) \equiv \{O_n(1), \ldots, O_n(r)\}$ denote the set of indices corresponding to the $r$ smallest adjusted $p$-values for a given gene ontology and testing scenario. Measures of agreement between testing scenarios are provided by the numbers of common GO terms among sets of ordered GO terms $\mathcal{O}_n(r)$ of various cardinality $r$, i.e., by the cardinality of intersections of sets $\mathcal{O}_n(r)$ for different testing scenarios. Figure 6 displays plots of numbers of common GO terms for pairs of testing scenarios. As expected, there is substantial overlap between the GO terms identified by Scenarios $\mathsf{MT}[t,t]$ and $\mathsf{MT}[d,t]$ for continuous DE gene-parameter profiles (blue plotting symbols in top panels). This suggests that, for the ALL dataset, standardized ($\lambda^t$) and unstandardized ($\lambda^d$) continuous measures of differential gene expression have similar properties. In contrast, there is much less overlap between the GO terms identified by Scenario $\mathsf{MT}[\neq,\chi]$, for binary

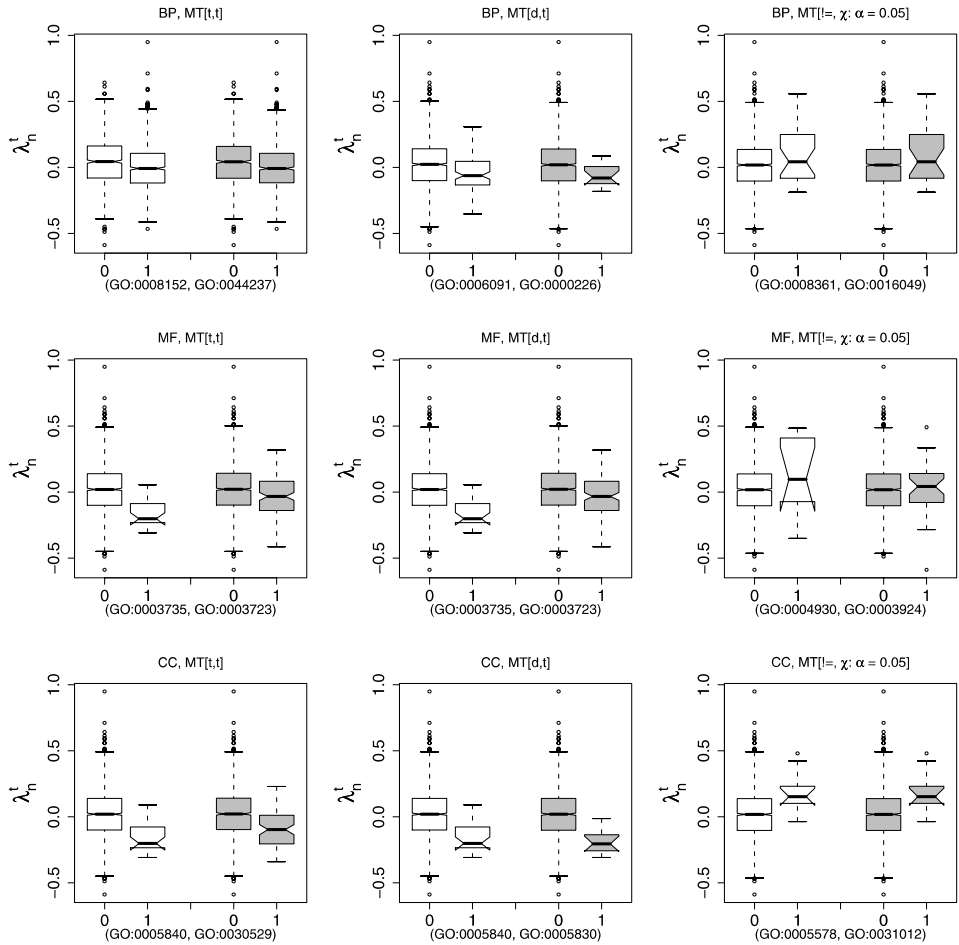

FIG 7. *GO terms associated with differential gene expression between BCR/ABL and NEG B-cell ALL, conditional distribution of $\lambda_n^t$ given A.* Conditional boxplots of the estimated continuous DE gene-parameter profile $\lambda_n^t$ given the gene-annotation profiles $A(\cdot, m)$ for the top two GO terms $m \in \{O_n(1), O_n(2)\}$ identified according to each of the three testing scenarios. Rows correspond to gene ontologies and columns to testing scenarios. In each panel, the white and gray boxplots correspond, respectively, to the GO terms with the smallest and second smallest adjusted $p$-values; boxplots for unannotated and annotated estimated gene-parameter profiles, $(\lambda_n^t(g) : A(g, m) = 0)$ and $(\lambda_n^t(g) : A(g, m) = 1)$, are labeled as 0 and 1, respectively. Non-overlapping notches (informally) represent large differences in medians.

DE gene-parameter profiles, and either Scenario $\mathsf{MT}[t, t]$ or $\mathsf{MT}[d, t]$. For example, for the MF gene ontology, among the top 10 GO terms $\mathcal{O}_n(10)$ identified by each testing scenario, 6 are common to Scenarios $\mathsf{MT}[t, t]$ and $\mathsf{MT}[d, t]$, whereas at most 3 are common to Scenarios $\mathsf{MT}[t, t]$ and $\mathsf{MT}[\neq, \chi]$. Again, note the near perfect agreement between Scenarios $\mathsf{MT}[\neq, \chi : \alpha = 0.05]$ and $\mathsf{MT}[\neq, \chi : \gamma G = 20]$ (purple plotting symbols in lower panels). Figure 6 again illustrates the lack of robustness of Scenario $\mathsf{MT}[\neq, \chi : \gamma G]$ to the choice of parameter $\gamma G$.

Moreover, examine graphical summaries of the joint distributions of the estimated continuous DE gene-parameter profile $\lambda_n^t$ and the gene-annotation profiles $A(\cdot, m)$ for the top two GO terms $m \in \{O_n(1), O_n(2)\}$ identified according to each testing scenario. Figure 7 displays conditional boxplots of $\lambda_n^t$ given $A(\cdot, m)$, that is, boxplots of the unannotated and annotated estimated gene-parameter profiles, $(\lambda_n^t(g) : A(g, m) = 0)$ and $(\lambda_n^t(g) : A(g, m) = 1)$, respectively. Although the boxplots reveal clear differences (non-overlapping notches) between unannotated and annotated profiles for some of the GO terms (e.g., MF term `GO:0003735`), the differences can be subtle for other terms (e.g., MF term `GO:0003924`). Not surprisingly, the most extreme differences are seen for Scenarios $\mathsf{MT}[t, t]$ and $\mathsf{MT}[d, t]$, and, to a lesser extent, for Scenario $\mathsf{MT}[\neq, \chi : \alpha = 0.05]$ for the CC ontology. The boxplots again illustrate differences between Scenario $\mathsf{MT}[\neq, \chi]$ and either Scenario $\mathsf{MT}[t, t]$ or $\mathsf{MT}[d, t]$.

Tables 5, 6, and 7 report various $p$-value-based measures of association between the estimated DE gene-parameter profiles $\lambda_n^t$ and $\lambda_{n,\alpha}^{\neq}$ and the gene-annotation profiles $A(\cdot, m)$ for the top two GO terms $m \in \{O_n(1), O_n(2)\}$ identified according to each testing scenario, in the BP, CC, and MF gene ontologies, respectively. The transformation to the $[0, 1]$ $p$-value scale allows a more direct comparison of the various testing scenarios. The tables again highlight the differences between Scenario $\mathsf{MT}[\neq, \chi]$, for binary DE gene-parameter profiles, and either Scenario $\mathsf{MT}[t, t]$ or $\mathsf{MT}[d, t]$, for continuous DE gene-parameter profiles. As expected, Scenarios $\mathsf{MT}[t, t]$ and $\mathsf{MT}[d, t]$ tend to identify GO terms with small $p$-values $P_{0n}^{t,t}(m)$ for $t$-tests of association between estimated continuous gene-parameter profiles $\lambda_n^t$ and gene-annotation profiles $A(\cdot, m)$. In contrast, and also as expected, Scenario $\mathsf{MT}[\neq, \chi]$ tends to identify GO terms with small $p$-values $P_{0n}^{\neq,\chi}(m)$ for $\chi^2$-tests of association between estimated binary gene-parameter profiles $\lambda_{n,\alpha}^{\neq}$ and gene-annotation profiles $A(\cdot, m)$. Furthermore, the tables corroborate our earlier observation that

TABLE 5. *GO terms associated with differential gene expression between BCR/ABL and NEG B-cell ALL, top two BP GO terms.* This table provides association measures between the estimated DE gene-parameter profiles $\lambda_n^t$ and $\lambda_{n,\alpha}^{\neq}$ and the gene-annotation profiles $A(\cdot, m)$ for the top two BP GO terms $m \in \{O_n(1), O_n(2)\}$ identified according to each of the three testing scenarios. $A_1(m) = \sum_g A(g, m)$: Number of genes directly or indirectly annotated with GO term $m$ (out of $G = 2{,}071$ genes, `GOALLLOCUSID` environment in `GO` package). $P_{0n}^{t,t}(m)$: Nominal unadjusted $p$-value for the two-sample $t$-test comparing the unannotated and annotated estimated continuous DE gene-parameter profiles, $(\lambda_n^t(g) : A(g, m) = 0)$ and $(\lambda_n^t(g) : A(g, m) = 1)$, respectively (`t.test` function from the R package `stats`, with default argument values). $P_{0n}^{\neq,\chi}(m)$: Unadjusted $p$-value for the $\chi^2$-test of independence between the estimated binary DE gene-parameter profile $\lambda_{n,\alpha}^{\neq}$, $\alpha = 0.05$, and the gene-annotation profile $A(\cdot, m)$ (`chisq.test` function from the R package `stats`, with arguments `simulate.p.value = TRUE, correct=FALSE`). $\widetilde{P}_{0n}(m)$: Bootstrap-based single-step maxT adjusted $p$-value, according to which the top two GO terms are identified for each testing scenario.

| | | BP | | | |
|---|---|---|---|---|---|
| **Scenario** | **GO term** | $A_1(m)$ | $P_{0n}^{t,t}(m)$ | $P_{0n}^{\neq,\chi}(m)$ | $\widetilde{P}_{0n}(m)$ |
| $\mathsf{MT}[t, t]$ | `GO:0008152` | 1076 | 2.5e-09 | 1.7e-01 | 2.6e-02 |
| | `GO:0044237` | 1045 | 3.8e-08 | 1.8e-01 | 4.3e-02 |
| $\mathsf{MT}[d, t]$ | `GO:0006091` | 98 | 5.2e-06 | 6.3e-01 | 3.7e-02 |
| | `GO:0000226` | 14 | 1.8e-03 | 1.0e+00 | 5.8e-02 |
| $\mathsf{MT}[\neq, \chi : \alpha = 0.05]$ | `GO:0008361` | 27 | 5.5e-02 | 3.5e-03 | 8.3e-02 |
| | `GO:0016049` | 27 | 5.5e-02 | 1.5e-03 | 8.3e-02 |
| $\mathsf{MT}[\neq, \chi : \gamma G = 20]$ | `GO:0008361` | 27 | 5.5e-02 | 4.0e-03 | 2.1e-01 |
| | `GO:0016049` | 27 | 5.5e-02 | 1.5e-03 | 2.1e-01 |
| $\mathsf{MT}[\neq, \chi : \gamma G = 50]$ | `GO:0048522` | 87 | 3.6e-02 | 6.5e-03 | 1.9e-01 |
| | `GO:0048518` | 96 | 4.4e-02 | 1.3e-02 | 2.3e-01 |
| $\mathsf{MT}[\neq, \chi : \gamma G = 100]$ | `GO:0050793` | 24 | 8.5e-02 | 1.7e-02 | 1.5e-01 |
| | `GO:0007155` | 59 | 5.9e-04 | 1.2e-01 | 2.0e-01 |

TABLE 6. *GO terms associated with differential gene expression between BCR/ABL and NEG B-cell ALL, top two CC GO terms.* Details in Table 5 caption.

| | | CC | | | |
|---|---|---|---|---|---|
| **Scenario** | **GO term** | $A_1(m)$ | $P_{0n}^{t,t}(m)$ | $P_{0n}^{\neq,\chi}(m)$ | $\widetilde{P}_{0n}(m)$ |
| MT$[t,t]$ | GO:0005840 | 25 | 1.3e-08 | 1.0e+00 | 5.6e-03 |
| | GO:0030529 | 77 | 3.1e-10 | 6.4e-01 | 1.4e-02 |
| MT$[d,t]$ | GO:0005840 | 25 | 1.3e-08 | 1.0e+00 | 4.0e-03 |
| | GO:0005830 | 11 | 2.8e-05 | 1.0e+00 | 5.2e-03 |
| MT$[\neq,\chi:\alpha=0.05]$ | GO:0005578 | 10 | 1.7e-02 | 1.0e-01 | 4.9e-01 |
| | GO:0031012 | 10 | 1.7e-02 | 1.1e-01 | 4.9e-01 |
| MT$[\neq,\chi:\gamma G=20]$ | GO:0005578 | 10 | 1.7e-02 | 1.0e-01 | 3.5e-01 |
| | GO:0031012 | 10 | 1.7e-02 | 9.2e-02 | 3.5e-01 |
| MT$[\neq,\chi:\gamma G=50]$ | GO:0005576 | 54 | 9.1e-04 | 1.0e+00 | 7.8e-03 |
| | GO:0005615 | 31 | 4.8e-02 | 2.4e-01 | 7.8e-03 |
| MT$[\neq,\chi:\gamma G=100]$ | GO:0005576 | 54 | 9.1e-04 | 1.0e+00 | 4.9e-02 |
| | GO:0005615 | 31 | 4.8e-02 | 2.6e-01 | 1.3e-01 |

TABLE 7. *GO terms associated with differential gene expression between BCR/ABL and NEG B-cell ALL, top two MF GO terms.* Details in Table 5 caption.

| | | MF | | | |
|---|---|---|---|---|---|
| **Scenario** | **GO term** | $A_1(m)$ | $P_{0n}^{t,t}(m)$ | $P_{0n}^{\neq,\chi}(m)$ | $\widetilde{P}_{0n}(m)$ |
| MT$[t,t]$ | GO:0003735 | 24 | 1.1e-09 | 1.0e+00 | 2.4e-03 |
| | GO:0003723 | 143 | 1.5e-06 | 4.1e-01 | 1.2e-01 |
| MT$[d,t]$ | GO:0003735 | 24 | 1.1e-09 | 1.0e+00 | 2.2e-03 |
| | GO:0003723 | 143 | 1.5e-06 | 3.8e-01 | 7.8e-02 |
| MT$[\neq,\chi:\alpha=0.05]$ | GO:0004930 | 10 | 2.2e-01 | 3.0e-03 | 3.7e-02 |
| | GO:0003924 | 34 | 6.5e-01 | 3.9e-02 | 7.0e-01 |
| MT$[\neq,\chi:\gamma G=20]$ | GO:0004930 | 10 | 2.2e-01 | 3.0e-03 | 1.7e-02 |
| | GO:0003924 | 34 | 6.5e-01 | 3.8e-02 | 6.2e-01 |
| MT$[\neq,\chi:\gamma G=50]$ | GO:0004930 | 10 | 2.2e-01 | 3.5e-03 | 4.1e-01 |
| | GO:0030246 | 22 | 8.6e-01 | 2.0e-01 | 4.8e-01 |
| MT$[\neq,\chi:\gamma G=100]$ | GO:0005509 | 69 | 3.8e-04 | 1.3e-01 | 3.1e-01 |
| | GO:0004930 | 10 | 2.2e-01 | 1.5e-03 | 3.3e-01 |

Scenario MT$[\neq,\chi]$ tends to be more conservative than either Scenario MT$[t,t]$ or MT$[d,t]$. Indeed, some of the GO terms with small $p$-values $P_{0n}^{t,t}(m)$ for continuous gene-parameter profiles have very large $p$-values $P_{0n}^{\neq,\chi}(m)$ for binary gene-parameter profiles (e.g., MF term GO:0003735 in Table 7). Such terms are likely to be identified by Scenarios MT$[t,t]$ and MT$[d,t]$, but missed by Scenario MT$[\neq,\chi]$. The converse phenomenon is not as striking. However, one should keep in mind that Scenario MT$[\neq,\chi]$ depends on the choice of estimator for the binary DE gene-parameter profile $\lambda^{\neq}$, i.e., on parameters such as $\alpha$ and $\gamma G$. In particular, with certain values of $\alpha$ (or $\gamma G$), binary Scenario MT$[\neq,\chi]$ may become more similar to either continuous Scenario MT$[t,t]$ or MT$[d,t]$. Column $A_1(m)$ in Tables 5–7 suggests that, compared to Scenario MT$[\neq,\chi]$, Scenarios MT$[t,t]$ and MT$[d,t]$ tend to identify GO terms annotating a greater number of genes (this observation also holds for the top 20 terms identified according to each testing scenario; data not shown).

Figure 8 displays a scatterplot matrix of the 50 smallest adjusted $p$-values, based on Scenario MT$[t,t]$, for each of the three gene ontologies. The plots indicate that more terms tend to be identified in the BP ontology compared to either the CC or MF ontologies, and fewer terms tend to be identified in the MF ontology compared to either the BP or CC ontologies. Note that comparisons based on adjusted $p$-values take into account differences in the numbers of tested hypotheses, $M_{BP} = 367$, $M_{CC} = 81$, and $M_{MF} = 185$, for each ontology.

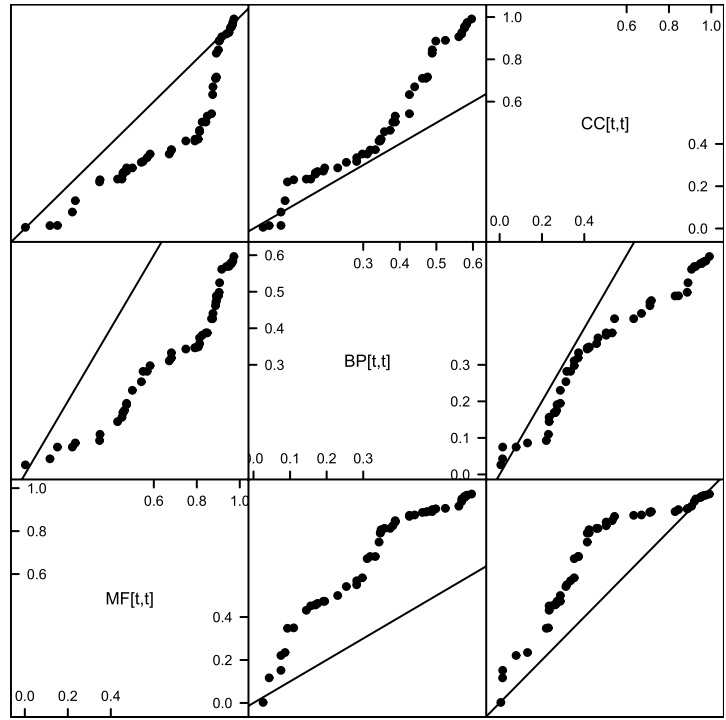

FIG 8. *GO terms associated with differential gene expression between BCR/ABL and NEG B-cell ALL, comparison of adjusted p-values for the three gene ontologies.* Scatterplot matrix of the 50 smallest adjusted *p*-values for each of the three gene ontologies, based on Scenario MT$[t, t]$. The identity line is drawn for reference.

Tables 8, 9, and 10 list the 20 GO terms with the smallest adjusted *p*-values for Scenario MT$[t, t]$, applied to the BP, CC, and MF gene ontologies, respectively. Figures 9, 10, and 11 display portions of the directed acyclic graphs for the top 20 GO terms in each ontology. The figures suggest that GO terms associated with BCR/ABL vs. NEG differential gene expression tend to concentrate in certain branches of the DAGs, i.e., differential expression is associated with related properties of gene products. Although it is known that many of the effects of the BCR/ABL fusion are mediated by tyrosine kinase activity, the MF GO term *protein-tyrosine kinase activity* (GO:0004713) does not appear to be significantly associated with differential gene expression between BCR/ABL and NEG B-cell ALL (adjusted *p*-value of 0.8890 for Scenario MT$[t, t]$).

For illustration purposes, we further investigate two of the GO terms from Tables 8 and 10: GO term *anti-apoptosis* (GO:0006916), with ninth smallest adjusted *p*-value for Scenario MT$[t, t]$ applied to the BP gene ontology, and GO term *structural constituent of ribosome* (GO:0003735), with the smallest adjusted *p*-value for Scenario MT$[t, t]$ applied to the MF gene ontology. Tables 11 and 12 list genes directly or indirectly annotated with GO terms GO:0006916 and GO:0003735, respectively. Figure 12 displays mean-difference plots of the average expression measures in BCR/ABL and NEG cell samples for genes annotated with GO terms GO:0006916 and GO:0003735.

TABLE 8. *GO terms associated with differential gene expression between BCR/ABL and NEG B-cell ALL, top 20 BP GO terms.* This table lists the 20 GO terms with the smallest adjusted $p$-values for Scenario MT$[t, t]$ applied to the BP gene ontology. $A_1(m) = \sum_g A(g, m)$: Number of genes directly or indirectly annotated with GO term $m$ (out of $G = 2,071$ genes, GOALLLOCUSID environment in GO package). $\widetilde{P}_{0n}(m)$: Bootstrap-based single-step maxT adjusted $p$-value for Scenario MT$[t, t]$.

**BP, Scenario MT$[t, t]$**

| GO term ID | GO term | $A_1(m)$ | $\widetilde{P}_{0n}(m)$ |
|---|---|---|---|
| GO:008152 | *metabolism* | 1076 | 2.6e-02 |
| GO:044237 | *cellular metabolism* | 1045 | 4.3e-02 |
| GO:009058 | *biosynthesis* | 187 | 7.5e-02 |
| GO:044238 | *primary metabolism* | 1002 | 7.5e-02 |
| GO:044249 | *cellular biosynthesis* | 169 | 8.6e-02 |
| GO:006091 | *generation of precursor metabolites and energy* | 98 | 9.3e-02 |
| GO:019882 | *antigen presentation* | 15 | 1.1e-01 |
| GO:030333 | *antigen processing* | 14 | 1.4e-01 |
| GO:006916 | *anti-apoptosis* | 21 | 1.6e-01 |
| GO:043066 | *negative regulation of apoptosis* | 26 | 1.7e-01 |
| GO:043069 | *negative regulation of programmed cell death* | 26 | 1.7e-01 |
| GO:007154 | *cell communication* | 390 | 1.8e-01 |
| GO:006457 | *protein folding* | 52 | 1.9e-01 |
| GO:007165 | *signal transduction* | 351 | 1.9e-01 |
| GO:000226 | *microtubule cytoskeleton organization and biogenesis* | 14 | 2.3e-01 |
| GO:006082 | *organic acid metabolism* | 65 | 2.5e-01 |
| GO:006163 | *purine nucleotide metabolism* | 29 | 2.8e-01 |
| GO:007155 | *cell adhesion* | 59 | 2.8e-01 |
| GO:007028 | *cytoplasm organization and biogenesis* | 10 | 3.0e-01 |
| GO:019752 | *carboxylic acid metabolism* | 63 | 3.1e-01 |

TABLE 9. *GO terms associated with differential gene expression between BCR/ABL and NEG B-cell ALL, top 20 CC GO terms.* Details in Table 8 caption.

**CC, Scenario MT$[t, t]$**

| GO term ID | GO term | $A_1(m)$ | $\widetilde{P}_{0n}(m)$ |
|---|---|---|---|
| GO:0005840 | *ribosome* | 25 | 5.6e-03 |
| GO:0030529 | *ribonucleoprotein complex* | 77 | 1.4e-02 |
| GO:0005830 | *cytosolic ribosome (sensu Eukaryota)* | 11 | 1.4e-02 |
| GO:0043234 | *protein complex* | 334 | 7.8e-02 |
| GO:0005886 | *plasma membrane* | 200 | 1.3e-01 |
| GO:0005829 | *cytosol* | 78 | 2.2e-01 |
| GO:0005737 | *cytoplasm* | 578 | 2.3e-01 |
| GO:0005887 | *integral to plasma membrane* | 125 | 2.3e-01 |
| GO:0031226 | *intrinsic to plasma membrane* | 125 | 2.3e-01 |
| GO:0019866 | *inner membrane* | 37 | 2.6e-01 |
| GO:0005743 | *mitochondrial inner membrane* | 28 | 2.6e-01 |
| GO:0005746 | *mitochondrial electron transport chain* | 11 | 2.7e-01 |
| GO:0000502 | *proteasome complex (sensu Eukaryota)* | 26 | 2.7e-01 |
| GO:0000323 | *lytic vacuole* | 28 | 2.9e-01 |
| GO:0005764 | *lysosome* | 28 | 2.9e-01 |
| GO:0005576 | *extracellular region* | 54 | 3.1e-01 |
| GO:0005773 | *vacuole* | 29 | 3.2e-01 |
| GO:0005622 | *intracellular* | 1152 | 3.4e-01 |
| GO:0043228 | *non-membrane-bound organelle* | 218 | 3.5e-01 |
| GO:0043232 | *intracellular non-membrane-bound organelle* | 218 | 3.5e-01 |

Table 10. *GO terms associated with differential gene expression between BCR/ABL and NEG B-cell ALL, top 20 MF GO terms.* Details in Table 8 caption.

**MF, Scenario MT[$t,t$]**

| GO term ID | GO term | $A_1(m)$ | $\widetilde{P}_{0n}(m)$ |
|---|---|---|---|
| GO:0003735 | *structural constituent of ribosome* | 24 | 2.4e-03 |
| GO:0003723 | *RNA binding* | 143 | 1.2e-01 |
| GO:0048037 | *cofactor binding* | 11 | 1.5e-01 |
| GO:0051082 | *unfolded protein binding* | 47 | 2.2e-01 |
| GO:0016853 | *isomerase activity* | 28 | 2.3e-01 |
| GO:0016491 | *oxidoreductase activity* | 89 | 3.5e-01 |
| GO:0005509 | *calcium ion binding* | 69 | 3.5e-01 |
| GO:0015399 | *primary active transporter activity* | 57 | 4.3e-01 |
| GO:0004872 | *receptor activity* | 101 | 4.5e-01 |
| GO:0004871 | *signal transducer activity* | 242 | 4.6e-01 |
| GO:0016765 | *transferase activity, transferring alkyl or aryl (other than methyl) groups* | 10 | 4.6e-01 |
| GO:0016860 | *intramolecular oxidoreductase activity* | 13 | 4.6e-01 |
| GO:0016614 | *oxidoreductase activity, acting on CH-OH group of donors* | 18 | 4.7e-01 |
| GO:0016616 | *oxidoreductase activity, acting on the CH-OH group of donors, NAD or NADP as acceptor* | 18 | 4.7e-01 |
| GO:0043169 | *cation binding* | 230 | 5.0e-01 |
| GO:0005489 | *electron transporter activity* | 47 | 5.4e-01 |
| GO:0005386 | *carrier activity* | 73 | 5.5e-01 |
| GO:0004888 | *transmembrane receptor activity* | 59 | 5.7e-01 |
| GO:0003824 | *catalytic activity* | 635 | 5.8e-01 |
| GO:0003676 | *nucleic acid binding* | 449 | 6.7e-01 |

Panel (a) in Figure 12 indicates that genes annotated with BP GO term *anti-apoptosis* (GO:0006916) tend to be over-expressed in BCR/ABL compared to NEG cell samples. Among these 21 genes, only SOCS2 is significantly differentially expressed between BCR/ABL and NEG B-cell ALL (nominal FWER level $\alpha = 0.05$, Table 3). However, a brief survey of the literature reveals that several of the genes in Table 11 interact with the BCR/ABL proto-oncogene. For instance, [27] investigate mechanisms for the BCR/ABL-mediated activation of the transcription factor NF-$\kappa$B/Rel encoded by the NFKB1 gene. Their findings suggest that NF-$\kappa$B/Rel may be a potential target for molecular therapies of leukemia. [30] demonstrate that ectopic expression of BCR/ABL interferes with the tumor necrosis factor (TNF) signaling pathway through the down-regulation of TNF receptors. The TNF gene encodes a multifunctional proinflammatory cytokine involved in the regulation of a wide spectrum of biological processes, including cell proliferation, differentiation, apoptosis, lipid metabolism, and coagulation. The TNF gene has been implicated in a variety of diseases, including autoimmune diseases, insulin resistance, and cancer.

As seen in Table 12, 22 of the 24 genes annotated with MF GO term *structural constituent of ribosome* (GO:0003735) code for ribosomal proteins. Although none of the 24 annotated genes is identified as being significantly differentially expressed between BCR/ABL and NEG B-cell ALL (nominal FWER level $\alpha = 0.05$, Table 3), Panel (b) in Figure 12 suggests that these genes tend to be under-expressed in BCR/ABL cell samples.

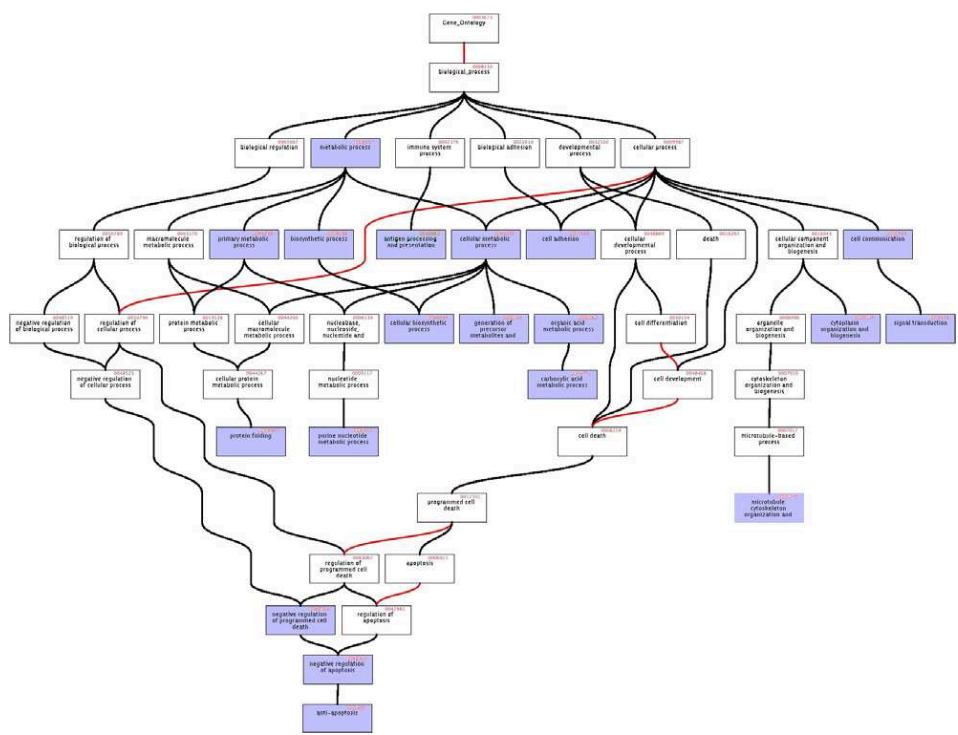

Fig 9. *GO terms associated with differential gene expression between BCR/ABL and NEG B-cell ALL, DAG for top 20 BP GO terms.* Portion of the directed acyclic graph for the 20 GO terms with the smallest adjusted $p$-values for Scenario $\mathsf{MT}[t,t]$ applied to the BP gene ontology. Nodes for the top 20 terms are shaded in lavender; black and red edges indicate, respectively, "is a" and "part of a" relationships among terms. The figure was produced using the QuickGO browser. According to QuickGO, the GO term IDs `GO:019882` and `GO:0030333` listed in Table 8 correspond to the same term, *antigen processing and presentation*. (Higher-resolution color version on website companion.)

## 6. Discussion

We have proposed a general and formal statistical framework for multiple tests of association with biological annotation metadata. A key component of our approach is the systematic and precise translation of a generic biological question into a corresponding multiple hypothesis testing problem, concerning association measures between known gene-annotation profiles and unknown gene-parameter profiles. This general and rigorous formulation of the statistical inference question allows us to apply the multiple hypothesis testing methodology developed in [14] and related articles, to control a broad class of Type I error rates, in testing problems involving general data generating distributions (with arbitrary dependence structures among variables), null hypotheses, and test statistics.

The flexibility of our approach was illustrated using the ALL microarray dataset of [13], with the aim of relating GO annotation to differential gene expression between BCR/ABL and NEG B-cell ALL. This analysis demonstrates the importance of selecting a suitable DE gene-parameter profile $\lambda$ and measure $\rho$ for the association between this gene-parameter profile and GO gene-annotation profiles $A$. Indeed, for the ALL dataset, the choice of gene-parameter profile for measur-

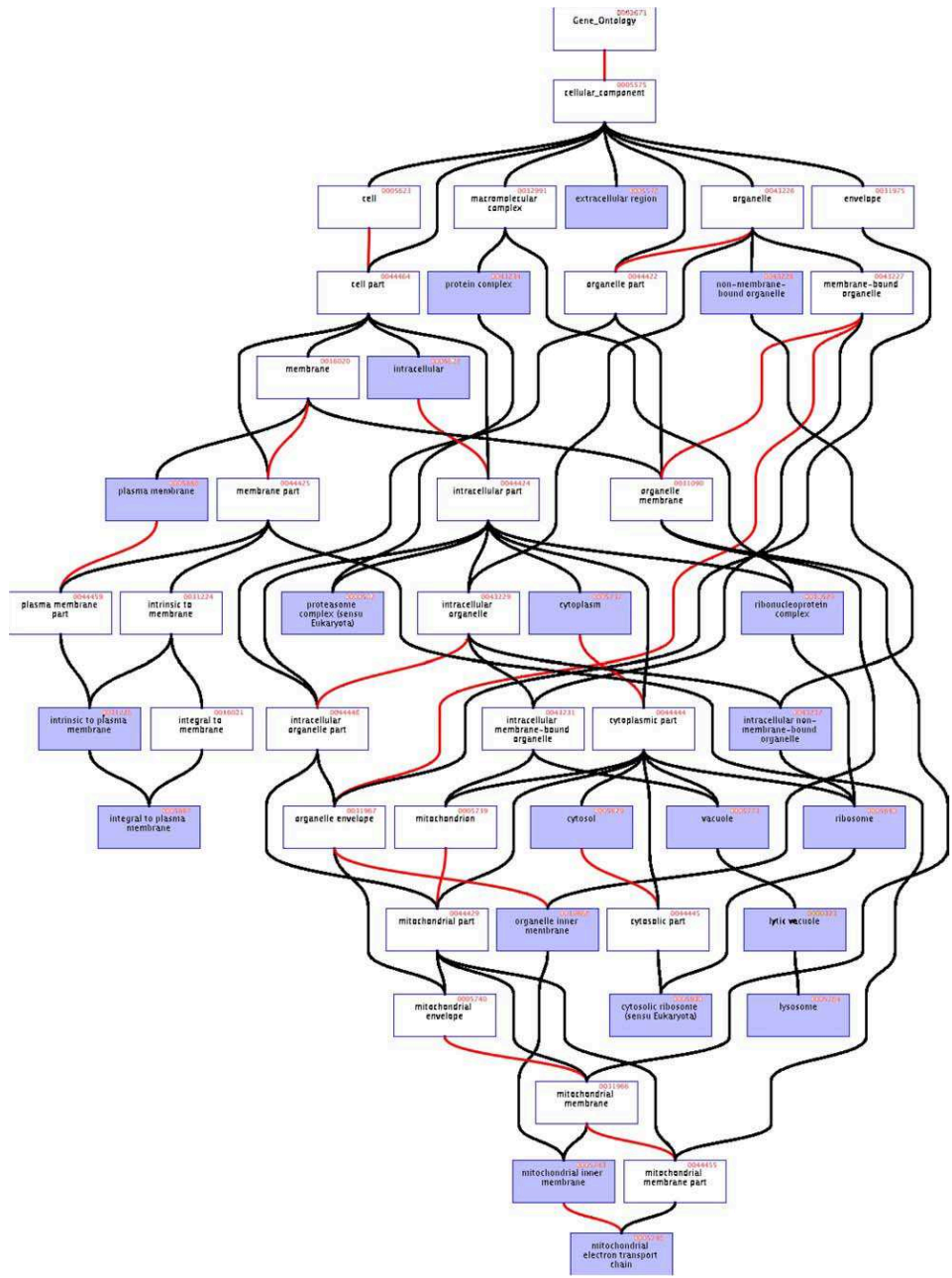

FIG 10. *GO terms associated with differential gene expression between BCR/ABL and NEG B-cell ALL, DAG for top 20 CC GO terms.* Portion of the directed acyclic graph for the 20 GO terms with the smallest adjusted $p$-values for Scenario $\mathsf{MT}[t,t]$ applied to the CC gene ontology. Nodes for the top 20 terms are shaded in lavender; black and red edges indicate, respectively, "is a" and "part of a" relationships among terms. The figure was produced using the QuickGO browser. (Higher-resolution color version on website companion.)

ing differential expression between BCR/ABL and NEG B-cell ALL has a large impact on the list of identified GO terms. Testing scenarios based on binary DE

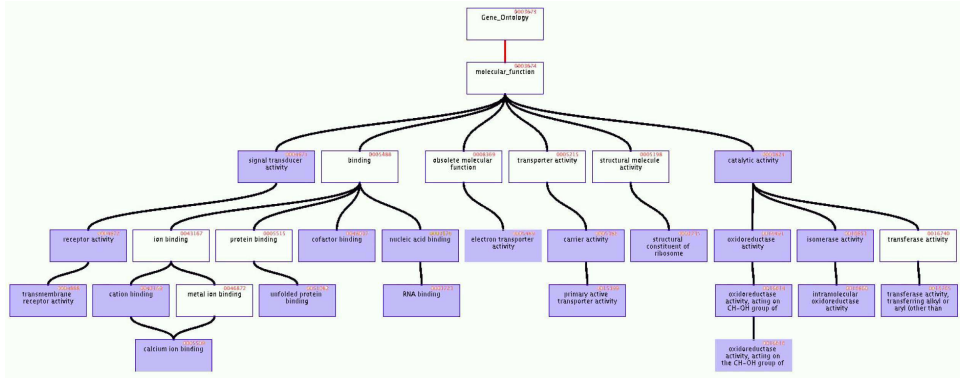

FIG 11. *GO terms associated with differential gene expression between BCR/ABL and NEG B-cell ALL, DAG for top 20 MF GO terms.* Portion of the directed acyclic graph for the 20 GO terms with the smallest adjusted $p$-values for Scenario $\mathsf{MT}[t,t]$ applied to the MF gene ontology. Nodes for the top 20 terms are shaded in lavender; black and red edges indicate, respectively, "is a" and "part of a" relationships among terms. The figure was produced using the QuickGO browser. (Higher-resolution color version on website companion.)

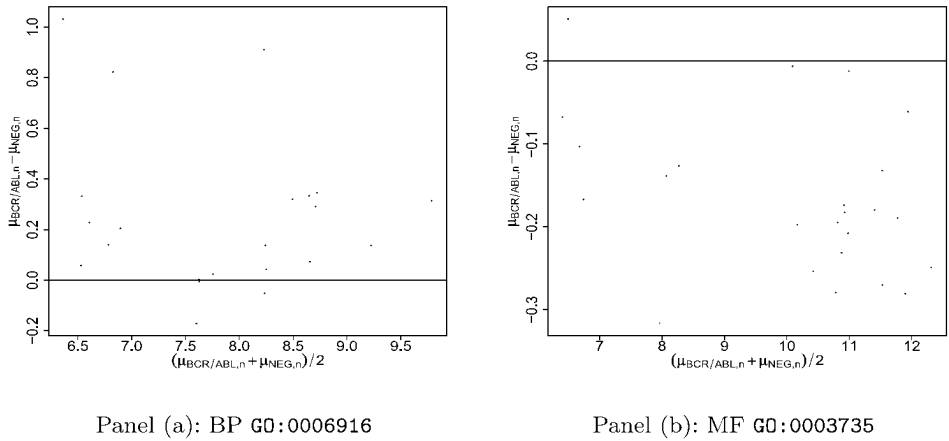

Panel (a): BP `GO:0006916`            Panel (b): MF `GO:0003735`

FIG 12. *GO terms associated with differential gene expression between BCR/ABL and NEG B-cell ALL, BP GO term `GO:0006916` and MF GO term `GO:0003735`.* This figure displays mean-difference plots of average expression measures in BCR/ABL and NEG cell samples, i.e., plots of $\mu_{BCR/ABL,n}(g) - \mu_{NEG,n}(g)$ vs. $(\mu_{BCR/ABL,n}(g) + \mu_{NEG,n}(g))/2$, for genes directly or indirectly annotated with GO terms `GO:0006916` (Panel (a)) and `GO:0003735` (Panel (b)). The term *anti-apoptosis* (`GO:0006916`) has the ninth smallest adjusted $p$-value for Scenario $\mathsf{MT}[t,t]$ applied to the BP gene ontology (Tables 8 and 11) and the term *structural constituent of ribosome* (`GO:0003735`) has the smallest adjusted $p$-value for Scenario $\mathsf{MT}[t,t]$ applied to the MF gene ontology (Tables 10 and 12).

gene-parameter profiles (Scenario $\mathsf{MT}[\neq,\chi]$) tended to be more conservative than scenarios based on continuous DE gene-parameter profiles (Scenarios $\mathsf{MT}[t,t]$ and $\mathsf{MT}[d,t]$), with little overlap between the lists of identified GO terms. Furthermore, testing scenarios based on binary gene-parameter profiles were sensitive to the somewhat arbitrary DE/non-DE gene dichotomization, that is, Scenario $\mathsf{MT}[\neq,\chi:\gamma G]$ lacked robustness with respect to the choice of parameter $\gamma G$ for the number of genes called differentially expressed according to the estimator $\lambda_{n,\gamma G}^{\neq}$. In contrast, continuous gene-parameter profiles based on standardized and unstandardized measures of differential gene expression lead to very similar results (Scenarios $\mathsf{MT}[t,t]$ and $\mathsf{MT}[d,t]$).

TABLE 11. *GO terms associated with differential gene expression between BCR/ABL and NEG B-cell ALL, BP GO term* `GO:0006916`. This table lists genes directly or indirectly annotated with GO term *anti-apoptosis* (out of $G = 2,071$ genes, `GOALLLOCUSID` environment in `GO` package). The term *anti-apoptosis* (`GO:0006916`) has the ninth smallest adjusted *p*-value for Scenario $\mathsf{MT}[t,t]$ applied to the BP gene ontology (Table 8).

| | **BP** `GO:0006916` | |
|---|---|---|
| **Probe ID** | **Symbol** | **Name** |
| 1237_at | IER3 | immediate early response 3 |
| 1295_at | RELA | v-rel reticuloendotheliosis viral oncogene homolog A, nuclear factor of kappa light polypeptide gene enhancer in B-cells 3, p65 (avian) |
| 1377_at | NFKB1 | nuclear factor of kappa light polypeptide gene enhancer in B-cells 1 (p105) |
| 1564_at | AKT1 | v-akt murine thymoma viral oncogene homolog 1 |
| 1830_s_at | TGFB1 | transforming growth factor, beta 1 (Camurati-Engelmann disease) |
| 1852_at | TNF | tumor necrosis factor (TNF superfamily, member 2) |
| 1997_s_at | BAX | BCL2-associated X protein |
| 277_at | MCL1 | myeloid cell leukemia sequence 1 (BCL2-related) |
| 31536_at | RTN4 | reticulon 4 |
| 32060_at | BNIP2 | BCL2/adenovirus E1B 19 kDa interacting protein 2 |
| 33284_at | MPO | myeloperoxidase |
| 36578_at | BIRC2 | baculoviral IAP repeat-containing 2 |
| 38578_at | TNFRSF7 | tumor necrosis factor receptor superfamily, member 7 |
| 38771_at | HDAC1 | histone deacetylase 1 |
| 38994_at | SOCS2 | suppressor of cytokine signaling 2 |
| 39097_at | SON | SON DNA binding protein |
| 39378_at | BECN1 | beclin 1 (coiled-coil, myosin-like BCL2 interacting protein) |
| 39436_at | BNIP3L | BCL2/adenovirus E1B 19 kDa interacting protein 3-like |
| 40570_at | FOXO1A | forkhead box O1A (rhabdomyosarcoma) |
| 595_at | TNFAIP3 | tumor necrosis factor, alpha-induced protein 3 |
| 641_at | PSEN1 | presenilin 1 (Alzheimer disease 3) |

Our analysis of the ALL microarray dataset clearly shows the limitations of binary gene-parameter profiles of differential expression indicators, which are still the norm for combined GO annotation and microarray data analyses. Our proposed statistical framework, with general definitions for the gene-annotation and gene-parameter profiles, allows consideration of a much broader class of inference problems, that extend beyond GO annotation and microarray data analysis. Gene-annotation profiles may be continuous or polychotomous and may correspond, for example, to exon/intron counts/lengths/nucleotide distributions, gene pathway membership, or gene regulation by particular transcription factors. Likewise, gene-parameter profiles may be continuous or polychotomous and may correspond, for example, to regression coefficients relating possibly censored biological and clinical outcomes to genome-wide transcript levels, DNA copy numbers, and other covariates.

This first application of our proposed methodology only considered control of the family-wise error rate using single-step common-cut-off maxT Procedure 1, based on the non-parametric bootstrap null shift and scale-transformed test statistics null distribution of Procedure 2. Adjusted *p*-values tended to be quite large, with only a handful of GO terms identified as being significantly associated with BCR/ABL vs. NEG differential gene expression. Joint augmentation and empirical Bayes procedures could be used for control of a broader and more biologically relevant class of Type I error rates, defined as generalized tail probabilities, $gTP(q,g) = \Pr(g(V_n, R_n) > q)$, and generalized expected values, $gEV(g) = \mathrm{E}[g(V_n, R_n)]$, for arbitrary functions $g(V_n, R_n)$ of the numbers of false positives $V_n$ and rejected hypotheses $R_n$ (Chapters 6 and 7 in [14], [15, 39, 40]). Error rates based on the proportion $V_n/R_n$ of false positives (e.g., TPPFP and FDR) are especially appealing for large-scale testing problems, compared to error rates based on the number $V_n$ of false positives (e.g., gFWER), as they do not increase exponentially with the number $M$ of tested hypotheses. More powerful analyses may also be achieved with the new null quantile-transformed test statistics null distribution of [42]. The multiple testing methodology developed in [14] and related articles is particularly well-suited to handle the variety of parameters of interest and the complex and unknown dependence structures among test statistics (e.g., implied by the DAG structure of GO terms) that are likely to be encountered in high-dimensional inference problems in biomedical and genomic research.

TABLE 12. *GO terms associated with differential gene expression between BCR/ABL and NEG B-cell ALL, MF GO term* `GO:0003735`. This table lists genes directly or indirectly annotated with GO term *structural constituent of ribosome* (out of $G = 2,071$ genes, `GOALLLOCUSID` environment in GO package). The term *structural constituent of ribosome* (`GO:0003735`) has the smallest adjusted $p$-value for Scenario MT$[t,t]$ applied to the MF gene ontology (Table 10).

| | MF `GO:0003735` | |
|---|---|---|
| **Probe ID** | **Symbol** | **Name** |
| `2016_s_at` | `RPL10` | `ribosomal protein L10` |
| `31511_at` | `RPS9` | `ribosomal protein S9` |
| `31546_at` | `RPL18` | `ribosomal protein L18` |
| `31955_at` | `FAU` | `Finkel-Biskis-Reilly murine sarcoma virus (FBR-MuSV)` |
| | | `ubiquitously expressed (fox derived)` |
| `32221_at` | `MRPS18B` | `mitochondrial ribosomal protein S18B` |
| `32315_at` | `RPS24` | `ribosomal protein S24` |
| `32394_s_at` | `RPL23` | `ribosomal protein L23` |
| `32433_at` | `RPL15` | `ribosomal protein L15` |
| `32437_at` | `RPS5` | `ribosomal protein S5` |
| `33117_r_at` | `RPS12` | `ribosomal protein S12` |
| `33485_at` | `RPL4` | `ribosomal protein L4` |
| `33614_at` | `RPL18A` | `ribosomal protein L18a` |
| `33661_at` | `RPL5` | `ribosomal protein L5` |
| `33668_at` | `RPL12` | `ribosomal protein L12` |
| `33674_at` | `RPL29` | `ribosomal protein L29` |
| `34316_at` | `RPS15A` | `ribosomal protein S15a` |
| `36358_at` | `RPL9` | `ribosomal protein L9` |
| `36572_r_at` | `ARL6IP` | `ADP-ribosylation factor-like 6 interacting protein` |
| `36786_at` | `RPL10A` | `ribosomal protein L10a` |
| `39856_at` | `RPL36AL` | `ribosomal protein L36a-like` |
| `39916_r_at` | `RPS15` | `ribosomal protein S15` |
| `41152_f_at` | `RPL36A` | `ribosomal protein L36a` |
| `41214_at` | `RPS4Y1` | `ribosomal protein S4, Y-linked 1` |
| `41746_at` | `NHP2L1` | `NHP2 non-histone chromosome protein 2-like 1` |
| | | `(S. cerevisiae)` |

Note that for asymptotic results, such as consistency or asymptotic linearity, the sample size $n$ refers to the number of observational units sampled from the population of interest to estimate the gene-parameter profiles, e.g., the number of patients in a cancer microarray study. While the sample size $n$ is typically much smaller than the dimension $J$ of the data structure $X$, sample sizes have considerably increased in recent genomic applications. In addition, simulation studies have indicated that our proposed MTPs have good finite sample properties in terms of both Type I error control and power.

Ongoing efforts include consideration of more general and biologically pertinent multivariate association measures $\rho$. For instance, for GO annotation metadata, the association parameter for a given GO term could take into account the structure of the DAG by considering the gene-annotation profiles of offspring or ancestor terms. We are also interested in developing better numerical and graphical approaches for representing and interpreting the multiple testing results, e.g., the lists of GO terms and associated adjusted $p$-values. Finally, we are planning on implementing the proposed methods in an R package to be released as part of the Bioconductor Project.

**Software and website companion**

The multiple testing procedures proposed in [14] and related articles [8, 15, 16, 31, 32, 33, 34, 39, 40, 41, 42] are implemented in the R package multtest, released as part of the Bioconductor Project, an open-source software project for the analysis of biomedical and genomic data ([14, Section 13.1]; [32]; www.bioconductor.org).

The experimental data (ALL) and annotation metadata (annaffy, annotate, GO, hgu95av2) packages used in the analysis of Section 5 may also be obtained from the Bioconductor Project website.

The website companion to [14] provides additional tables, figures, code, and references: www.stat.berkeley.edu/~sandrine/MTBook.

**Acknowledgments.** We are most grateful to Robert Gentleman (Fred Hutchinson Cancer Research Center) for many stimulating discussions on statistical and computational methods for the analysis of biological annotation metadata. We have also much appreciated Evelyn Camon's (European Bioinformatics Institute) advice regarding the QuickGO browser. Finally, we would like to thank an anonymous referee for constructive comments on an earlier version of this manuscript.